\documentclass{aa}  

\usepackage{natbib}
\bibpunct{(}{)}{;}{a}{}{,} 
\usepackage{graphics}   
\usepackage{graphicx}   
\usepackage{epstopdf}
\usepackage{amsmath} 
\usepackage[varg]{txfonts}
\usepackage{pdflscape}
\usepackage{hyperref}   
\usepackage{longtable}
\usepackage{balance}
\usepackage{color}
\usepackage[normalem]{ulem}

\hypersetup{
    bookmarks=true,         
    colorlinks=true,       
    linkcolor=blue,          
    citecolor=blue,        
    filecolor=blue,      
    urlcolor=blue           
}

\def\ga{\,\,\raise0.14em\hbox{$>$}\kern-0.76em\lower0.28em\hbox{$\sim$}\,\,}
\def\la{\,\,\raise0.14em\hbox{$<$}\kern-0.76em\lower0.28em\hbox{$\sim$}\,\,}

\def\iso#1{$^{#1}$}
\def\Msun{$M_{\odot}$}

\def\msun{M_{\odot}}

\def\cm3{cm$^{-3}$}

\def\chem#1#2{$\mathrm{^{#2}\kern-0.8pt#1}$}
\def\reac#1#2#3#4#5#6{$\mathrm{\, ^{#2}\kern-0.8pt{#1}\, ({#3}\, ,{#4})\, {}^{#6}\kern-0.8pt{#5}\, }$}

\def\nuc{\mathrm{nuc}}

\def\be{\begin{equation}} 
\def\ee{\end{equation}}
\def\beqy{\begin{eqnarray}}
\def\eeqy{\end{eqnarray}}
\def\bmlet{\begin{mathletters}}
\def\emlet{\end{mathletters}}

\begin{document}

\title{The intermediate neutron capture process}
\subtitle{III. The i-process in AGB stars of different masses and metallicities without overshoot}

\author{A. Choplin   
\and L. Siess
\and S. Goriely}
\offprints{arthur.choplin@ulb.be}

\institute{
Institut d'Astronomie et d'Astrophysique, Universit\'e Libre de Bruxelles,  CP 226, B-1050 Brussels, Belgium
}

\date{Received --; accepted --}

\abstract
{
Alongside the slow (s) and rapid (r) neutron capture processes, an intermediate neutron capture process (i-process) is thought to exist. It happens when protons are mixed in a convective helium-burning zone, and is referred to as proton ingestion event (PIE), however, the astrophysical site of the i-process is still a matter of debate. The asymptotic giant branch (AGB) phase of low-mass low-metallicity stars is among the promising sites in this regard. 
}
{
For the first time, we provide i-process yields of a grid of AGB stars experiencing PIEs. 
}
{
We computed 12 models with initial masses of 1, 2, and 3~$M_{\odot}$ and metallicities of [Fe/H]~$=-3.0$, $-2.5$ $-2.3,$ and $-2.0, $  with the stellar evolution code {\sf STAREVOL}. 
We used a nuclear network of 1160 species at maximum, coupled to the chemical transport equations. These simulations do not include any extra mixing process.
}
{
Proton ingestion takes place preferentially in low-mass and low-metallicity models, arising in six out of our 12 AGB models: the $1$~\Msun\, models with [Fe/H]~$=-3$, $-3,$  and $\alpha-$enhancement, 
$-2.5$, $-2.3,$ and the $2$~\Msun\, models with [Fe/H]~$=-3$ and $-2.5$. 
These models experience i-process nucleosynthesis characterized by neutron densities of $\simeq 10^{14} -10^{15}$~cm$^{-3}$. Depending on the PIE properties two different evolution paths follow: either the stellar envelope is quickly lost and no more thermal pulses develop or the AGB phase resumes with  additional thermal pulses. This behaviour critically depends on the pulse number when the PIE occurs, the mass of the ingested protons, and the extent to which the pulse material is diluted in the convective envelope. 
We show that the surface enrichment after a PIE is a robust feature of our models and it persists under various convective assumptions. 
In our i-process models, elements above iodine ($Z=53$) are the most overproduced, particularly Xe, Yb, Ta, Pb, and Bi. 
Our 3~\Msun\, models do not experience any i-process, but instead go through a convective s-process in the thermal pulse with a clear signature on their yields. 
}
{
 Thus, AGB stars at low-mass and low-metallicity are expected to contribute to the chemical evolution of heavy elements through the s- and i-processes. 
Our models can synthesise heavy elements up to Pb without any parametrized extra mixing process such as overshoot or inclusion of a $^{13}$C-pocket. 
Nevertheless, it remains to be explored how the i-process depends on mixing processes, such as overshoot, thermohaline, or rotation.}


\titlerunning{Development of the i-process in low-metallicity low-mass AGB stars}

\authorrunning{A. Choplin et al. }

\maketitle


\section{Introduction}
\label{sect:intro}

\begin{table*}[t]
\scriptsize{
\caption{Main characteristics of the models computed in this work, with the initial mass $M_{\rm ini}$, [Fe/H] ratio, initial metallicity in mass fraction $Z$, initial chemical composition (solar-scaled or $\alpha$-enhanced), total lifetime $\tau_{\rm star}$, duration of the AGB phase $\tau_{\rm AGB}$,  final mass $M_{\rm fin}$, and final mass of the convective envelope $M^{\rm env}_{\rm fin}$, number of pulses $N_{\rm pulse}$, along with information on whether the model experiences a PIE or not and the pulse number during which the PIE arises.
\label{table:1}
}
\begin{center}
\resizebox{16.5cm}{!} {
\begin{tabular}{lccccccccccc} 
\hline
Model & $M_{\rm ini}$  & [Fe/H] &$Z$  & Initial & $\tau_{\rm star}$ & $\tau_{\rm AGB}$ & $M_{\rm fin}$ &   $M^{\rm env}_{\rm fin}$             & $N_{\rm pulse}$ & PIE  &  PIE     \\
 label  &    [\Msun]        &        & & composition  &   [$10^9$~yr] & [$10^6$~yr] &   [\Msun] & [\Msun] &    &      &    PULSE                        \\
\hline
M1.0z3.0   &   1.0  & $-3.0$     &  $1.4\times 10^{-5}$  &  Solar      &   6.30    &   2.96   &    0.84      &           0.04             & 42  &  YES & 1    \\
M1.0z3.0$\alpha$   &  1.0   & $-3.0$    &  $6.4\times 10^{-5}$    & $\alpha$-enhanced &  6.30     &   0.39   &   0.55  & 0.02   &  2 & YES & 2       \\
M1.0z2.5   &  1.0    & $-2.5$    &  $4.3\times 10^{-5}$  & Solar &  6.28    &   0.66   & 0.53   & 0.01                        & 2  & YES & 2      \\
M1.0z2.3   &  1.0   & $-2.3$    &  $6.8\times 10^{-5}$   & Solar &  6.31    &   0.56   & 0.56  & 0.03                         &  2 & YES &  2  \\
M1.0z2.0   &  1.0  & $-2.0$     &   $1.4\times 10^{-4}$  & Solar &   6.34   &   2.24   & 0.69  & 0.01                         & 16  & NO  & $-$      \\
\hline
M2.0z3.0   &  2.0  &  $-3.0$    &  $1.4\times 10^{-5}$  & Solar &  0.75    &   1.09   & 0.76  & 0.04                  & 16 & YES & 2       \\
M2.0z3.0$\alpha$   &  2.0  &  $-3.0$    &  $6.4\times 10^{-5}$  & $\alpha$-enhanced &  0.76   &   1.94   & 0.73  &    0.02               &   20  & NO & $-$       \\
M2.0z2.5   & 2.0  &  $-2.5$    &  $4.3\times 10^{-5}$  &  Solar &  0.77    &   0.87   & 0.73 & 0.02                        & 12 & YES & 2     \\
M2.0z2.0   &   2.0  &  $-2.0$   &   $1.4\times 10^{-4}$  & Solar &  0.76    &  1.70    &  0.74 & 0.03                         &  19 & NO & $-$     \\
\hline
M3.0z3.0   &  3.0   &  $-3.0$    &  $1.4\times 10^{-5}$  & Solar &   0.25   & 0.61     &0.94 & 0.10                         &  23  & NO & $-$   \\
M3.0z2.5   &   3.0    &$-2.5$    &  $4.3\times 10^{-5}$  & Solar &  0.26    &   0.69   & 0.84 & 0.02                         &  25 & NO & $-$       \\
M3.0z2.0   & 3.0     & $-2.0$    &  $1.4\times 10^{-4}$  & Solar &   0.27   &   0.74   & 1.01  & 0.18                         & 25  & NO & $-$     \\
\hline 
\end{tabular}
}
\end{center}
}
\end{table*}


Exploring the origins of elements heavier than Fe is among the most topical questions in modern astrophysics \citep[e.g.][]{arnould20}. 
We do know that most of them come from neutron capture processes. Furthermore, besides the slow (s) and rapid (r) neutron capture processes, an intermediate (i) neutron capture process \citep[first suggested by][]{cowan77} is expected to operate at neutron densities intermediate between the s- and r-processes.

Different observations support the existence of the i-process. 
The so-called carbon-enhanced metal-poor (CEMP) -r/s stars, whose chemical composition is difficult to reconcile with s- or r-process models, point towards the existence of an i-process \citep[e.g.][]{jonsell06, lugaro12,dardelet14,roederer16,karinkuzhi21,goswami22}. 
Sometimes, the chemical composition is best reproduced if considering a combination of an s- and an i-process \citep{koch19}. 
Another clue suggesting the existence of the i-process is the puzzling Ba overabundance in open cluster stars \citep{mishenina15}. 
Several works have also shown that pre-solar grains may bear the isotopic signature of i-process nucleosynthesis \citep{fujiya13, jadhav13, liu14}.

The i-process is triggered when some hydrogen is mixed in a convective helium-burning zone. 
This mixing process is either referred to as proton ingestion event \citep[PIE, e.g.][]{cristallo09b}, flash-driven mixing \citep[e.g.][]{lau09}, or dual shell flash \citep[e.g.][]{campbell08}. 
In this work, we use the term PIE. 
When a PIE arises, protons are transported down in the convective helium-burning zone and burn on the way via the $^{12}$C($p,\gamma$)$^{13}$N reaction. The decay of $^{13}$N to $^{13}$C (in about 10 min) is followed by  $^{13}$C($\alpha,n$)$^{16}$O, operating mostly at the bottom of the convective helium-burning zone, where the temperature is high ($> 2\times 10^8$K).
Under these conditions, neutron densities as high as $10^{12} - 10^{16}$~cm$^{-3}$ are reached for typically $\sim 1$~yr, triggering an i-process nucleosynthesis. 
Because $^{12}$C($p,\gamma$)$^{13}$N releases a lot of energy in the middle of the convective helium-burning zone, a temperature inversion develops in the 1D models and the convective zone splits.

Proton ingestion events have been studied parametrically using one-zone models, for instance, \cite{hampel16,hampel19}, where the ability of the i-process to reproduce the chemical composition of CEMP-r/s stars is explained.
Nevertheless, the astrophysical site(s) hosting PIEs, hence the i-process, remain(s) actively debated. 
Shortly after \cite{schwarzschild65} first noticed the existence of thermal pulses in asymptotic giant branch (AGB) stars, it has been suggested that PIEs could appear during thermal pulses of these stars \citep[e.g.][]{schwarzschild67, despain76, fujimoto77,scalo79, fujimoto84}, despite the fact that the entropy barrier between the H- and He-rich layers tends to prevent such events from occurring \citep[e.g.][]{iben76}. 
Since then, various astrophysical sites hosting PIEs (i.e. the i-process) were suggested: 
the early AGB phase of metal-poor low-mass stars \citep{cassisi96,fujimoto00,chieffi2001,siess02,iwamoto04,cristallo09b,lau09,suda10,stancliffe11,cristallo16,gilpons18,choplin21,goriely21}, the core helium flash of very low-metallicity low-mass stars \citep{fujimoto90,fujimoto00,schlattl2001,suda10,campbell10, cruz13}, the very late thermal pulses of post-AGB stars \citep{herwig11}, rapidly accreting carbon-oxygen (C-O) or oxygen-neon (O-Ne) white dwarfs (RAWDs) in close binary systems \citep{denissenkov17,denissenkov19,denissenkov21,cote18, stephens21}, super-AGB stars \citep[7~$\msun \lesssim M_{\rm ini} \lesssim$~10~$\msun$,][]{siess2007,jones16b}, 
or the helium shell of very low- or zero-metallicity massive stars \citep[$M_{\rm ini} > $~10~$\msun$,][]{banerjee18, clarkson18, clarkson20}. 

In many of these studies, the i-process nucleosynthesis accompanying a PIE is not investigated and when it is, often just one model is considered. 
Indeed, such models can be difficult to compute and require large nuclear reaction networks.
So far, only \cite{denissenkov19} have produced a grid of seven RAWD models with i-process nucleosynthesis calculated in post-processing.
There is a clear need of computing grids of i-process models from the various possible sites. 

In \citep[][hereafter Paper I]{choplin21} we studied in details the development of the i-process during the AGB phase of a a 1~\Msun\, model at a metallicity of [Fe/H]~$=-2.5$. 
In \citep[][hereafter Paper II]{goriely21} we focused on the nuclear physics uncertainties affecting the i-process during the AGB phase.
In this third paper, we investigate the evolution and nucleosynthesis of a grid of 12 AGB models computed with various initial masses and metallicities. 
We highlight and explain the different evolutions, discuss the different nucleosynthesis processes at work, with a special emphasis on the i-process. 

Section~\ref{sect:inputs} presents the input physics. Section~\ref{sect:evol} discusses structure and evolution aspects, while Sect.~\ref{sect:nucleo} focuses on nucleosynthesis, surface enrichment, and stellar yields. Our summary and conclusions are given in Sect.~\ref{sect:concl}.


\begin{table*}[t]
\scriptsize{
\caption{
Important characteristics of our models experiencing a PIE. Given are the model label (column 1) initial mass $M_{\rm ini}$ (column 2), [Fe/H] ratio (column 3), the maximal neutron density $N_{\rm n,max}$ (column 4), the neutron exposure at the bottom of the convective pulse during the PIE $\tau_{\rm bot}$ (column 5), the mean neutron exposure in the pulse during the PIE $\langle \tau \rangle$ (Eq.~\ref{eq:meanexp}, column 6), the mass of hydrogen engulfed from the start of the PIE to the point where the neutron density is maximum $M^{\rm NEUT}_{\rm H}$ (Eq.~\ref{eq:mpie}, column 7), the mass of hydrogen engulfed from the start of the PIE to the splitting of the convective zone $M^{\rm SPLIT}_{\rm H}$ (column 8), whether the convective pulse splits or not (column 9), and the enrichment ratio $e$ (Eq.~\ref{eq:eratio}, column 10).
\label{table:2}
}
\begin{center}
\resizebox{14.5cm}{!} {
\begin{tabular}{lccccccccc} 
\hline
Model & $M_{\rm ini}$  & [Fe/H] & $\log(N_{\rm n,max}$) & $\tau_{\rm bot}$ &  $\langle \tau \rangle$    & $M^{\rm NEUT}_{\rm H}$ & $M^{\rm SPLIT}_{\rm H}$ & splitting & $e$  \\
 label  &    [\Msun]        &               &                                     &    mbarn$^{-1}$       &    mbarn$^{-1}$       & [$10^{-5} \, M_{\odot}$]  & [$10^{-5} \, M_{\odot}$]     &    &   \\
\hline
M1.0z3.0              &   1.0  & $-3.0$     &   14.67 &  137  &  5.50  &  0.27  & $-$ &NO  & $-$  \\
M1.0z3.0$\alpha$ &   1.0 & $-3.0$     &   15.27 &  190  &  5.39  &  0.64  & 26.4  & YES & 68 \\
M1.0z2.5              &  1.0   & $-2.5$     &   15.31 & 218   &  6.46  &  2.82  & 34.4  & YES& 74 \\
M1.0z2.3              &  1.0   & $-2.3$     &  15.34  & 121   &  3.60  &  0.67  & 24.1  & YES& 53 \\
\hline 
M2.0z3.0              &  2.0   &  $-3.0$    &   13.83 & 21.4  & 1.01  &   1.42 & 1.83 & YES&  503  \\
M2.0z2.5              & 2.0    &  $-2.5$    &  15.16  & 164   & 6.48  &  1.68 &  32.1 & YES& 101   \\
\hline 
\end{tabular}
}
\end{center}
}
\end{table*}



\section{Physical ingredients, numerical aspects, and important quantities}
\label{sect:inputs}

The models presented in this paper are computed with the stellar evolution code  {\sf STAREVOL} \citep[][and references therein]{siess00, siess06, goriely18c}. 
We computed models with initial masses of 1, 2, 3~\Msun. For each mass, we investigated the metallicities of [Fe/H]~$=-3$, $-2.5$ and $-2$, corresponding to $Z=1.4\times 10^{-5}$, $4.3\times 10^{-5}$,  and $1.4\times 10^{-4}$ in mass fractions, respectively. 
We also computed a model of 1~\Msun~at [Fe/H]~$=-2.3$ ($Z=6.8\times 10^{-5}$). 
These ten models were computed with the solar mixture of \cite{asplund09}. 
To investigate the effect of an $\alpha$-enhancement, we computed a 1~\Msun~and a 2~\Msun~model at [Fe/H]~$=-3$ with an $\alpha$-enhanced mixture. 
We raised the initial abundance of the isotopes $^{12}$C, $^{16}$O, $^{20}$Ne, $^{24}$Mg, $^{28}$Si, and $^{32}$S following \cite{ritter18b}. 
Their Table~1 reports the mass fractions of $\alpha$-enhanced isotopes according to halo and disc stars \citep[][and references therein]{reddy06}. For the elements Ne and S, they refer to \cite{kobayashi06}. 
In the end, our $\alpha$-enhanced models have a metallicity $Z=6.4\times 10^{-5}$ in mass fraction (i.e. similar to the [Fe/H]~$-2.3$ model). 

Other input physics are the same as in Paper I and II. 
Especially, we use the mass-loss rate from \cite{reimers75} from the main sequence up to the beginning of the AGB and then switch to the \cite{vassiliadis93} prescription.
The opacity change due to the formation of molecules when the star becomes carbon rich is taken into account \citep{marigo02}. 
The mixing length parameter $\alpha$ is set to 1.75 and we do not consider extra mixing (e.g. overshoot, thermohaline). 
Table~\ref{table:1} summarizes the main characteristics of the models computed.

\subsection{The coupling of diffusion and nucleosynthesis}
\label{sect:coupling}
Particular attention is paid to the equation describing the abundances change of chemicals. 
The abundance $X_i$ of a nucleus $i$ is followed by the equation:
\begin{equation}
\frac{\partial X_i}{\partial t}= \frac{\partial}{\partial m_r}  \left[\,(4 \pi r^2 \rho)^2\,D\ \frac{\partial X_i}{\partial m_r}\right] + \frac{\partial X_i}{\partial t}\bigg|_\nuc \ ,
\label{eq_dif}
\end{equation}
where $D$ is the diffusion coefficient associated with convection only, since other mixing mechanisms are not included here. 

The first and second terms in the right hand side of Eq.~\ref{eq_dif} account for the changes due to transport and nuclear burning, respectively.
In general these two terms are treated separately, one after the other. 
However, during a PIE, the mixing timescale becomes comparable to the nuclear timescale, so that dealing with the transport and the nuclear burning independently may no longer be valid. 
Indeed, a chemical (especially protons) may be transported in a zone of the star where it should not exist since it would be burnt before reaching this zone.
For this reason, it is necessary to fully couple the transport and nuclear burning. 
In this series of papers, during the PIE, the nucleosynthesis and transport equations are solved simultaneously once the structure has converged. 
This method requires many inversions of the Jacobian matrix of size ($K \times N$) $\times$ ($K \times N$), where $K$ is the number of nuclei and $N$ the number of shells. 
In our cases, it corresponds to matrices of about ($ 2 \times 10^6$) $\times$ ($ 2 \times 10^6$), which leads to a total of more than $10^{12}$ matrix elements. Fortunately, these matrices are very sparse. The fill factor can be defined as:
\begin{equation}
\eta = \frac{N_{\rm non\_zero}}{(K \times N)^2}
,\end{equation}
where $N_{\rm non\_zero}$ is the number of non zero value in the matrix and $(K \times N)^2$ the total number of elements. In our models, $\eta \sim 2 \times 10^{-6}$. 
The sparsity of the matrix makes it interesting to use the compressed sparse row format, which reduces the matrix information into two one-dimensional (1D) arrays of size $N_{\rm non\_zero}$ and one 1D array of size $K\times N + 1$. To solve the system of abundance equations (Eq.~\ref{eq_dif}), we use the PARDISO solver\footnote{http://pardiso-project.org/} \citep{schenk04}.

\subsection{Convection}
\label{sect:convMLT}

In a convective zone and in the present work, the coefficient $D$ in Eq.~\ref{eq_dif} is equal to the convective diffusion coefficient $D_{\rm conv}$. 
This coefficient is computed according to the mixing-length theory and can be expressed as:
\begin{equation}
D_{\rm conv} = \alpha_{\rm mlt}\frac{v_{\rm conv} H_P}{3}
\label{eq:dconv}
,\end{equation}
where $\alpha_{\rm mlt}$ is the mixing length parameter, $H_P$ the pressure scale height, and $v_{\rm conv}$ the convective velocity. The convective turnover timescale $\tau_{\rm conv}$ is defined as:
\begin{equation}
\tau_{\rm conv} = \int_{r_1}^{r_2} \frac{1}{v_{\rm conv} (r)} \, dr
\label{eq:tconv}
,\end{equation}
with $r_1$ and $r_2$ the boundaries of the convective region.
The turnover timescale expresses the approximate time it takes for a cell of matter to travel across the entire convection zone.

\subsection{Nuclear reaction network}
\label{sect:network}

The {\sf STAREVOL} stellar evolution code includes two reaction networks made of either 411 or 1160 nuclei. The largest network includes all species with a half-life greater than about 1~second and the 2123 nuclear reactions ($n$-, $p$-, $\alpha$-captures and $\alpha$-decays), weak (electron captures, $\beta$-decays), and electromagnetic interactions of relevance to properly follow neutron capture processes up to neutron densities of $\sim 10^{17}$~cm$^{-3}$.  
In our models, as soon as the neutron density exceeds $10^{13}$~cm$^{-3}$, the large network is adopted.

Nuclear reaction rates are taken from the Nuclear Astrophysics Library of the Universit\'e Libre de Bruxelles\footnote{available at http://www.astro.ulb.ac.be/Bruslib} \citep{arnould06}. 
The latest experimental rates are extracted through the NETGEN interface tool \citep{Xu13}. When not available experimentally, the Maxwellian-averaged cross sections are calculated with the {\sf TALYS} reaction code \citep{Koning12,Goriely08a}. 
Additional details can be found in Papers I and II.

\subsection{Neutron exposure}
\label{sect:expo}

Besides the neutron density, a useful indicator to quantify the efficiency of a neutron capture process is the neutron exposure. We define below some quantities that are relevant for the present work.   
The neutron exposure $\tau$ at a given mass coordinate $M_{\rm r}$, between times $t_1$ and $t_2$ is defined as

\begin{equation}
\tau \, (M_{\rm r}) = \int_{t_1}^{t_2}  N_{\rm n}(t) \, v_{\rm T}(t) \, \text{d} t
\label{eq:exposure}
,\end{equation}
where $N_{\rm n}$ is the neutron density and $v_{\rm T} = \sqrt{ \, 2 \, k_B \, T(t) /m_{\rm n}}$ the neutron thermal velocity with $k_B$ the Boltzmann constant, $T(t)$ the temperature at time, $t,$ and $m_n$ the neutron mass. Since the neutron density $N_{\rm n}$ varies with time and mass coordinate in the model, it is also interesting to define the neutron exposure averaged over the convective pulse and over the duration of the PIE:

\begin{equation}
\langle \tau \rangle = \int_{t_1}^{t_2} \langle \, N_{\rm n}(t) \, v_{\rm T}(t)  \rangle_{M} \, \text{d} t
\label{eq:meanexp}
,\end{equation}
where the neutron density times thermal velocity averaged over the mass of the convective pulse, at time, $t$, is defined as:
\begin{equation}
\langle N_{\rm n}(t) \, v_{\rm T} (t) \rangle_{M} = \frac{1}{M_{\rm r,2 }(t)-M_{\rm r,1}(t)} \int_{M_{\rm r,1}(t)}^{M_{\rm r,2}(t)} N_n (M_{\rm r}) \, v_{\rm T} (M_{\rm r}) \, \text{d}M_{\rm r}
\label{meandens}
.\end{equation}
with $M_{\rm r,1} (t)$ and $M_{\rm r,2} (t)$ the mass coordinate boundaries of the convective pulse at time, $t$, $N_n(M_{\rm r}),$ and $v_T(M_{\rm r})$ the neutron density and thermal velocity at mass coordinate, $M_{\rm r}$, respectively. 
We also define $\tau_{\rm bot}$, which is the neutron exposure at the bottom of the convective pulse, where the neutron density is the highest. It is expressed as Eq.~\ref{eq:exposure} but $M_{\rm r}$ can vary so as to follow the bottom of the convective pulse. 
In this case, the first temporal boundary $t_1$ in Eq.~\ref{eq:exposure} corresponds to the beginning of the PIE. The second temporal boundary $t_2$ is chosen to be either just before the split, if any, or at the end of the PIE, when the maximal neutron density in the star has dropped below $\sim 10^{10}$~cm$^{-3}$.
These quantities are given in Table~\ref{table:2}.

 \begin{figure}[h!]
\includegraphics[scale=0.47, trim = 0cm 1.0cm 0cm 0cm]{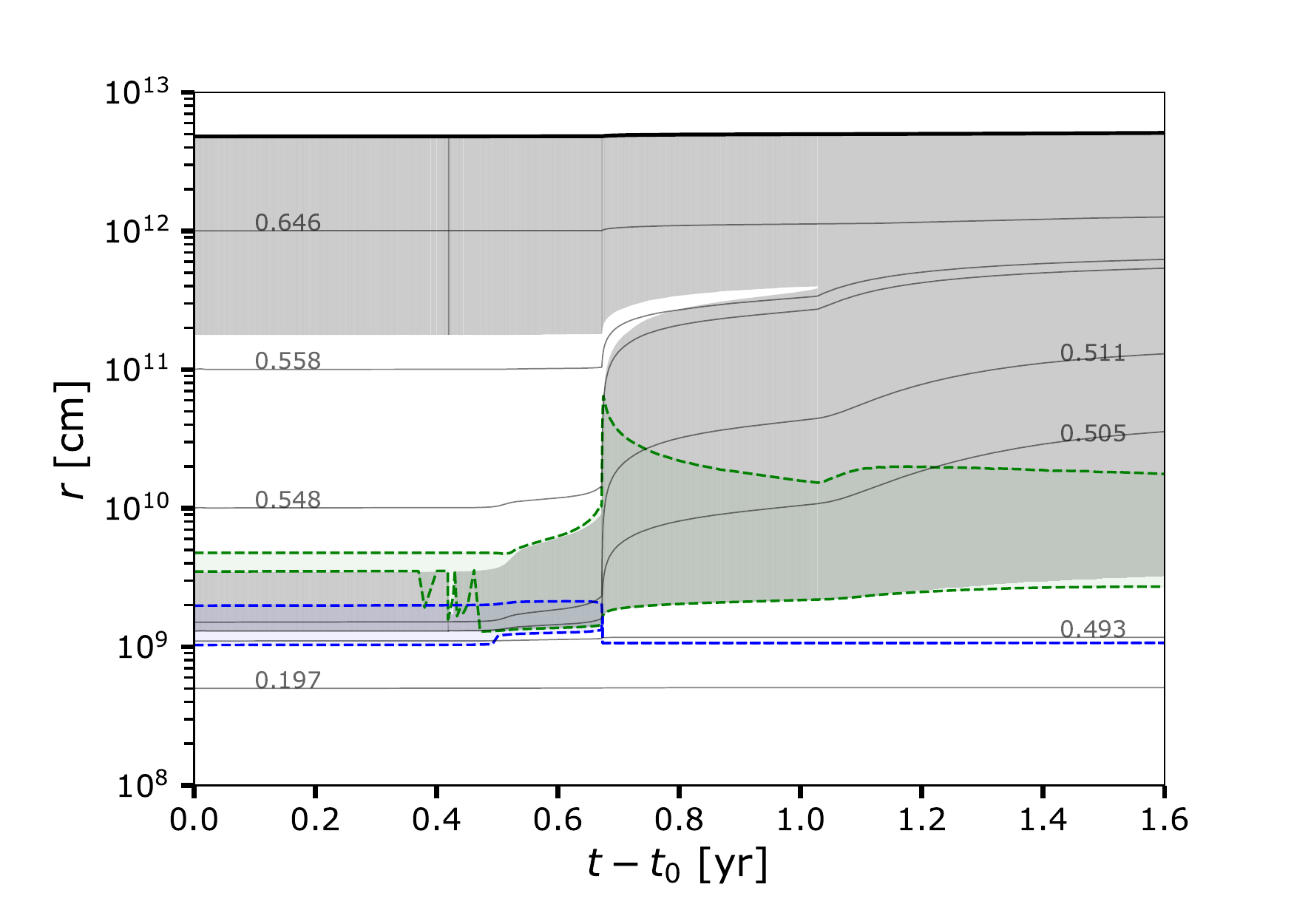}
\includegraphics[scale=0.47, trim = 0cm 0.4cm 0cm 0.8cm]{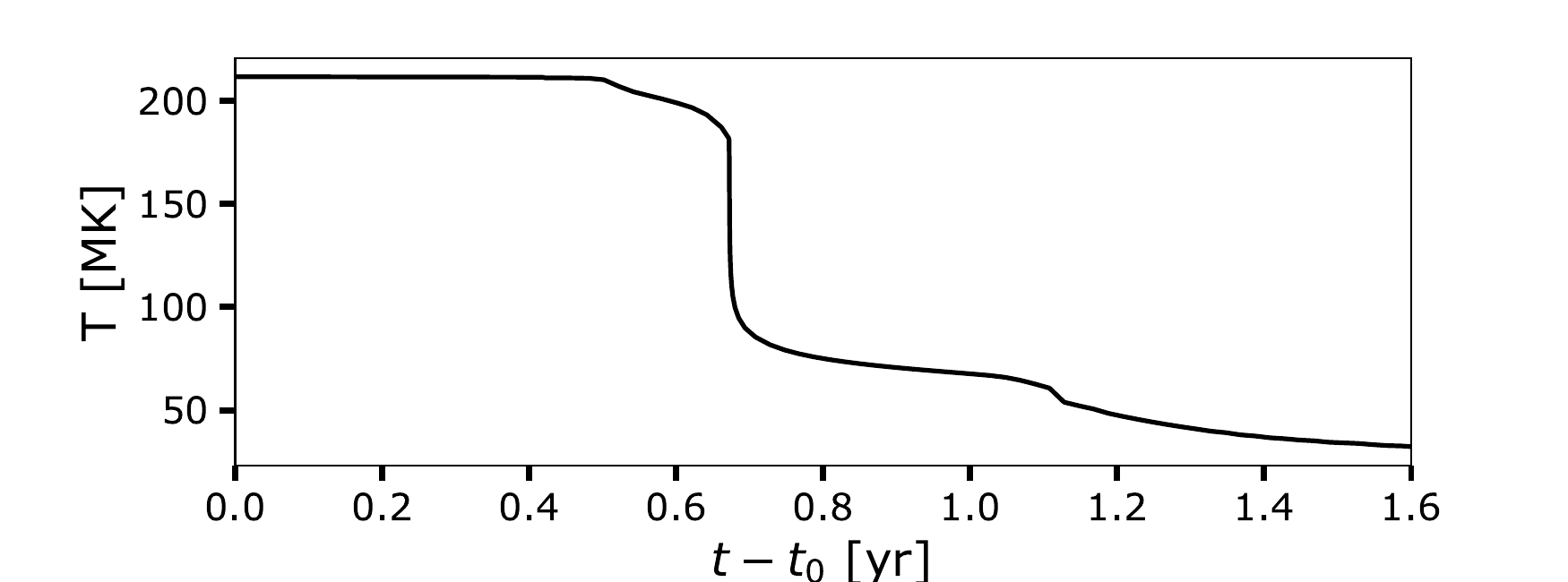}
\includegraphics[scale=0.47, trim = 0cm 1.1cm 0cm 0.8cm]{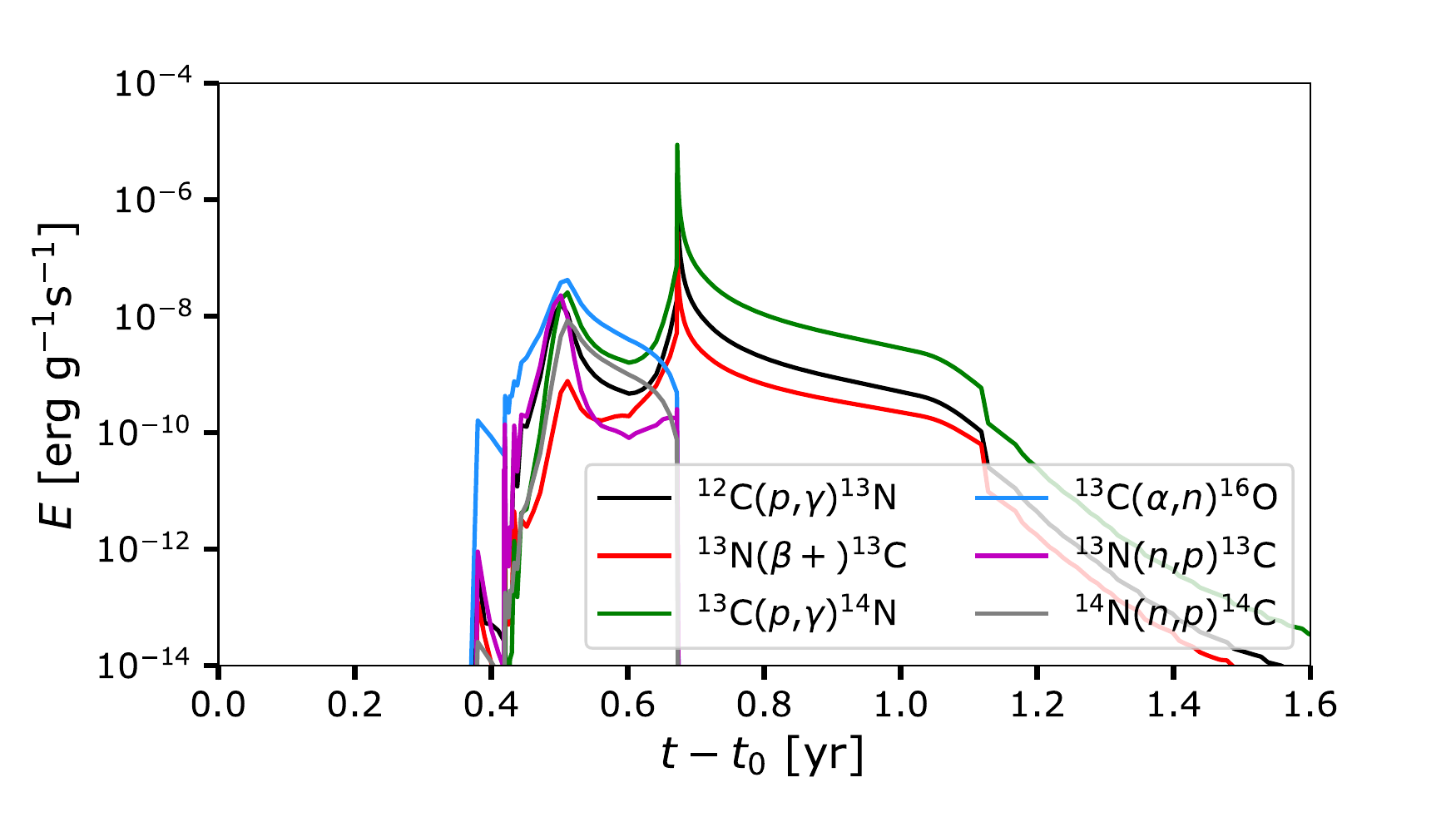}
\includegraphics[scale=0.47, trim = 0cm 0.7cm 0cm 0.8cm]{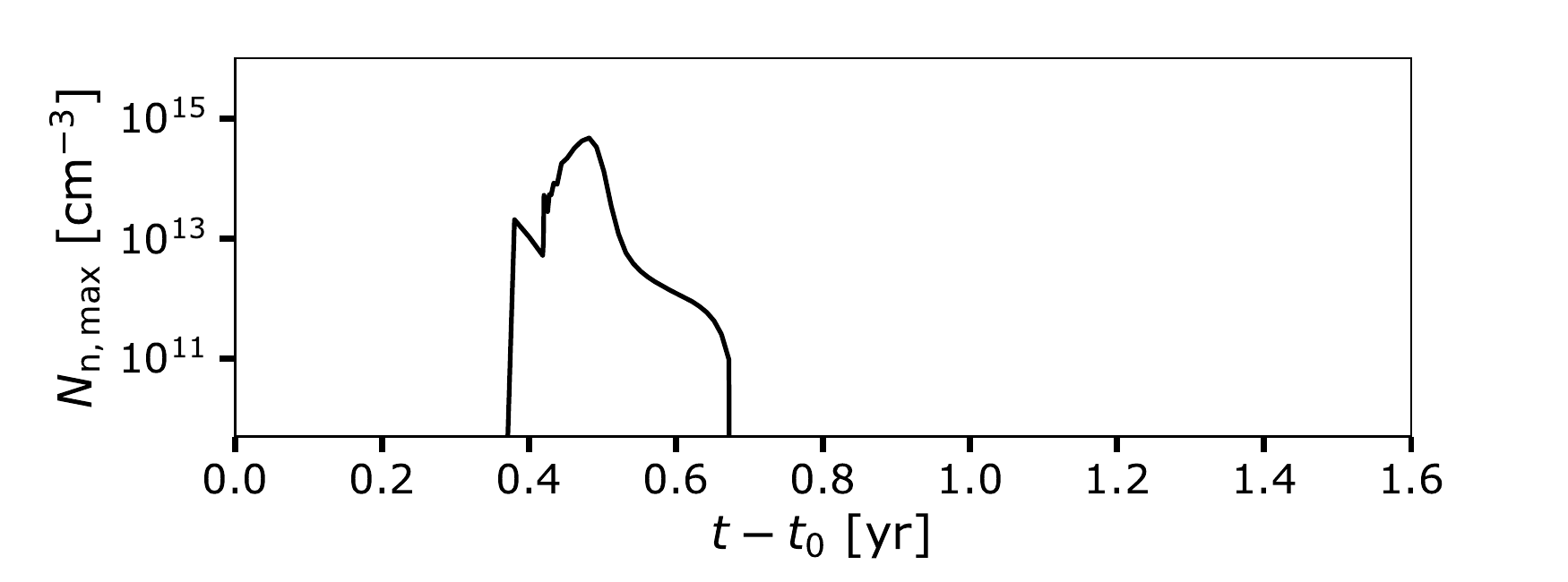}
\includegraphics[scale=0.47, trim = 0cm 0cm 0cm 0.8cm]{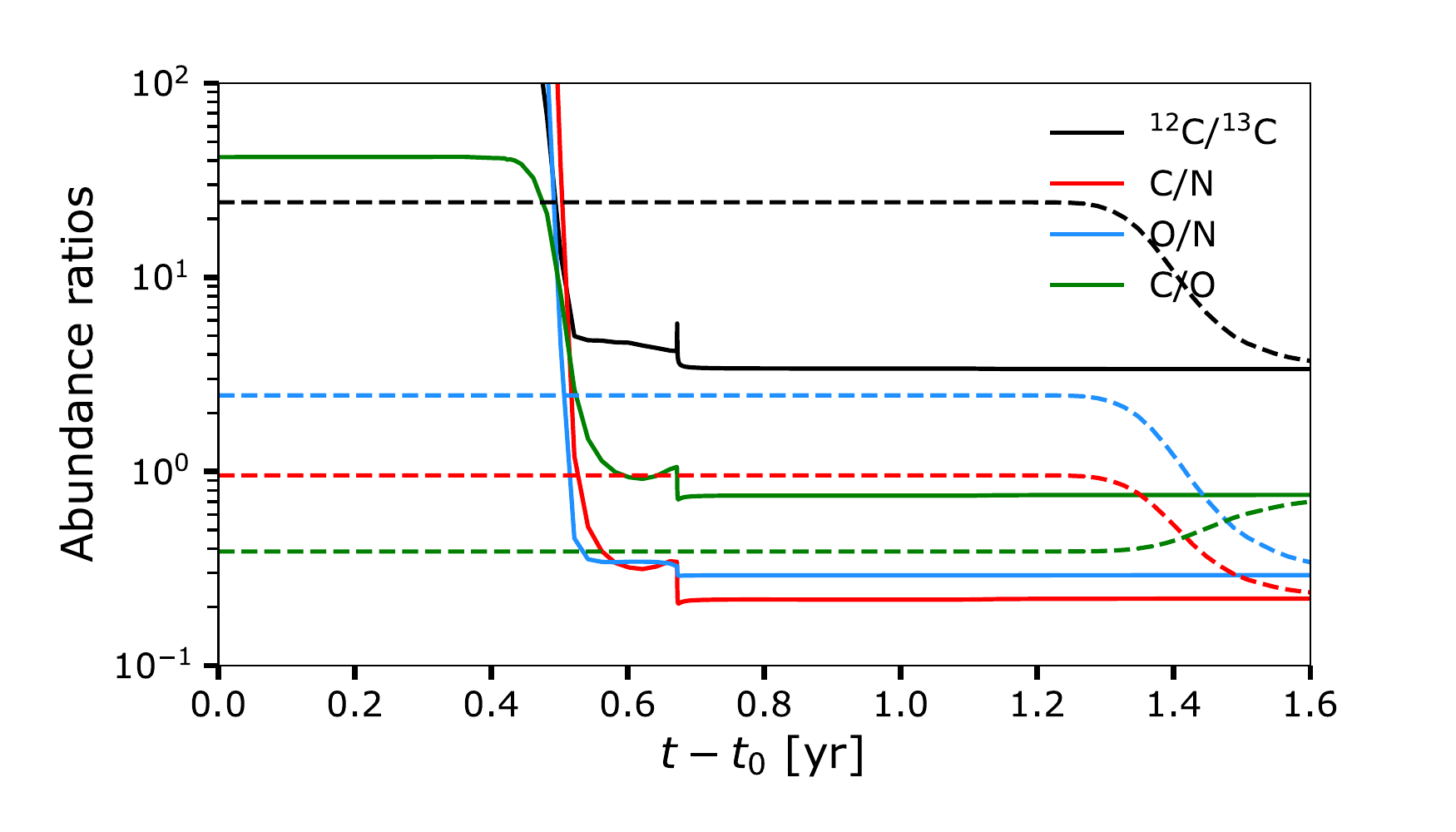}
\caption{ Evolution of the M1.0z3.0 model during the PIE: 
Kippenhahn diagram (top panel), temperature at the bottom of the convective thermal pulse (or envelope for $t-t_0 \gtrsim 1$~yr, second panel), energy released by the main reactions at the bottom of the pulse (or envelope, third panel), maximal neutron density (fourth panel). 
The bottom panel shows the $^{12}$C/$^{13}$C, C/N, and O/N ratios at the bottom of the convective zone (solid lines) and at the stellar surface (dashed lines). 
The dashed green and blue lines in the Kippenhahn diagram (top panel) delineate the hydrogen and helium-burning zones (where the nuclear energy production by H- and He-burning exceeds 10~erg~g$^{-1}$~s$^{-1}$). The thin light-grey lines indicate iso-masses expressed in \Msun.
}
\label{fig:ratio_bot}
\end{figure}

 \begin{figure*}[h!]
  \begin{minipage}[c]{2\columnwidth}
\includegraphics[width=0.34\columnwidth]{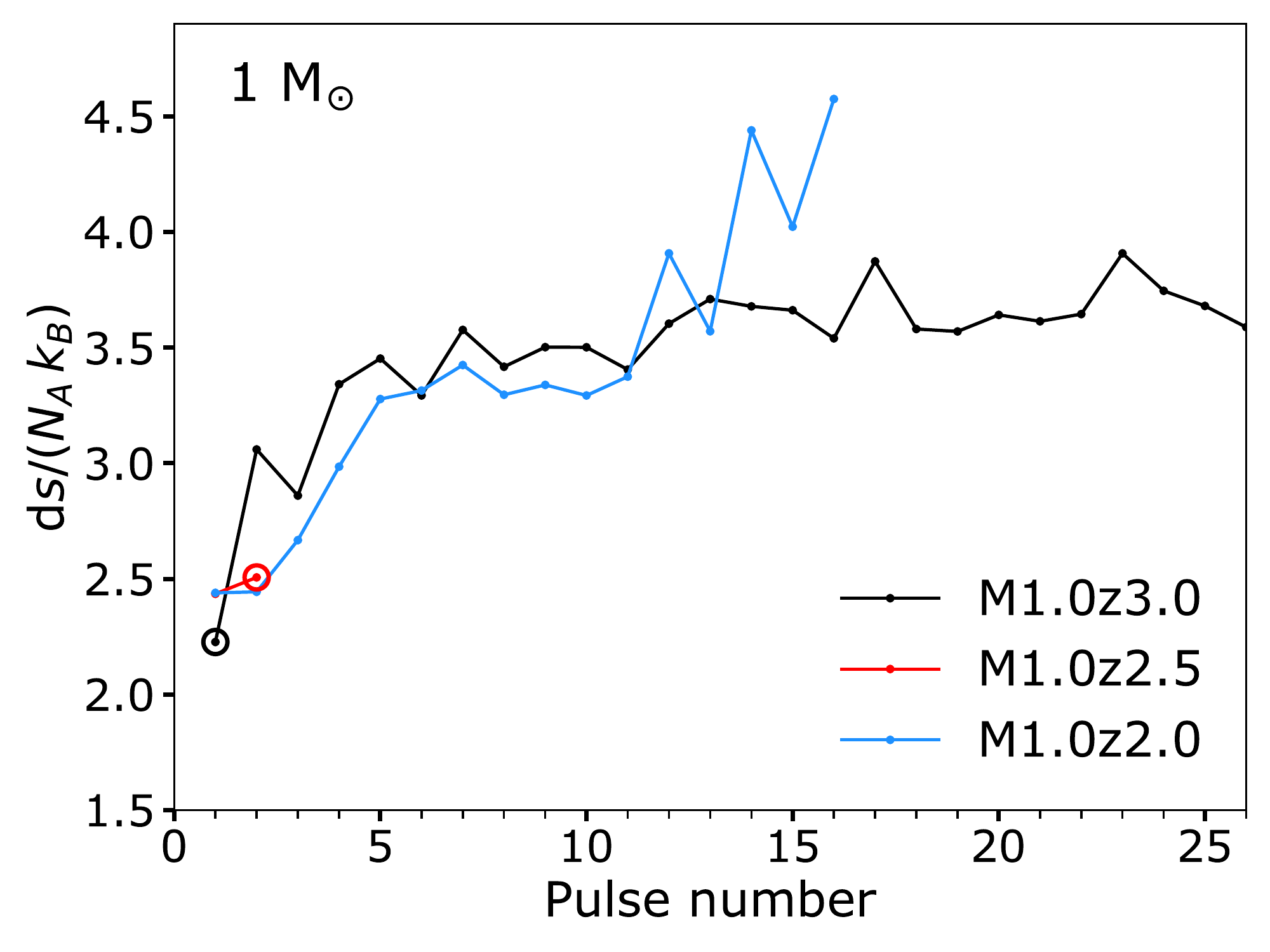}
\includegraphics[width=0.34\columnwidth]{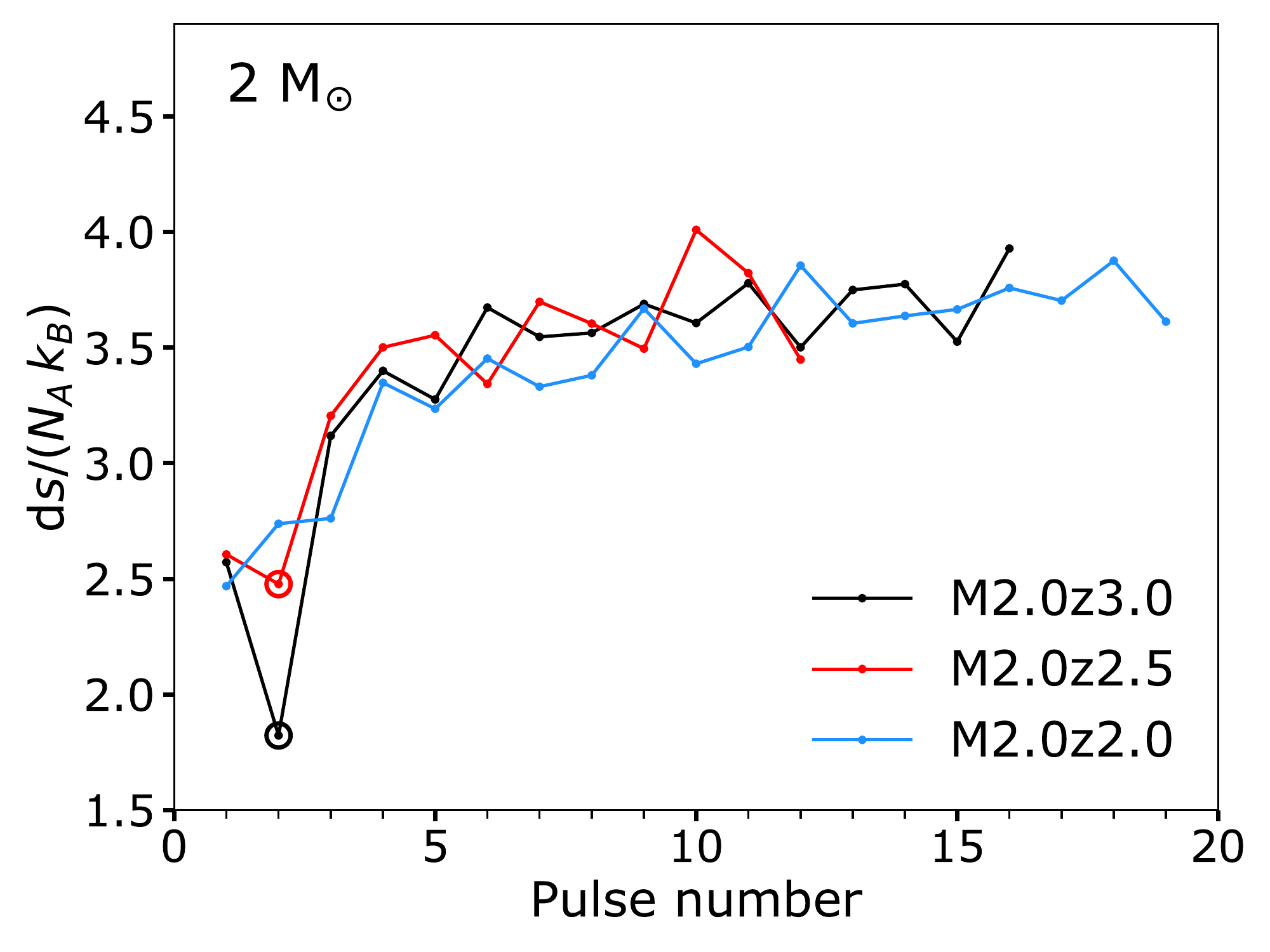}
\includegraphics[width=0.34\columnwidth]{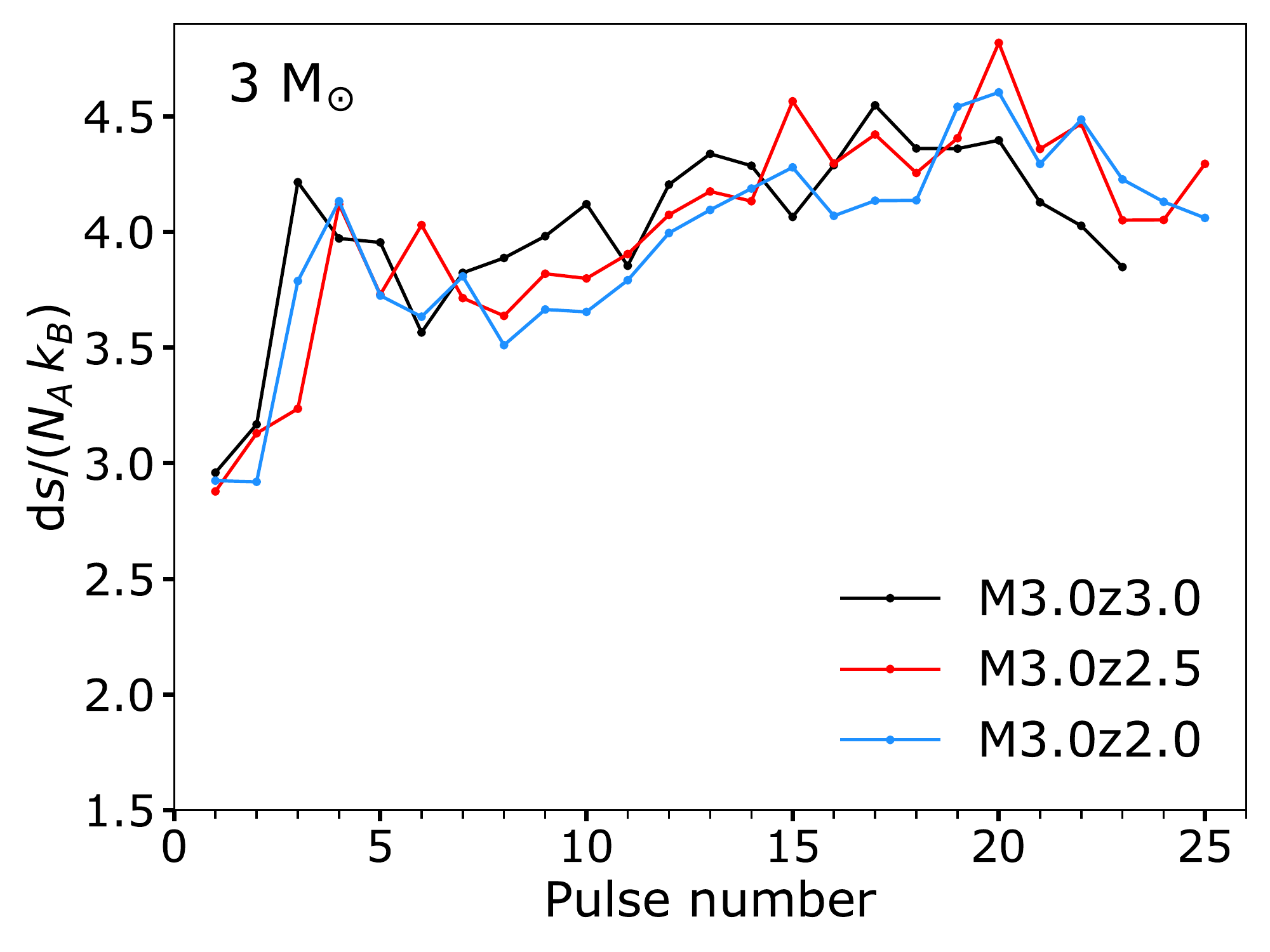}
  \end{minipage}
\caption{
Entropy barrier between the top of the convective thermal pulse and the bottom of the H-rich layers as a function of the pulse number (see text for details on the computation of d$s$). 
The empty circles mark the occurrence of the PIE.
}
\label{fig:ds_epulse}
\end{figure*}

\subsection{Yields and overproduction factors}

The yield $\mathcal{Y}_i$ of a nucleus $i$ is computed according to the relation:
\begin{equation}
\mathcal{Y}_i =  \int_{0}^{\tau_{\rm star}} \dot{M}(t) \, X_{\rm i,s}(t) \, \text{d}t
\label{eq:yie}
,\end{equation}
where $\tau_{\rm star}$ is total lifetime of the model star (given in Table~\ref{table:1}), 
and $X_{\rm i,s}(t)$ and $\dot{M}(t)$ are the surface mass fraction of nucleus, $i,$ and the mass-loss rate at time, $t$, respectively.

We also use the overproduction factor, $f_i$, associated with an element, $i,$ defined as:
\begin{equation}
f_i =  \frac{\mathcal{Y}_i}{(M_{\rm ini} - M_{\rm fin} ) \, X_{\rm i,ini}} 
\label{eq:op}
,\end{equation}
with $M_{\rm ini}$ and $M_{\rm fin}$ the initial and final mass of the model, respectively, and $X_{\rm i,ini}$ the initial mass fraction of element, $i$.

\subsection{Mass of hydrogen ingested during a PIE}

The mass of hydrogen ingested during a PIE is critical. It can have dramatic consequences on  the evolution, structure, and nucleosynthesis. 
In our models, protons are naturally ingested, so this mass is not a free parameter. . 

We define $M^{\rm NEUT}_{\rm H}$ which is the mass of hydrogen engulfed from the start of the PIE to the point where the neutron density is maximum. It can be written as:
\begin{equation}
M^{\rm NEUT}_{\rm H} = M_{\rm H} \, (t_{\rm Nnmax}) - M_{\rm H} \, (t_0)
\label{eq:mpie}
,\end{equation}
where $M_{\rm H}$ is the total mass of hydrogen contained in the star, evaluated just before the start of the PIE at $t = t_0$ and when the neutron density reaches its maximum at $t=t_{\rm Nnmax}$. We typically have $t_{\rm Nnmax} - t_0 \sim 1$~yr. 
Similarly, we define $M^{\rm SPLIT}_{\rm H}$, which is the hydrogen mass difference between the start of the PIE and the time of splitting of the convective pulse at $t=t_{\rm split}$. 
We checked that the mass of hydrogen lost through winds during these intervals of time is negligible compared to the mass of the ingested hydrogen. 
We also considered that the H-burning shell is switched off just before the PIE, which is a reasonable assumption given that most of the energy comes from He-burning during a thermal pulse. 
We note that neither $M^{\rm NEUT}_{\rm H}$  nor $M^{\rm SPLIT}_{\rm H}$ correspond to the total mass of hydrogen ingested, since more hydrogen is engulfed after the splitting of the convective pulse, when the upper part grows in mass before merging with the convective envelope. 



\section{Structure and evolution aspects}
\label{sect:evol}

We routinely checked during the AGB phase of our models whether a PIE occurs by recomputing a given pulse with higher spatial and temporal resolutions. 
As mentioned in Paper I, we noticed in our models that a PIE can be missed if adopting a too large time step.
 
We can see from Table~\ref{table:1} that 6 out of our 12 models experience a PIE: 
the 1~\Msun{} models with [Fe/H]~$=-3.0$, $-2.5$ and $-2.3$, the 1~\Msun{} model with $\alpha$-enhancement and the 2~\Msun{} models with [Fe/H]~$=-3.0$ and $-2.5$.
Such events are more likely to take place at lower initial masses and metallicities, as already noticed in previous works \citep[e.g.][]{iwamoto04, campbell08, suda10}. 
Also, PIEs always take place during the first or second thermal pulses\footnote{Pulses arising at the very early stage of the TP-AGB phase, that are not fully developed, are not counted as proper pulses. A pulse is referred to as such if the maximal temperature at the bottom of the He-driven convective zone is greater than $2 \times 10^8$~K.} (Table~\ref{table:1}). 
In the models with a PIE, two different categories emerge: (i) the models in which the TP-AGB phase ends after the PIE because of strong mass loss or (ii) the models in which the TP-AGB phase resumes after the PIE. 
The M1.0z2.5, M1.0z3.0$\alpha$ and M1.0z2.3 models belong to the first category while the M1.0z3.0, M2.0z3.0 and M2.0z2.5 to the second category.
The physical reasoning behind to these different evolutionary pathways are discussed in the next sections.

\subsection{The AGB phase of the 1~\Msun{}, [Fe/H]~$=-3.0$ model}
\label{sect:M1.0z3.0}

The M1.0z3.0 model experiences a PIE during the very first thermal pulse and a normal AGB phase afterwards with 41 thermal pulses without any further PIE. 
We discuss below the peculiar structure and evolutionary aspects of this model during and after the PIE.

\subsubsection{Evolution during the PIE}

Once the PIE starts (Fig.~\ref{fig:ratio_bot} top panel, at $t-t_0 \sim 0.4$~yr), protons are transported down in the convective thermal pulse (the turnover timescale is about 1~hr) and burnt on the fly by $^{12}$C($p,\gamma$)$^{13}$N. 
With a half-life of about 10 min, $^{13}$N decays into $^{13}$C which is transported at the bottom of the pulse where $T>200$~MK (Fig.~\ref{fig:ratio_bot}, second panel). At this temperature, $^{13}$C($\alpha,n$)$^{16}$O is efficiently activated (Fig.~\ref{fig:ratio_bot}, third panel) and releases neutrons. It gives rise to a maximum neutron density of about $10^{15}$~cm$^{-3}$ (Fig.~\ref{fig:ratio_bot}, fourth panel).
Contrary to most of our models, this model does not experience a proper split of the pulse, as explained in Sect.~\ref{sect:splitnosplit}. 
The energy released by nuclear burning produces an outward expansion of the convective pulse which engulfs more protons, as seen in the top panel of Fig.~\ref{fig:ratio_bot}, for $t-t_0 \ga 0.5$~yr.
At the same time, the pulse dilates (see the iso-mass contours in top panel of Fig.~\ref{fig:ratio_bot}), leading to a decrease in the temperature at the bottom of the convective region. 
This reduces the efficiency of \iso{13}C($\alpha$,$n$)\iso{16}O (blue line in Fig.~\ref{fig:ratio_bot}, third panel) and the neutron production.
Because a copious amount of protons has been engulfed, reactions from the CNO cycle become dominant (black, red, and green line in Fig.~\ref{fig:ratio_bot}, third panel). 

Reactions from the CNO cycle starts to be active in the thermal pulse as soon as protons are ingested. 
At first, they operate together with He-burning reactions when the temperature is high enough. In a second step they operate without He-burning when the temperature drops below about 150~MK.
The operation of the CNO cycle in the pulse dramatically reduces the C/N, O/N, and $^{12}$C/$^{13}$C ratios (Fig.~\ref{fig:ratio_bot}, bottom panel). 
At the end of the PIE, at $t-t_0 \sim 0.7$~yr , the $^{12}$C/$^{13}$C ratio at the bottom of the pulse is 3.4 which corresponds to the CNO equilibrium value. 
At $t-t_0 \sim 1$~yr, the pulse eventually merges with the convective envelope. The surface ratios (Fig.~\ref{fig:ratio_bot}, bottom panel, dashed lines) are progressively modified by the large amount of metals coming from the pulse.

 \begin{figure*}[t]
\includegraphics[width=2\columnwidth]{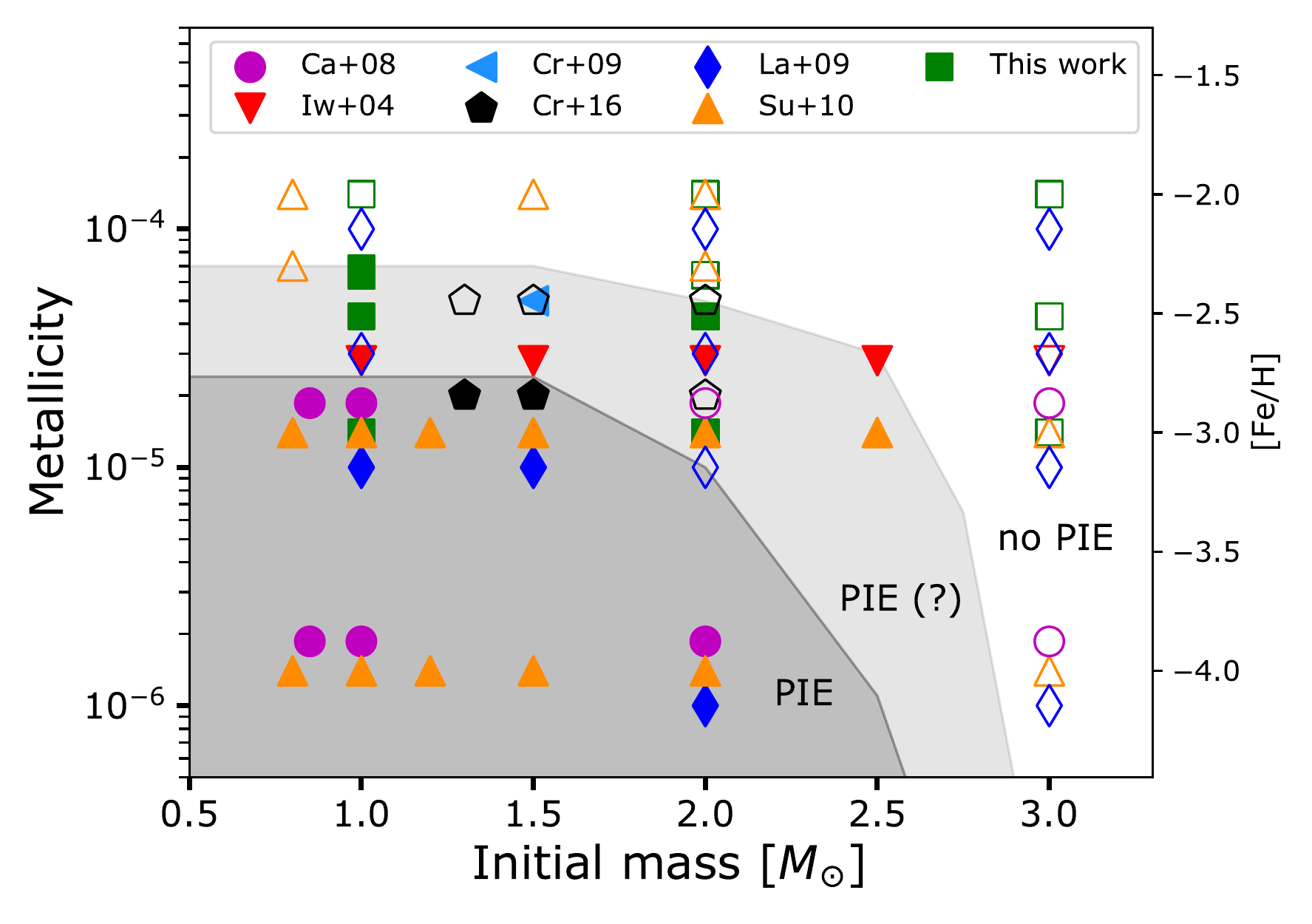}
\caption{
Mass-metallicity diagram showing the occurrence of PIEs during the early AGB phase of models from various authors. 
Filled symbols show models experiencing a PIE while empty symbols are for models that do not experience a PIE. 
The dark grey zone shows the approximate region where PIEs happen in all of the models and the light grey one where PIEs happen in most of the models. 
The corresponding [Fe/H] ratios are indicated on the right axis assuming solar-scaled mixtures. 
Models are from \citet[][red triangles]{iwamoto04}, \citet[][magenta circles]{campbell08}, \citet[][blue triangle]{cristallo09b}, \citet[][blue diamonds]{lau09}, \citet[][orange triangles]{suda10}, \citet[][black pentagons.]{cristallo16} Model results from this work are shown as green squares. All models were computed without extra mixing processes, except the models from \cite{cristallo09b,cristallo16}, which  consider overshooting below the convective envelope.
}
\label{fig:mzplot}
\end{figure*}

\subsubsection{Evolution after the PIE}
\label{sect:evaftpie}

It's reasonable to wonder why this model does not enter a standard thermally pulsating AGB phase (TP-AGB), while the M1.0z2.5 model does (cf. Paper I for more details on the M1.0z2.5 model).
As explained below, the difference between the M1.0z2.5 and M1.0z3.0 models is due to the fact that the M1.0z3.0 PIE occurs in the very first thermal pulse, but not until the second pulse for the M1.0z2.5 model (Table~\ref{table:1}). 
 
During the early AGB phase, the maximal temperature at the bottom of the pulse quickly increases with the pulse number. 
The maximal temperature at the bottom of the pulse during the PIE reaches 213~MK and 257~MK for the M1.0z3.0 and M1.0z2.5 models, respectively. 
The lower temperature in the M1.0z3.0 models leads to slower nuclear burning, especially the $3 \alpha$ reaction. 
The amount of metals synthesized in the M1.0z3.0 model is consequently lower\footnote{The $^{12}$C mass fractions at the bottom of the PIE pulse and at peak of neutron density are 0.06 and 0.17 in the M1.0z3.0 and M1.0z2.5 models, respectively. For $^{16}$O, these are $4.1 \times 10^{-3}$ and 0.016} and the convective envelope is therefore less enriched in metals after the PIE. 
Just after the PIE, the surface metallicity in mass fraction rises to $7.5 \times 10^{-3}$ ($1.9 \times 10^{-2}$) in the M1.0z3.0 (M1.0z2.5) model. 
In particular, the surface $^{12}$C and $^{16}$O mass fractions after the PIE are $0.7 \times 10^{-3}$ and $1.6 \times 10^{-3}$ for the M1.0z3.0 model while they are $7.7 \times 10^{-3}$ and $4.9 \times 10^{-3}$ for the M1.0z2.5 model. 
This leads to surface C/O ratios in number after the PIE of 0.75 and 2.55 for the for M1.0z3.0 and M1.0z2.5 models, respectively. 
This results in a less dramatic increase in CO molecular opacities in the stellar envelope of the M1.0z3.0 model, which, in turn, implies a weaker mass loss (more details in Sect.~\ref{sect:opa}). 
In the end, because of the lower surface metallicity and C/O ratio, the M1.0z3.0 model does not lose its envelope as fast as the M1.0z2.5 model and the AGB phase consequently resumes after the PIE. 
We also note that the PIE starts 220 yrs (305 yrs) after the beginning of the thermal pulse in the M1.0z3.0 (M1.0z2.5) model. It means that helium-burning has less time to process in the M1.0z3.0 model, which also contributes to the smaller C and O production and surface enrichment in this star. The temperature at the bottom of the pulse and the interval of time between the start of the pulse and the PIE are two critical aspects that dictate the subsequent AGB evolution. 

We confirmed our arguments by doing a numerical experiment on our M1.0z3.0 model. 
We intentionally missed the PIE happening in the first pulse by setting up time steps that were too large. 
Then, at the second pulse, a PIE happens. This time, the maximal temperature in the pulse is higher (around 250~MK), so that the model follows the same behaviour as the M1.0z2.5 model: high surface enrichment, strong mass loss, and the end to the AGB phase. 

Ultimately, we see that PIEs tend to happen earlier in the AGB phase (during the very first thermal pulse) of lower metallicity models. 
If this happens, an AGB phase with (many) thermal pulses can follow the PIE, in contrast to higher metallicity models where the AGB phase ends right after the PIE. 

We may also wonder why no other PIE happen in the M1.0z3.0 model after the first one. As mentioned in \cite{iwamoto04}, once a PIE has taken place, it becomes more difficult to have a second one. Indeed, following a PIE, the envelope is enriched in metals which results in a larger entropy barrier between the He- and H-rich layers and, thus, this tends to prevent any further PIE. 
After a PIE, the situation resembles that of a star of higher metallicity, where PIEs are less prone to arise.

 \begin{figure}[t]
\includegraphics[scale=0.55, trim = 0cm 0cm 0cm 0cm]{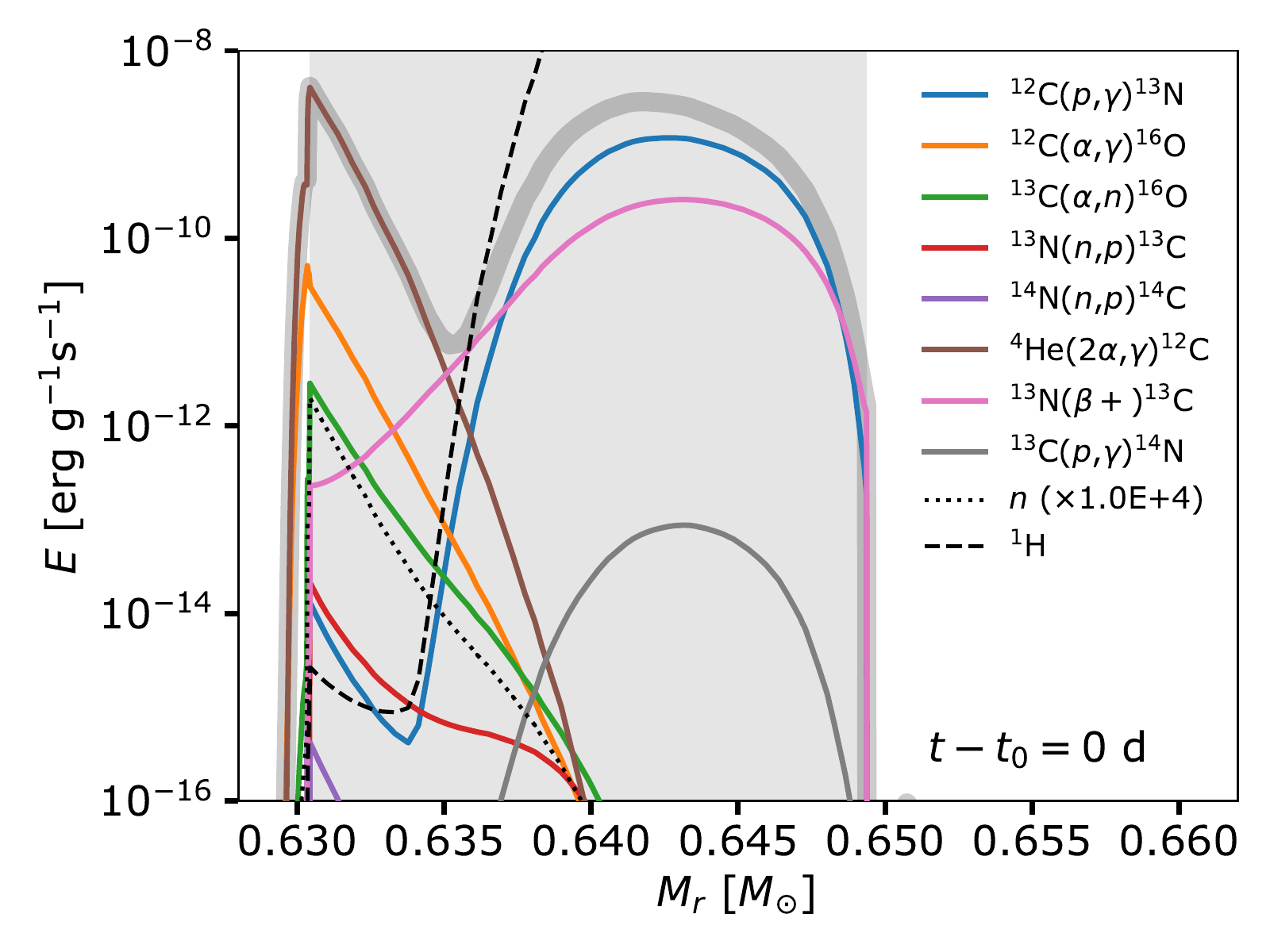}
\includegraphics[scale=0.55, trim = 0cm 0cm 0cm 0cm]{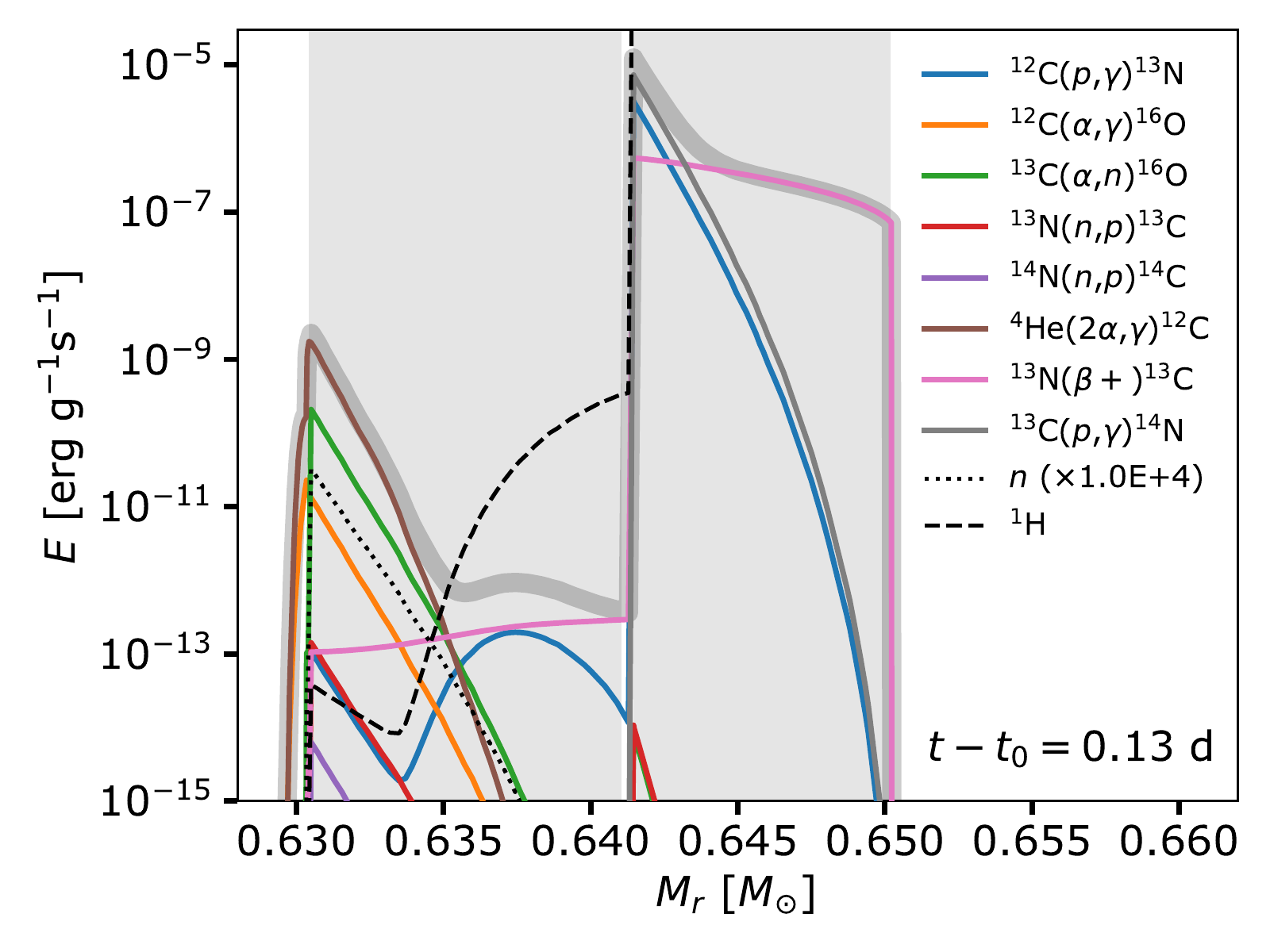}
\caption{ 
Energy generation from the main reactions during the PIE in the M2.0z3.0 model just before the splitting (top panel), and just after the splitting (bottom panel). The total nuclear energy production rate is shown by the thick grey line.  
The protons and neutrons mass fractions are shown.
}
\label{fig:fluxes4}
\end{figure}

 \begin{figure}[ht!]
\includegraphics[scale=0.55, trim = 0cm 0cm 0cm 0cm]{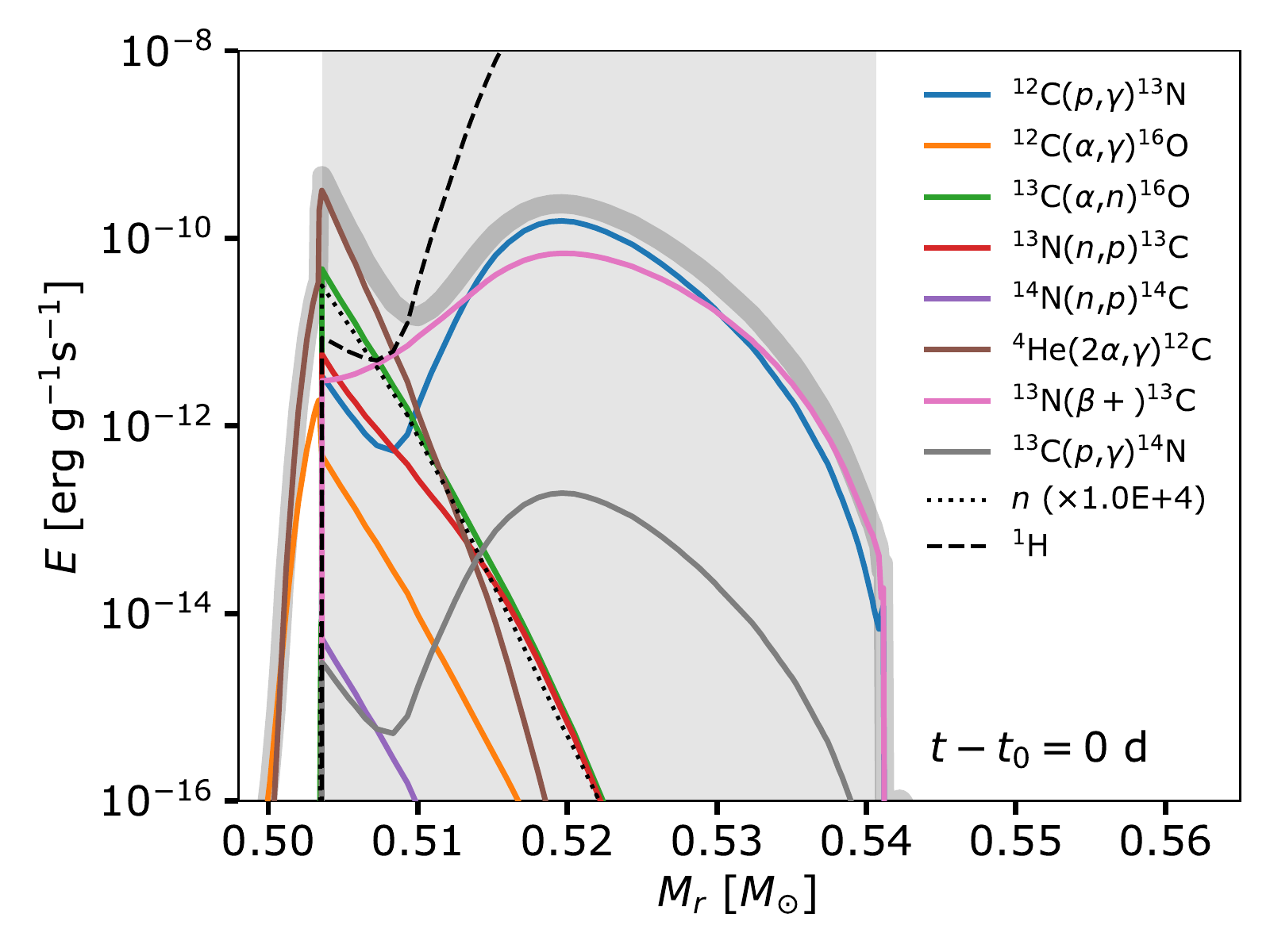}
\includegraphics[scale=0.55, trim = 0cm 0cm 0cm 0cm]{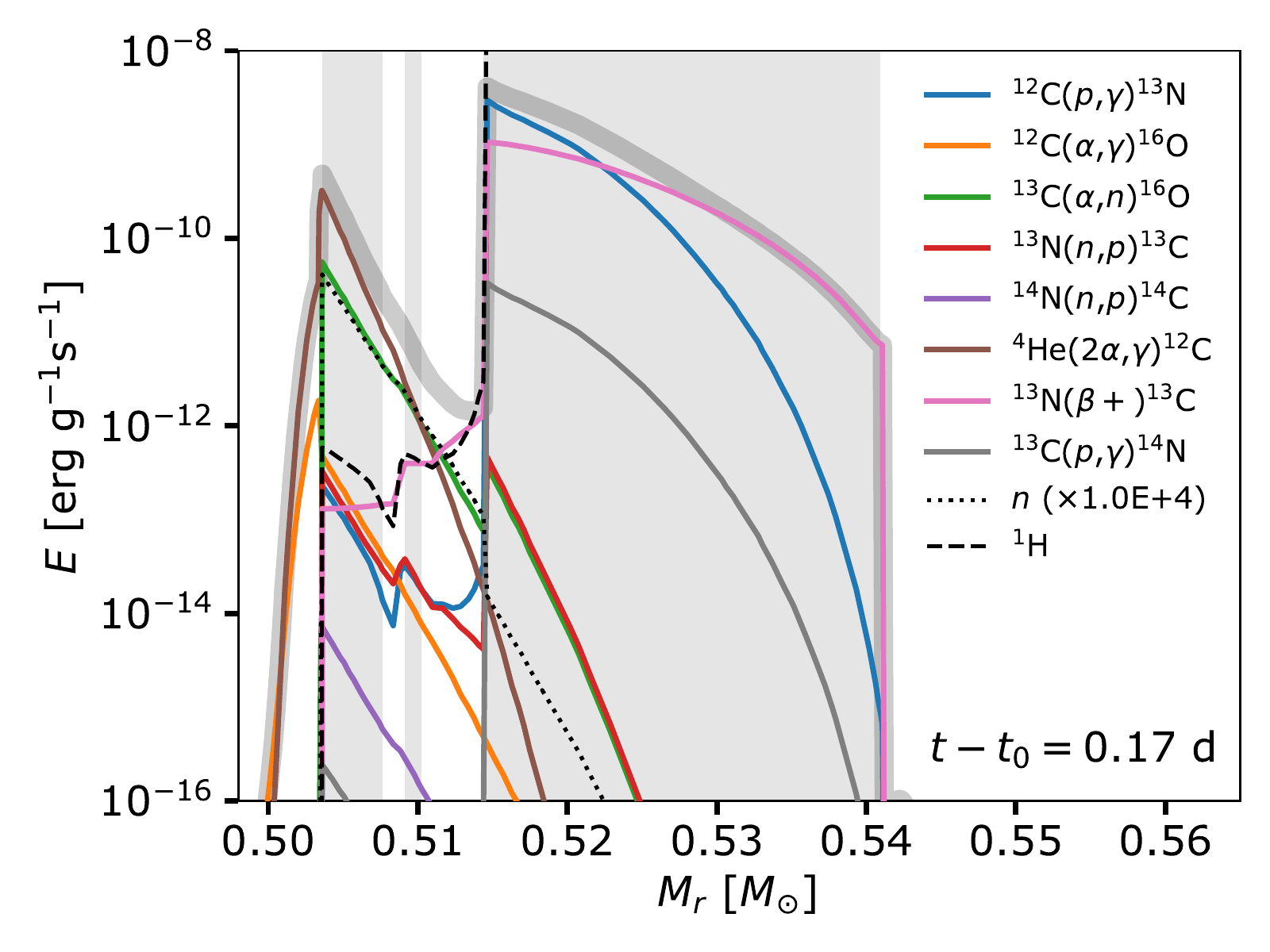}
\includegraphics[scale=0.55, trim = 0cm 0cm 0cm 0cm]{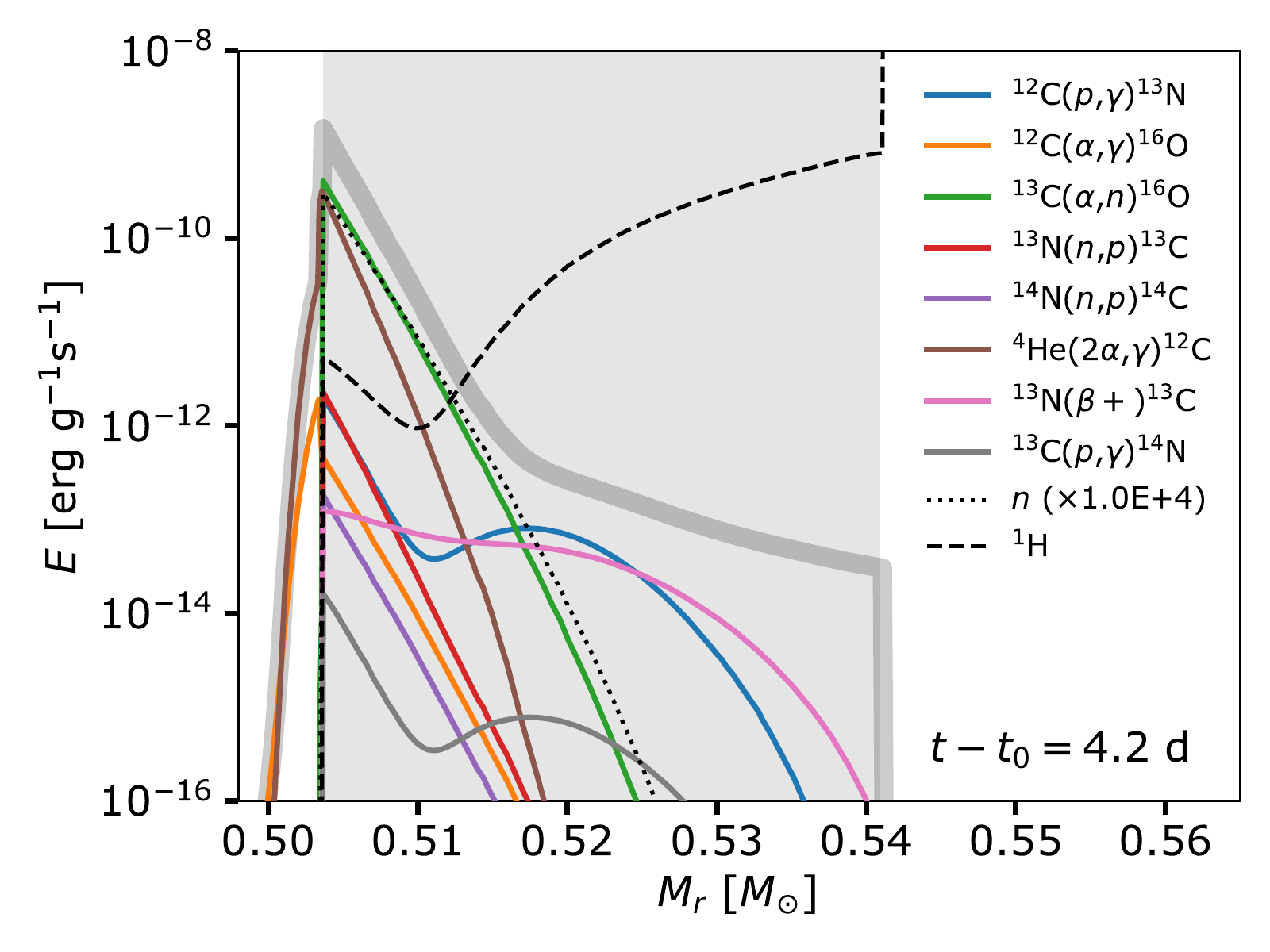}
\caption{
Energy generation from the main reactions during the PIE in the M1.0z3.0 model just before the splitting (top panel), just after the splitting (middle panel) and just after both parts of the pulse have merged  again (bottom panel).  
The total nuclear energy production rate is shown by the thick grey line.  
The protons and neutrons mass fractions are shown.
}
\label{fig:fluxes5}
\end{figure}

\subsection{The AGB phase of the 2~\Msun{}, [Fe/H]~$=-2.5,$ and $-3.0$ models}
\label{sect:M2.0z2.5}

The M2.0z3.0 and M2.0z2.5 models experience a PIE followed by about 10 standard thermal pulses (Table~\ref{table:1}). 
They follow a similar evolutionary pathway as the M1.0z3.0 model but the physical reasons leading to it are different.

The maximal temperature at the bottom of the PIE pulse is similar or even higher in the M2.0z3.0 and M2.0z2.5 models than in the M1.0z2.5 model. 
In the pulse, carbon and oxygen are produced in similar amounts than in the M1.0z2.5 model. 
An important difference however is higher dilution factors in the 2~\Msun\, stars compared to the 1~\Msun\, stars. The ratios between the pulse and envelope masses are about $0.1$ and $0.01$ 
for the 1 and 2 \Msun\, models, respectively\footnote{The envelope and pulse masses are about 0.35 and 0.04 \Msun\ for the 1 \Msun\ models and 1.31 and 0.017 \Msun\ for the 2 \Msun\ models, respectively.}. The material from the pulse in the 2 \Msun\, models is therefore about ten times more diluted in the envelope than in the 1 \Msun\ and the surface metallicity does not increase that much in this model. In the M1.0z2.5 model, the surface metallicity rises to $1.9 \times 10^{-2}$ while it increases only to $1.2 \times 10^{-3}$ and $2.6 \times 10^{-3}$ in the M2.0z3.0 and M2.0z2.5 models, respectively. The high surface enrichment in the M1.0z2.5 model leads to high surface opacities and a large expansion of the stellar radius from $\sim 80$ to $\sim 400$~$R_{\odot}$. The effective temperature drops to 2700 K.
In the M2.0z2.5 and M2.0z3.0 models, the radius ranges from $\sim 120$ to $\sim 200$~$R_{\odot}$ and the effective temperature experiences a milder drop, down to about 4000~K. 
In the end, the mass loss rate after the PIE does not increase dramatically in the M2.0z2.5 and M2.0z3.0 models (unlike the M1.0z2.5 model), so  the AGB phase resumes with additional thermal pulses.

\subsection{The key role of molecular opacities}
\label{sect:opa}

As mentioned in Sect.~\ref{sect:inputs}, molecular CO opacities are taken into account in our models. 
Molecular CO opacities become important when the effective temperature drops below 5000~K and when the C/O ratio becomes larger than about 0.8 in number. 
These opacities play a key role in the models that lose all their envelope quickly after the PIE (e.g. in the M1.0z2.5 model). 
The convective envelope of these models becomes highly enriched in C, N, and O after the PIE with a C/O ratio greater than 1. 
This leads to high CO molecular opacities and enhanced mass loss which prevents the further occurrence of thermal pulses. 

We confirmed the critical role of CO molecular opacities by recomputing our M1.0z2.5 model starting from just after the PIE with the CO molecular opacities switched off. 
In this case, the surface opacity (hence, the stellar radius) does not rise dramatically, the mass loss rate modestly increases, and the star keeps its envelope for a much longer time. 
The PIE is consequently followed by a normal AGB phase with regular thermal pulses (as in the M1.0z3.0 model, see Sect.~\ref{sect:M1.0z3.0}).

\subsection{The entropy barrier} 

To our knowledge, there is no simple criterion at hand to predict whether a model will experience a PIE or not. 
However, as noticed in previous works \citep[e.g.][]{fujimoto90, iwamoto04}, the height of the entropy barrier at the bottom of the H-rich layers plays an important role in the occurrence of a PIE. 
The higher the barrier, the less likely the convective pulse is to reach the H-rich layers, hence, the less likely it is for a PIE to occur. 

In the present work, we estimate this barrier at the maximal development of each thermal pulse for each model. 
We define the entropy barrier as d$s = s_2 - s_1$ where $s_1$ is the entropy at the top of the convective pulse and $s_2$ is the entropy at the base of the H-rich zone (defined where the mass fraction of hydrogen drops below $10^{-1}$). 
For models experiencing a PIE, the entropy barrier was derived just before the PIE.  
The results are reported in Fig.~\ref{fig:ds_epulse}. 

The entropy barrier overall increases with the pulse number. This is consistent with the fact that PIE preferentially happen during the first thermal pulses (Table~\ref{table:1}).
The entropy barrier tends to be higher in higher mass models which makes a PIE less likely to occur as mass increases \citep[as recognized early by e.g.][especially his Fig.~3]{iben77}. 
This is clearly visible, at least, in our 3 \Msun\, models, as compared to the less massive models, especially during the first pulses. 
Higher entropy barriers in more massive AGB stars is consistent with the fact that PIEs take place in some of our 1 and 2 \Msun\, models, but not in our 3 \Msun\, models. 

It was also recognized that PIEs preferentially occur at low metallicities because of the smaller entropy barrier at the base of the hydrogen-burning shell \citep[e.g.][]{fujimoto90}. 
The metallicity dependence can be understood by the fact that less CNO catalysts are present in the H-burning shell of low metallicity AGB stars. 
The entropy barrier between the H- and He-burning zones consequently decreases with metallicity, thus facilitating PIEs. 
In our models, as soon as d$s \, / \, N_{A}\,k_B \lesssim 2.5$, a PIE is triggered (Fig.~\ref{fig:ds_epulse}).
Nevertheless, the metallicity dependence is not clear
as models with the same initial mass have overall similar entropy barriers (Fig.~\ref{fig:ds_epulse}). 
To explain this, we first note that although the height of the entropy barrier gives a good hint on the possible occurrence of a PIE, it should be taken with  caution because the definition of this barrier is not absolute and can be estimated using different criteria. 
Secondly, our models lie in the metallicity range separating models with and without PIE. 
Figure~\ref{fig:mzplot} shows a compilation of AGB models from different works investigating PIEs in AGB stars. 
All the models were computed without extra mixing processes except the model from \cite{cristallo09b,cristallo16} that includes overshooting below the convective envelope following the time dependent formalism described in \cite{chieffi01}. 
 Overall, PIEs occur in all the models shown in the dark grey region, while the light grey region is subject to some uncertainties. 
Our models (especially the 1 and 2 \Msun\, models) lie close to the outer grey border, where very slight stellar structural differences can be responsible for the presence or absence of PIEs.

 \begin{figure*}
  \begin{minipage}[c]{2\columnwidth}
\includegraphics[width=0.5\columnwidth]{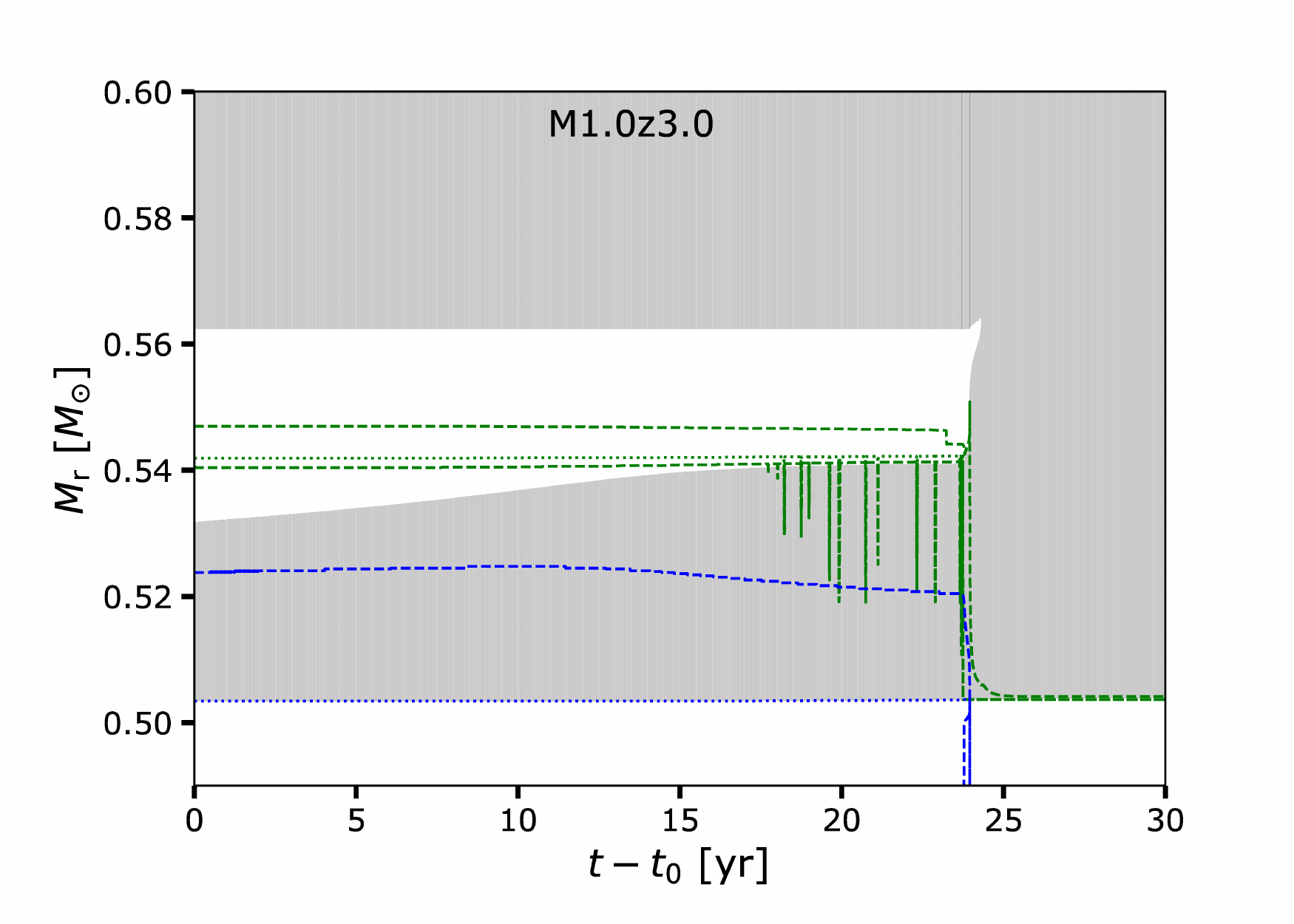}
\includegraphics[width=0.5\columnwidth]{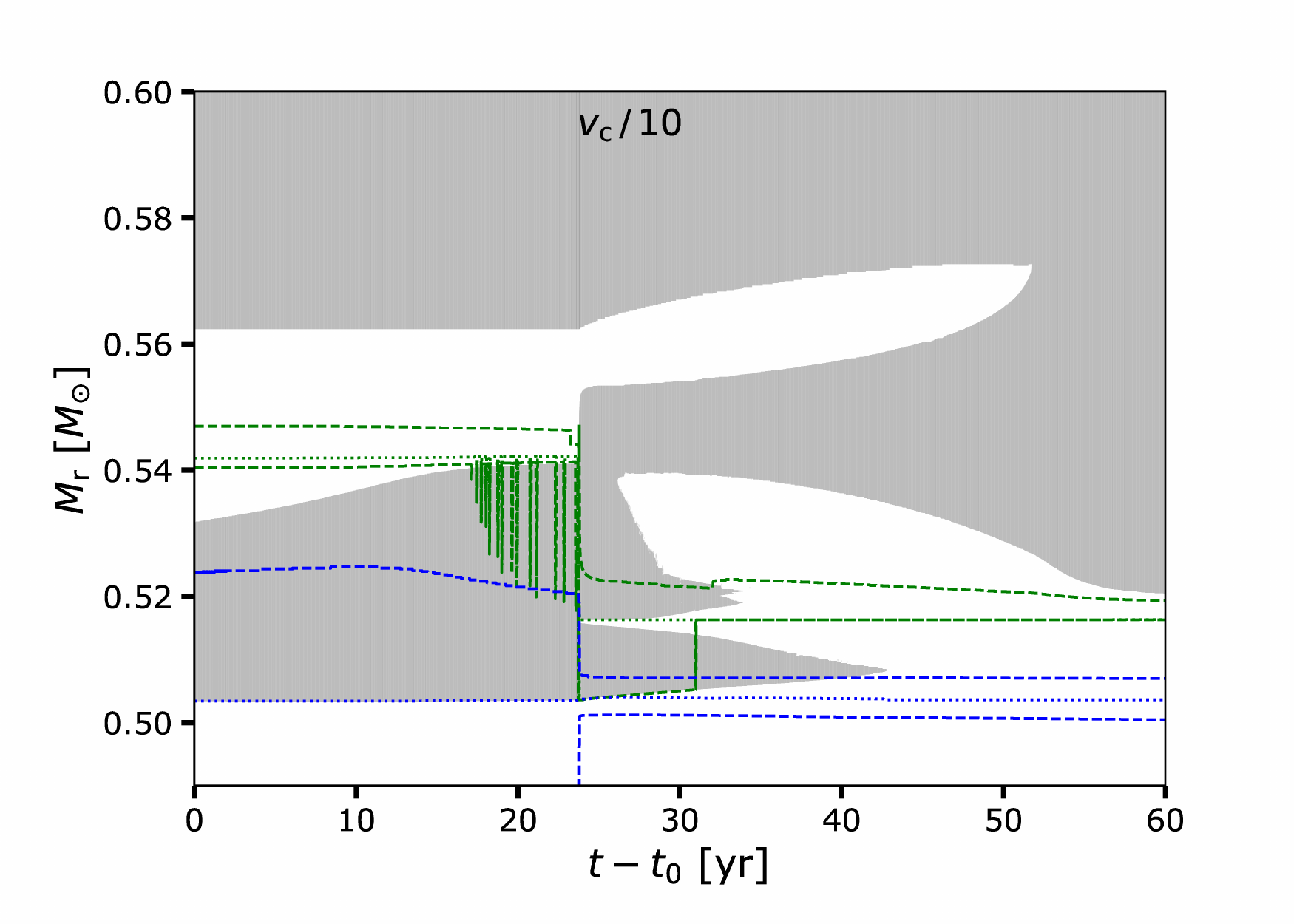}
  \end{minipage}
  \begin{minipage}[c]{2\columnwidth}
  \vspace{-1.0cm}
\includegraphics[width=0.5\columnwidth]{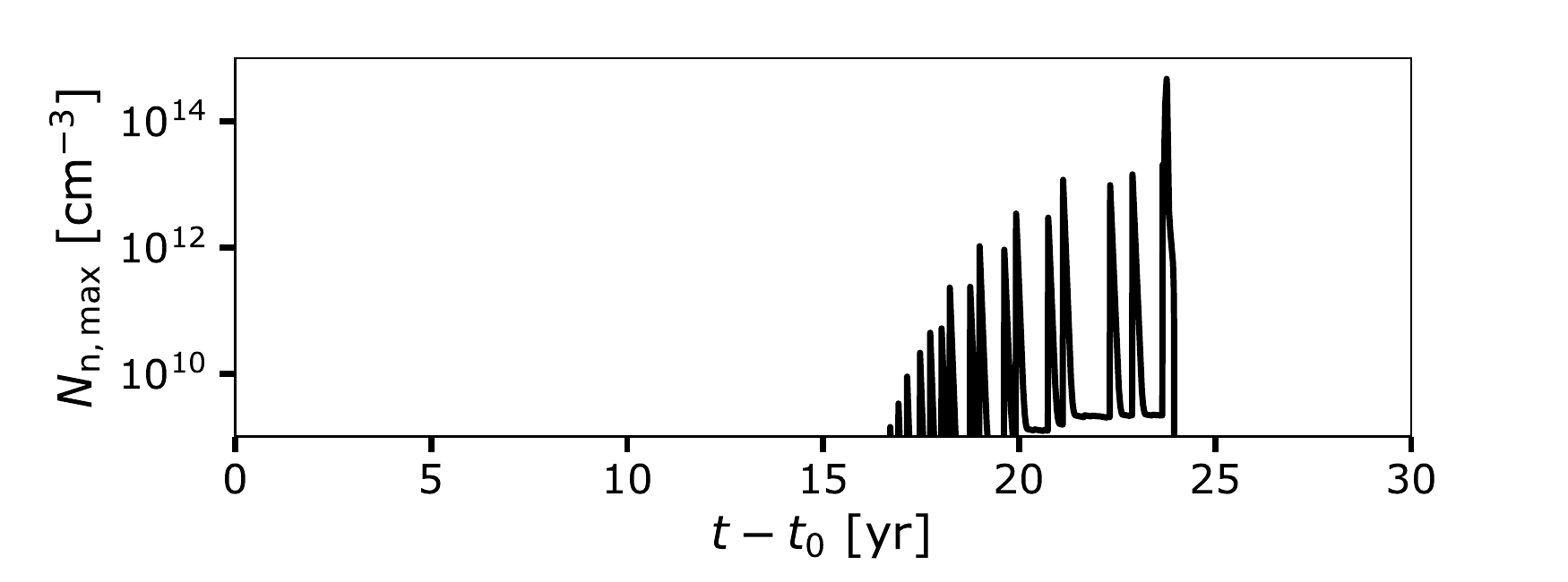}
\includegraphics[width=0.5\columnwidth]{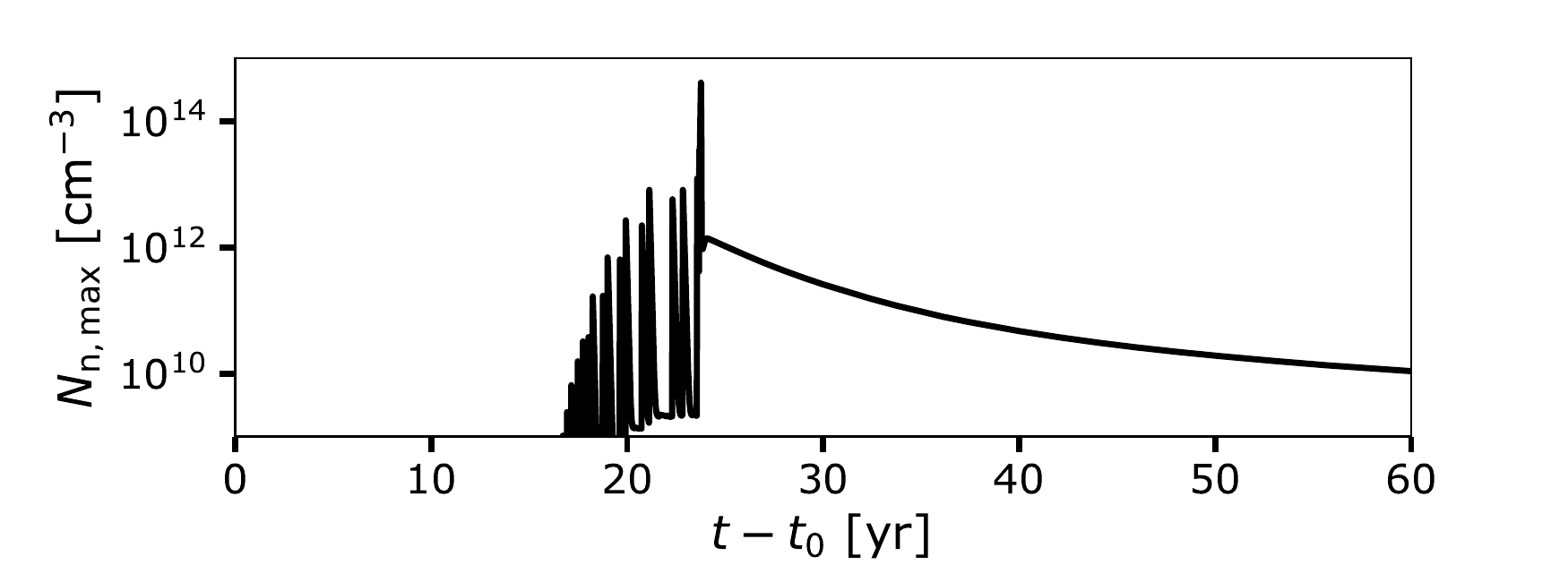}
  \end{minipage}
  \begin{minipage}[c]{2\columnwidth}
\includegraphics[width=0.5\columnwidth]{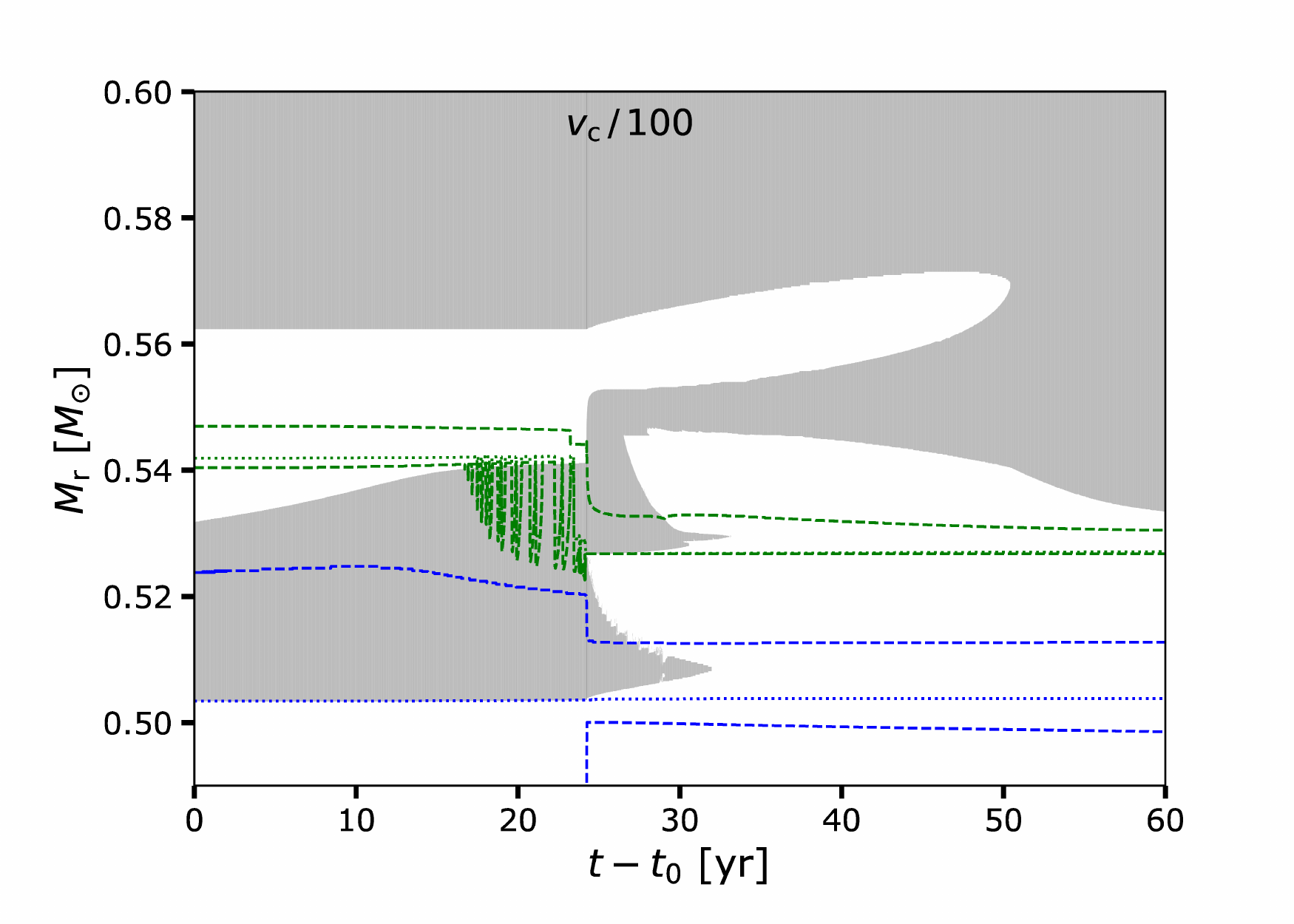}
\includegraphics[width=0.5\columnwidth]{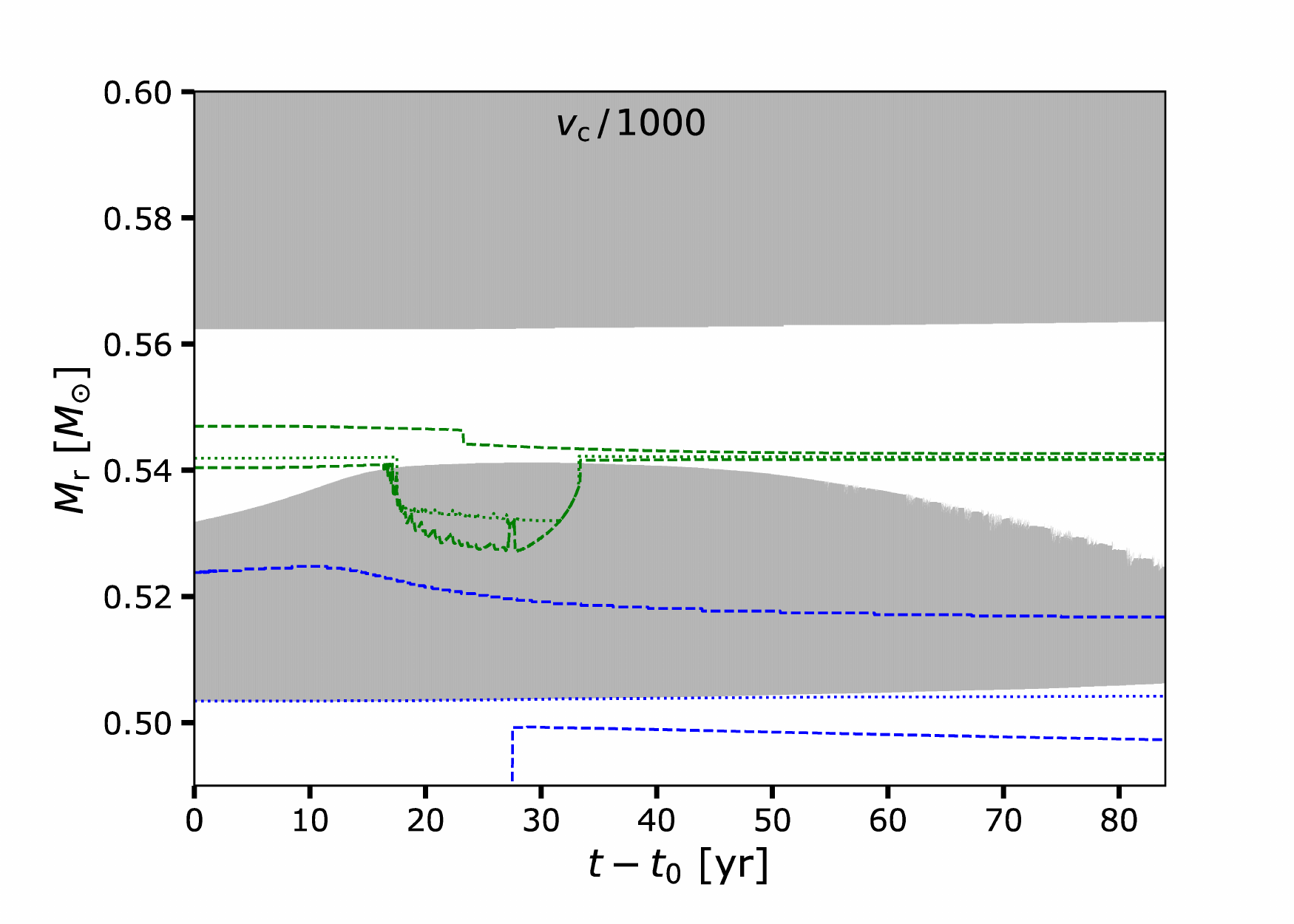}
  \end{minipage}
  \begin{minipage}[c]{2\columnwidth}
  \vspace{-1.0cm}
\includegraphics[width=0.5\columnwidth]{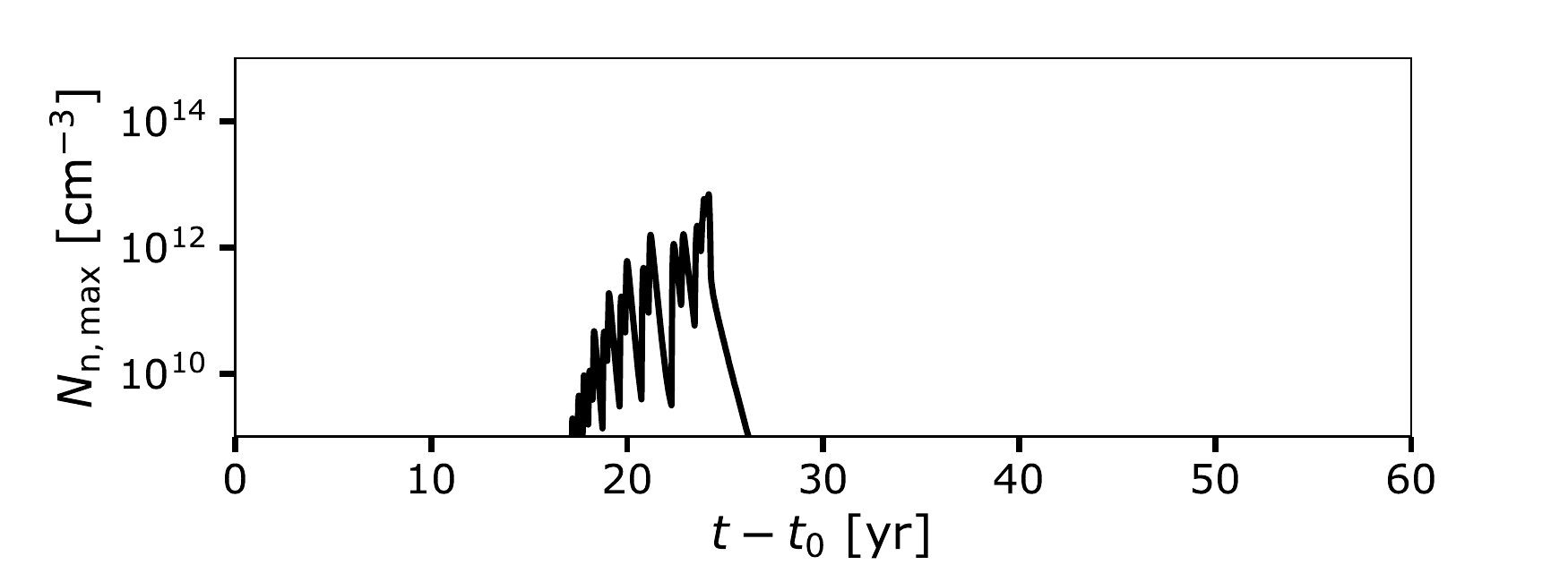}
\includegraphics[width=0.5\columnwidth]{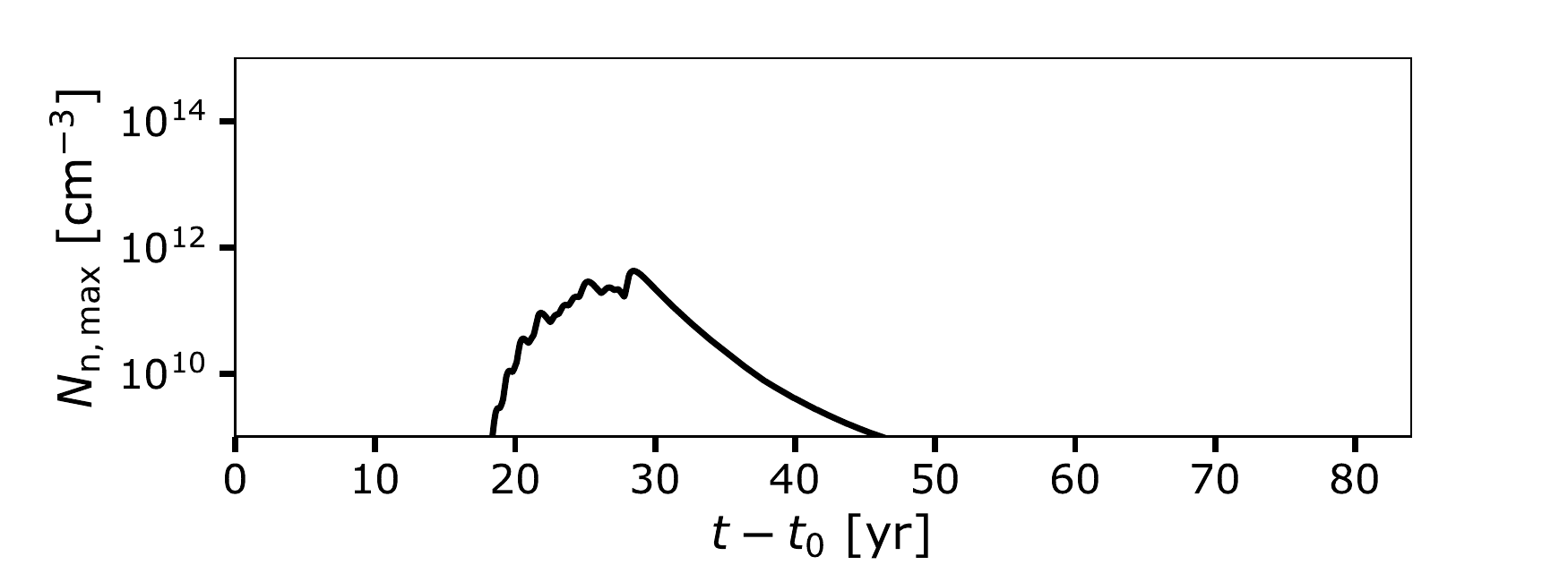}
  \end{minipage}
\caption{ Kippenhahn diagram and maximal neutron density of the 1~\Msun\, model with [Fe/H]~$=-3.0$ during the PIE while considering different values for the convective velocity: the standard case (top left), $v_{\rm conv}$ divided by 10 (top right), 100 (bottom left), and 1000 (bottom right). 
The dotted green and blue lines in the Kippenhahn diagram correspond to the maximum nuclear energy production by hydrogen- and helium-burning, respectively. The  dashed green and blue lines delineate the H- and He-burning zones (when the production of energy by H- and He-burning exceeds 10~erg~g$^{-1}$~s$^{-1}$).
}
\label{fig:vc_kip}
\end{figure*}

 \begin{figure}[h!]
\includegraphics[scale=0.5, trim = 0cm 1cm 0cm 0cm]{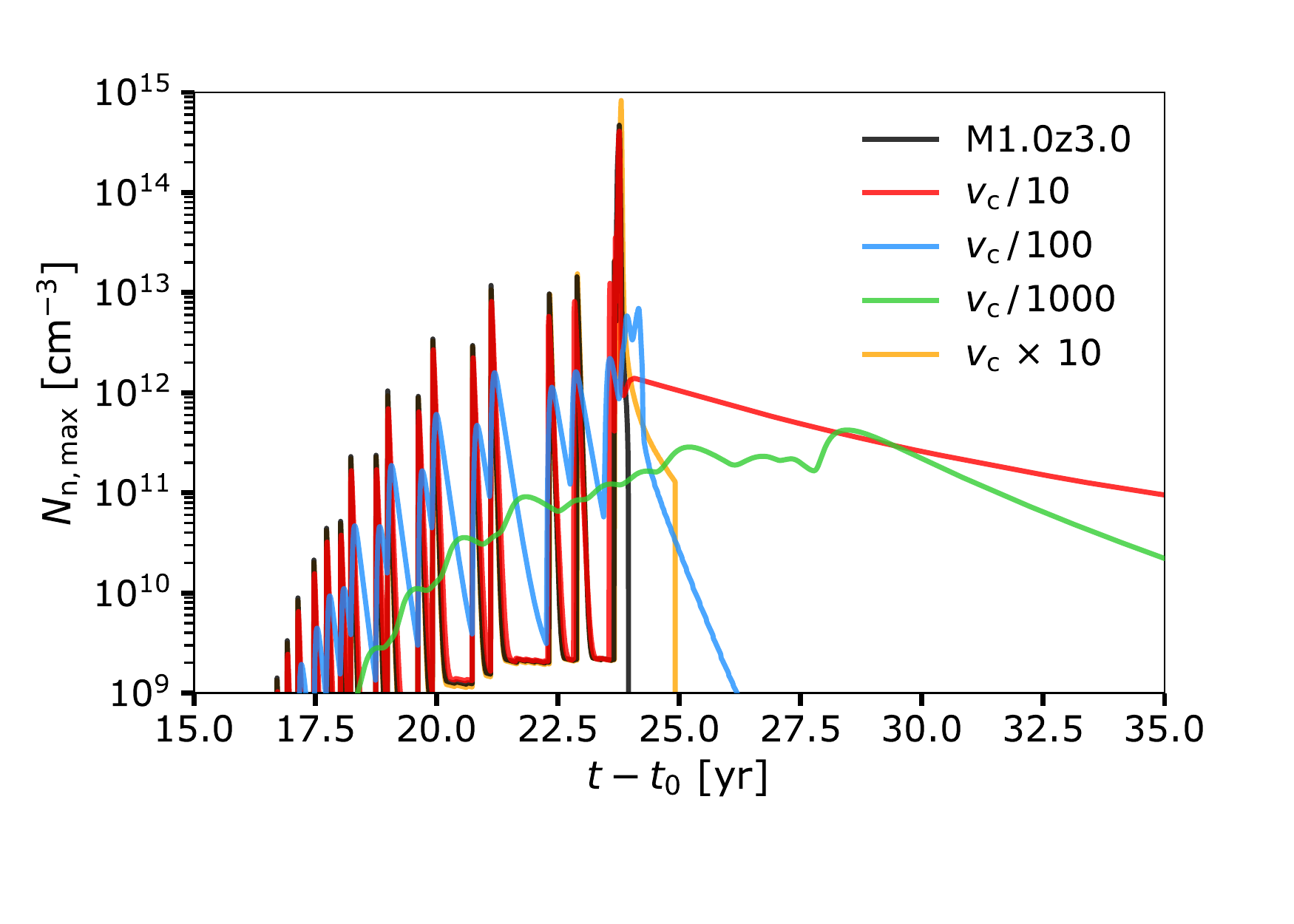}
\caption{ Maximum neutron density for the M1.0z3.0 model during the PIE for different values for the convective velocity. }
\label{fig:vc_Nn}
\end{figure}

 \begin{figure}[h!]
\includegraphics[scale=0.5, trim = 0cm 1cm 0cm 0cm]{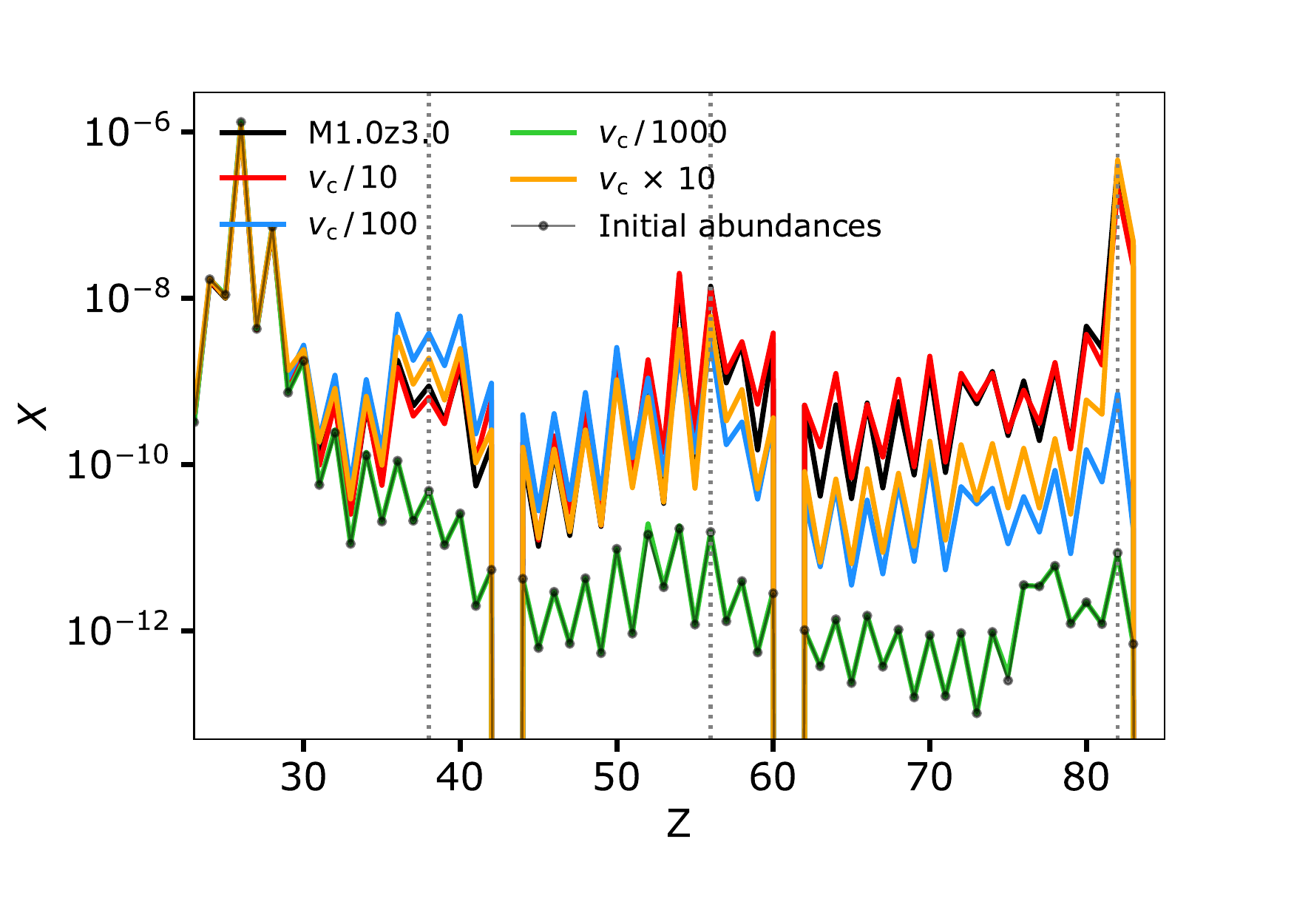}
\caption{ Surface mass fractions after the PIE for the M1.0z3.0 model for  different values for the convective velocity. }
\label{fig:vc_ab}
\end{figure}

\subsection{The split of the convective pulse during a PIE}
\label{sect:splitnosplit}

In 1D simulations, during a PIE, proton are transported down in the convective helium-burning zone until the timescale associated with the reaction $^{12}$C($p,\gamma$)$^{13}$N becomes similar to the local transport timescale of protons. 
At this location, protons accumulate, the nuclear energy production from $^{12}$C($p,\gamma$)$^{13}$N is maximal, a temperature inversion develops, and the helium-burning convection zone splits. 
From this point, protons cannot be mixed in the hottest bottom part of the pulse anymore and the chemicals synthesized in the lower part of the now-split convective region (especially through the i-process) remain locked in the star unless further thermal pulses and dredge-up events take place later on. 
The split is thus an important aspect of a PIE since it can prevent the enrichment of the AGB surface in heavy elements if it happens before neutrons are released. 

In our M2.0z3.0 model the energy released from the main reactions just before (top panel) and after (bottom panel) the splitting is shown in Fig.~\ref{fig:fluxes4}. 
We clearly see the two peaks of nuclear energy at the bottom of the pulse and at $M_{\rm r} \simeq 0.642$~\Msun\,. The first peak is mainly due to the 3$\alpha$ reaction while the second one is attributed to $^{12}$C($p,\gamma$)$^{13}$N. 
A fraction of a day later, the pulse has split and the bottom part becomes much less active compared to the upper part, driven by hot hydrogen-burning. 
In this model, the split takes place after the  neutron density has reached its maximum, so that the upper part of the pulse is fully enriched in i-process products before the splitting (Sect.~\ref{sect:surfenr} for a more detailed discussion).

All our models with a successful PIE (except for M1.0z3.0) follow a similar behaviour, although the split takes place at different depths in the pulse. 
Our M1.0z3.0 model experiences an early splitting of the convective pulse but both parts of the pulse merge again after a few days, as shown in Fig.~\ref{fig:fluxes5}. 
Also, contrary to the other models, this early splitting occurs before the neutron density reaches its maximum. 
Below we develop the reasoning behind this peculiar behaviour. 

Firstly, as discussed in Sect.~\ref{sect:evaftpie}, the M1.0z3.0 model experiences a PIE during the very first thermal pulse, contrary to all other models (cf. Table~\ref{table:1}). 
The temperature in the first pulse is lower than in subsequent pulses. In such conditions, protons can be transported deeper in the convective pulse before being completely burnt by $^{12}$C($p,\gamma$)$^{13}$N. 
This makes the two energy peaks closer in the M1.0z3.0 model compared to the M2.0z3.0 model 
(top panels of Figs.~\ref{fig:fluxes4} and \ref{fig:fluxes5}) and does not favor the split. 
The reaction $^{13}$N($\beta+$)$^{13}$C (pink) is mostly responsible for the smaller contrast between both peaks (this point is discussed in detail in Sect.~\ref{sect:reac}). 

Secondly, the M1.0z3.0 model is the model that ingests the smallest mass of protons ($2 \times 10^{-7}$~\Msun, see Table~\ref{table:2}). 
This makes a split less prone to happen.
Hydrogen-burning in the pulse is consequently weaker in the M1.0z3.0 model while helium-burning operates similarly in all models (Figs.~\ref{fig:fluxes4} and \ref{fig:fluxes5}, during the split).
This leads to a much lower entropy barrier in the M1.0z3.0 model: the entropy barrier at $M_{\rm r} = 0.515$~\Msun\ is $s\,/\,N_A\,k_B = 10^{-3}$ while the barrier at $M_{\rm r} = 0.641$~\Msun\ in the M2.0z3.0 model is $s\,/\,N_A\,k_B = 2.3$.
In the M1.0z3.0 model, the energy from the bottom part of the pulse is enough to make this zone fully convective again and overcomes the small entropy barrier separating both parts of the pulse. It results in the merging of the bottom and upper parts of the pulse (Fig.~\ref{fig:fluxes5}, bottom panel).

In the end, all our models with PIEs experience a split of the convective pulse but in one case (M1.0z3.0), both parts of the pulse quickly merge back again. 
This model behaves as if no split had developed. 
This can happen when a small amount of protons are engulfed or if the temperature in the pulse is sufficiently low (or both), so that the protons can be transported close enough to the bottom of the pulse.

As a final remark, we note that other PIE models in the literature are not always accompanied by a splitting. 
For instance, it is absent in the 1D RAWD models of \cite{denissenkov19} or in the 3D AGB simulation of \cite{stancliffe11} but 
reported in the 3D simulation of \citet{herwig11}.

\begin{figure*}[t]
\includegraphics[width=\columnwidth]{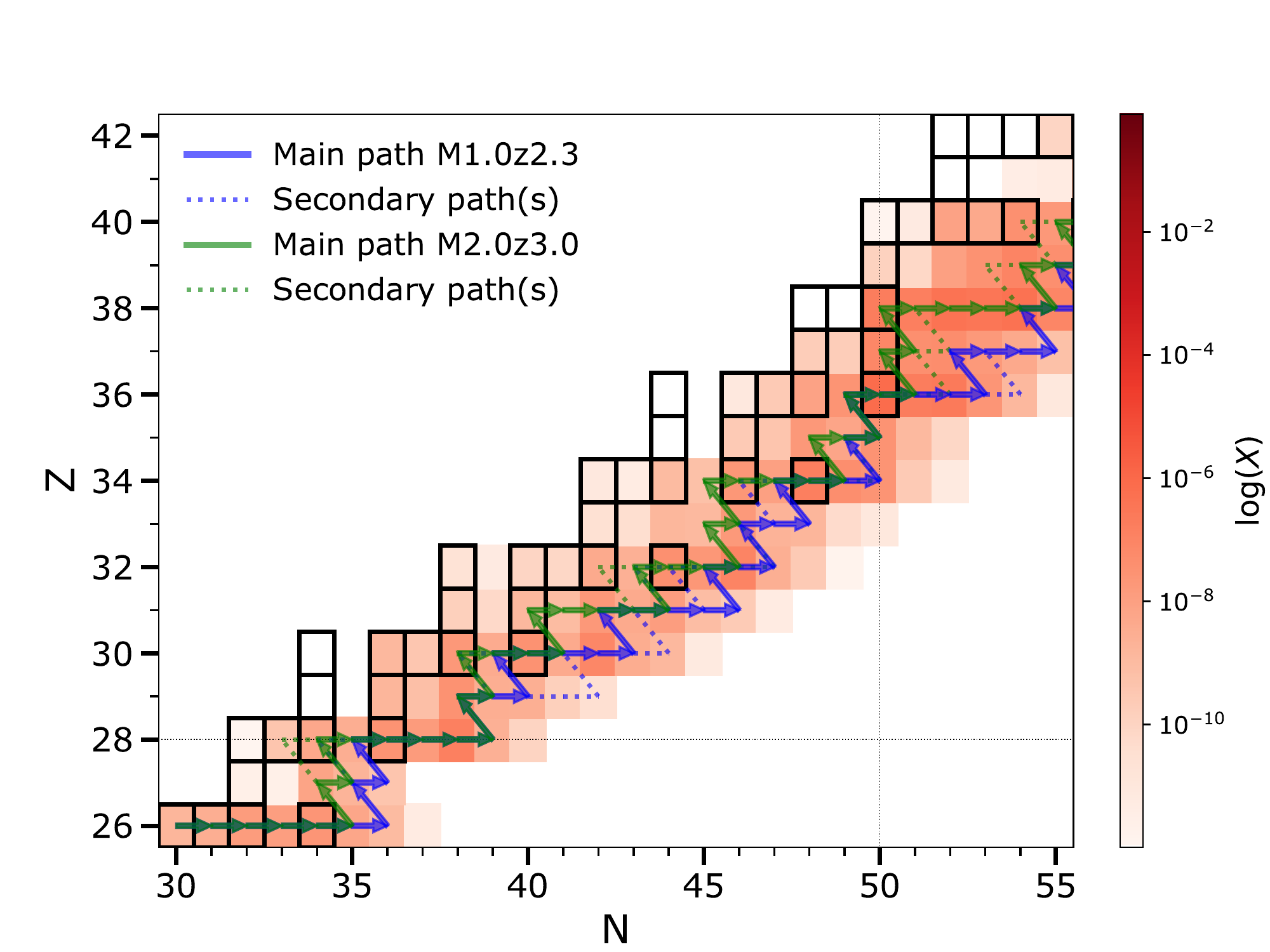}
\includegraphics[width=\columnwidth]{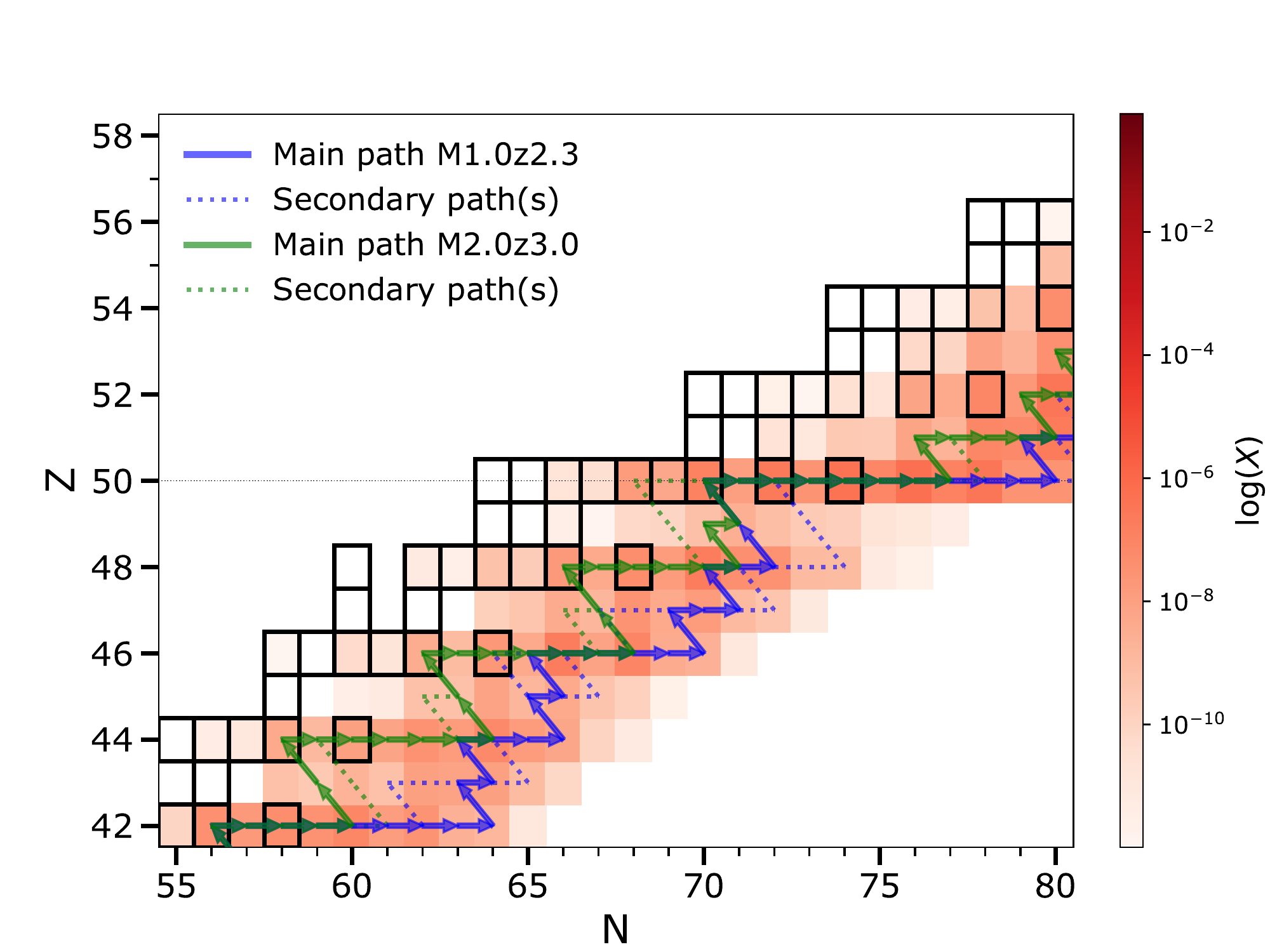}
\includegraphics[width=\columnwidth]{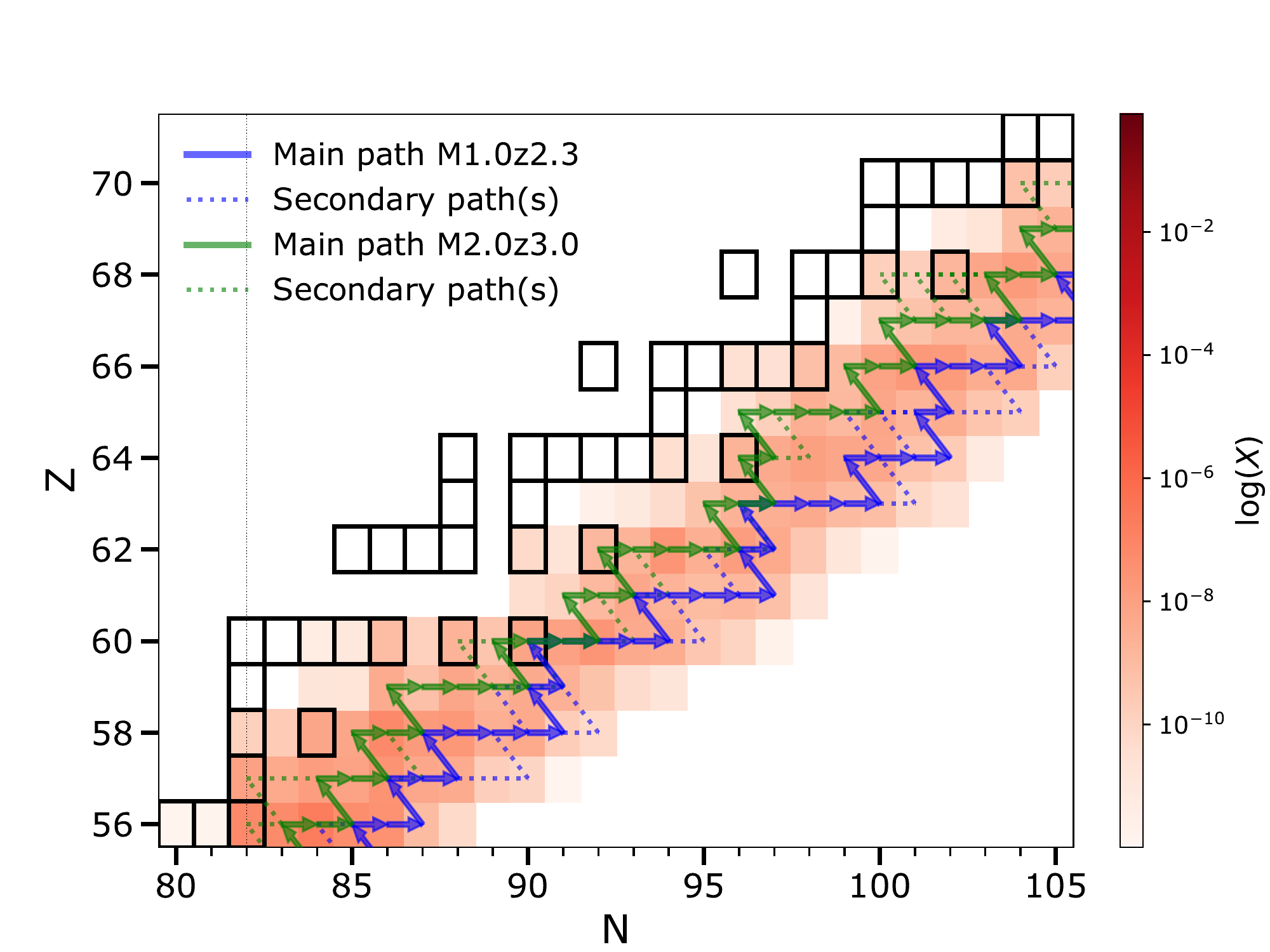}
\includegraphics[width=\columnwidth]{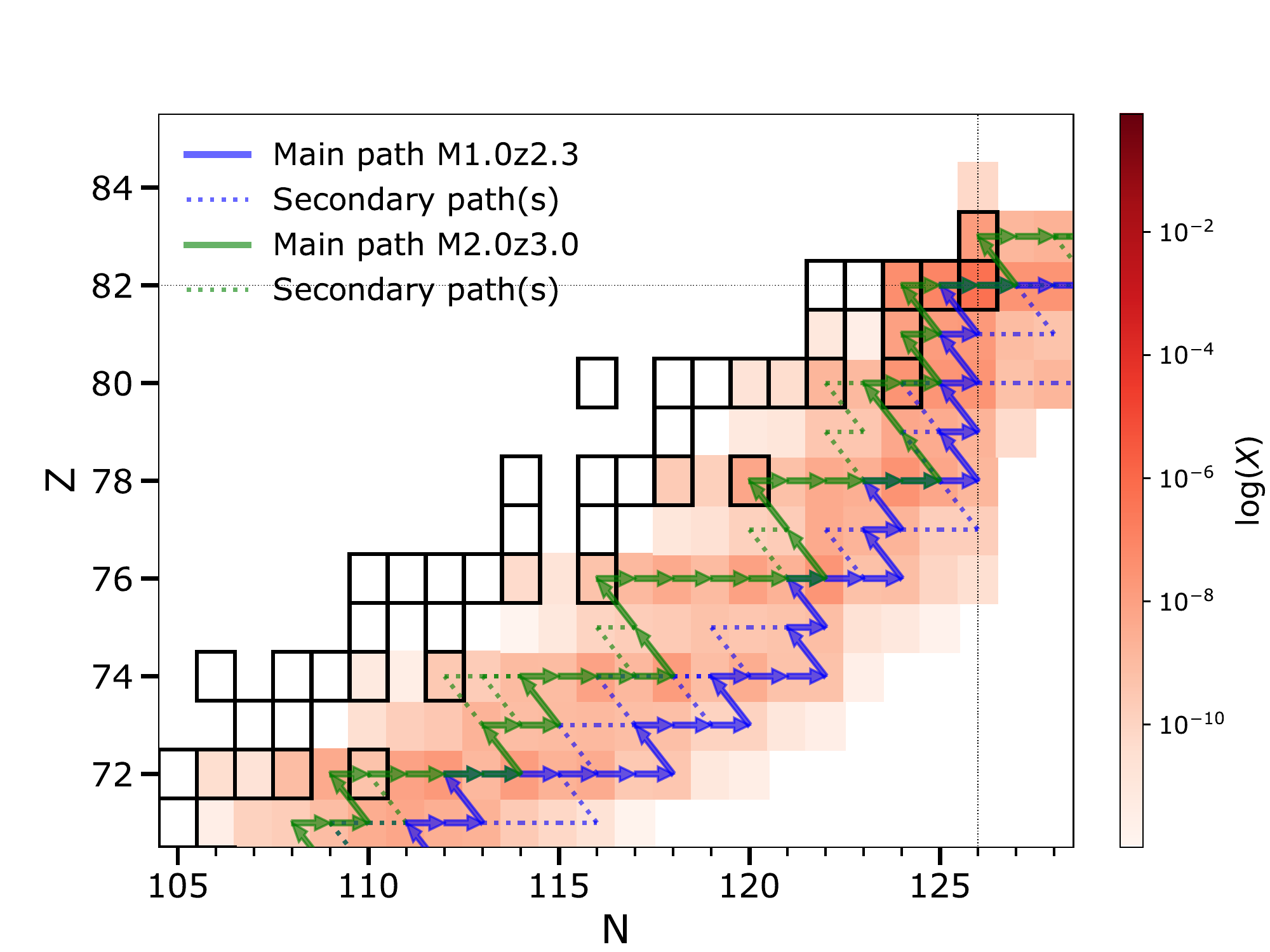}
\caption{
Main i-process path (starting from $^{56}$Fe) in the M1.0z2.3 (blue arrows) and M2.0z3.0 model (green arrows), at the bottom of the pulse, at the time of maximum neutron density ($N_n = 2.19 \times 10^{15}$ and $6.76 \times 10^{13}$~cm$^{-3}$, respectively).
The dashed lines show the secondary paths, where at least 30~\% of the total flux goes. 
The four panels corresponds to four different zones in the (N, Z) plane. 
The black squares highlight the stable nuclei. 
The abundances of the M1.0z2.3 model are shown by the red colour scale in mass fraction. 
}
\label{fig:flow}
\end{figure*}

 \begin{figure*}[h!]
  \begin{minipage}[c]{2\columnwidth}
\includegraphics[width=0.33\columnwidth]{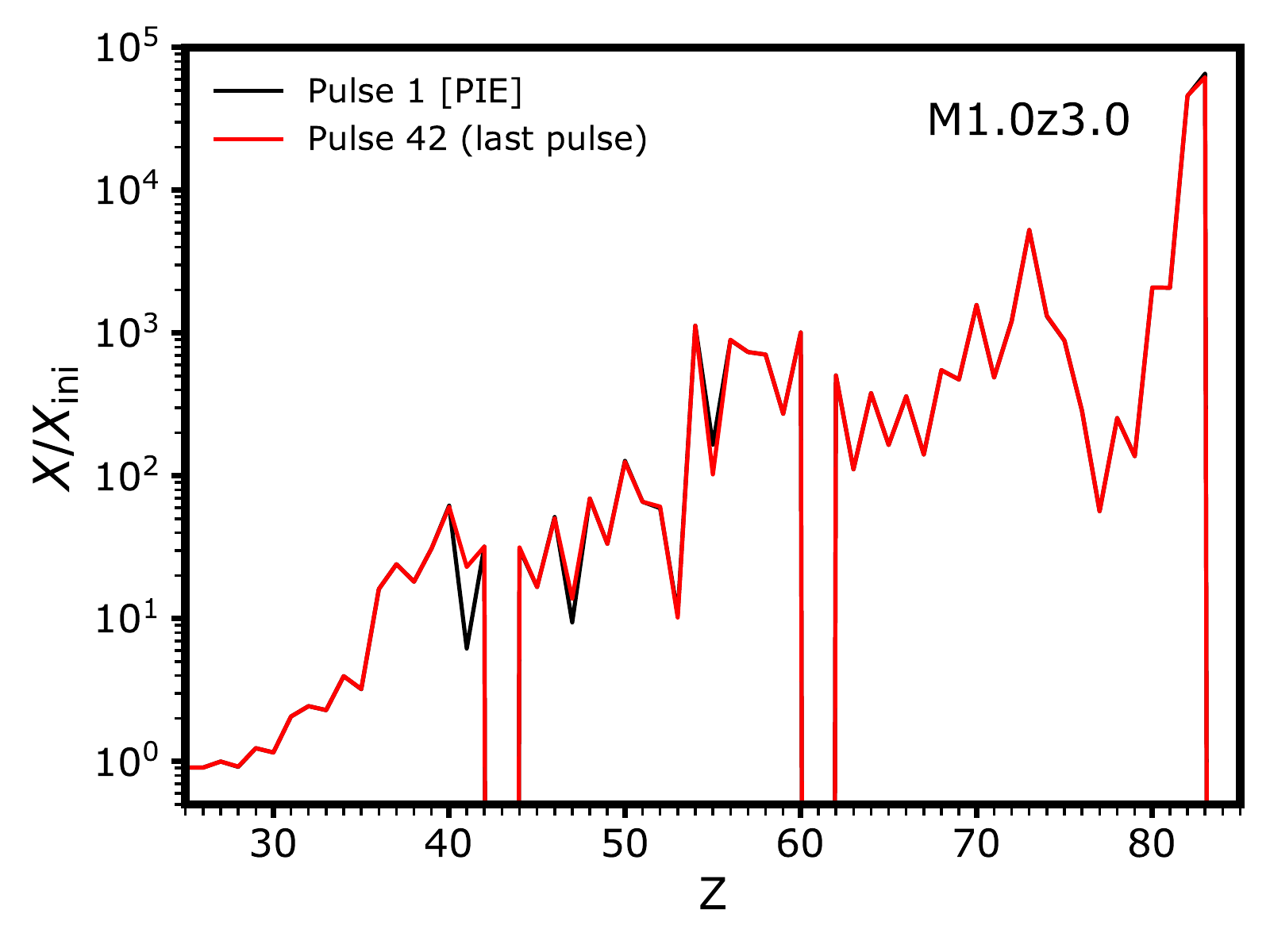}
\includegraphics[width=0.33\columnwidth]{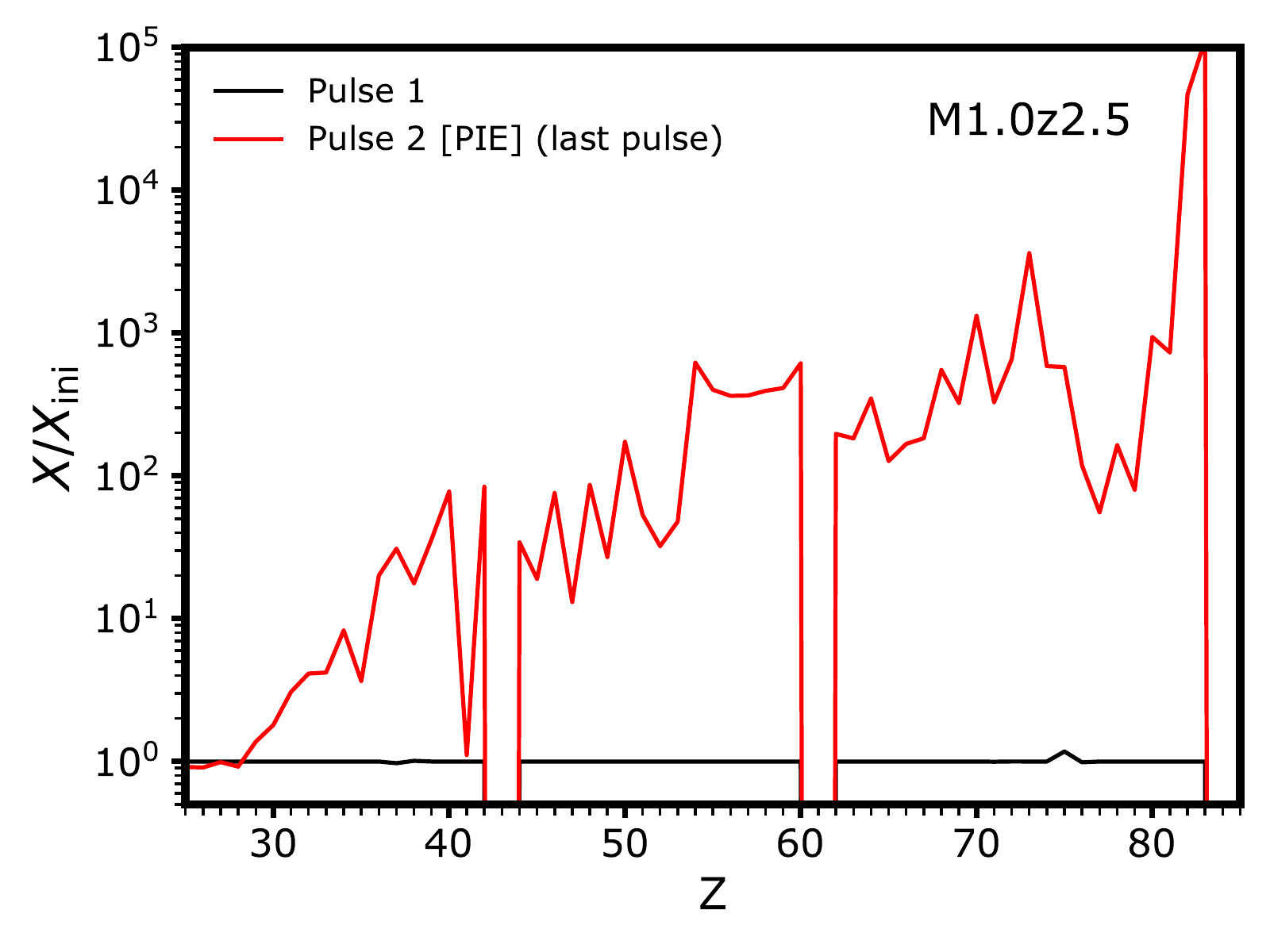}
\includegraphics[width=0.33\columnwidth]{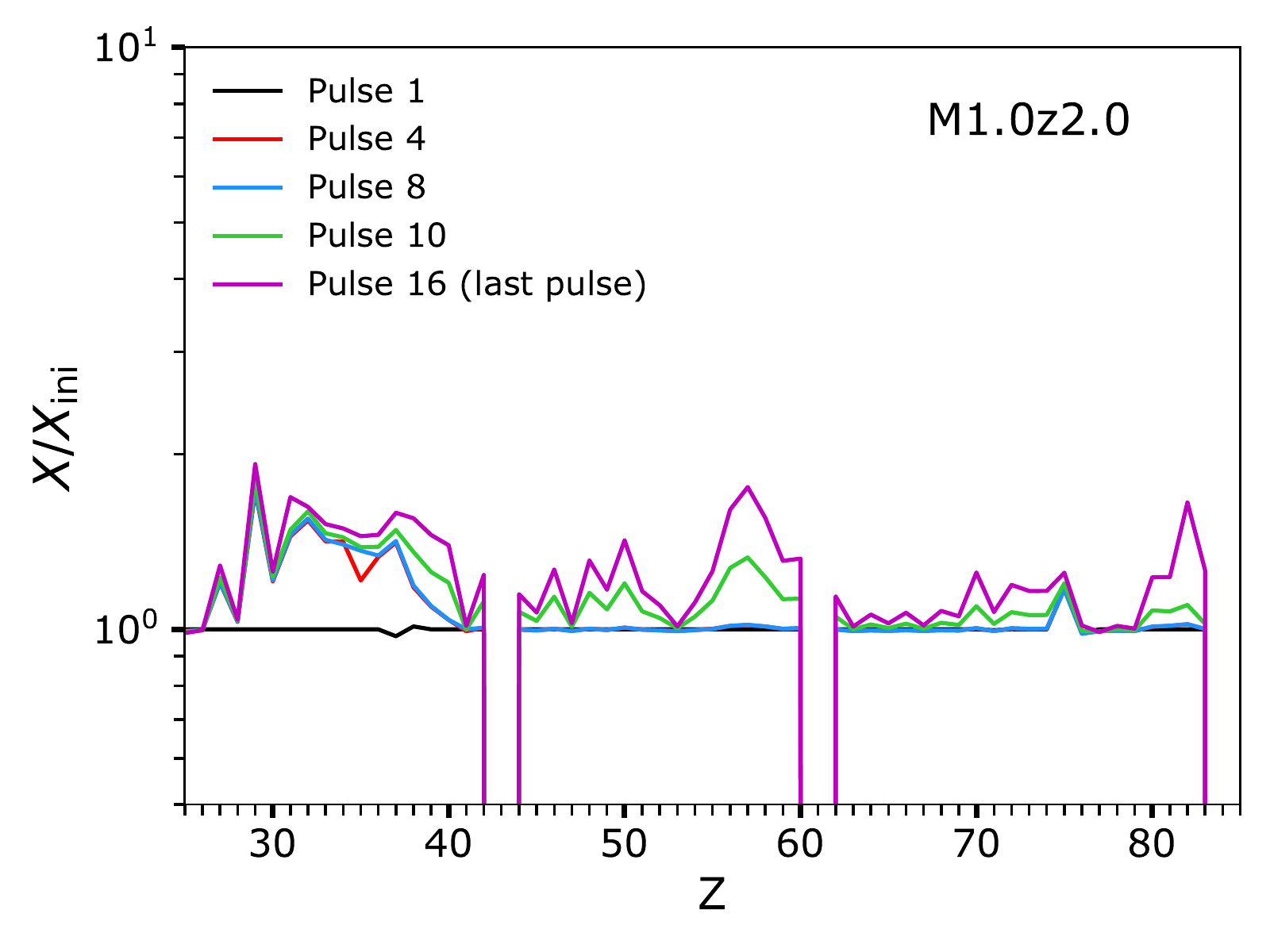}
  \end{minipage}
  \begin{minipage}[c]{2\columnwidth}
\includegraphics[width=0.33\columnwidth]{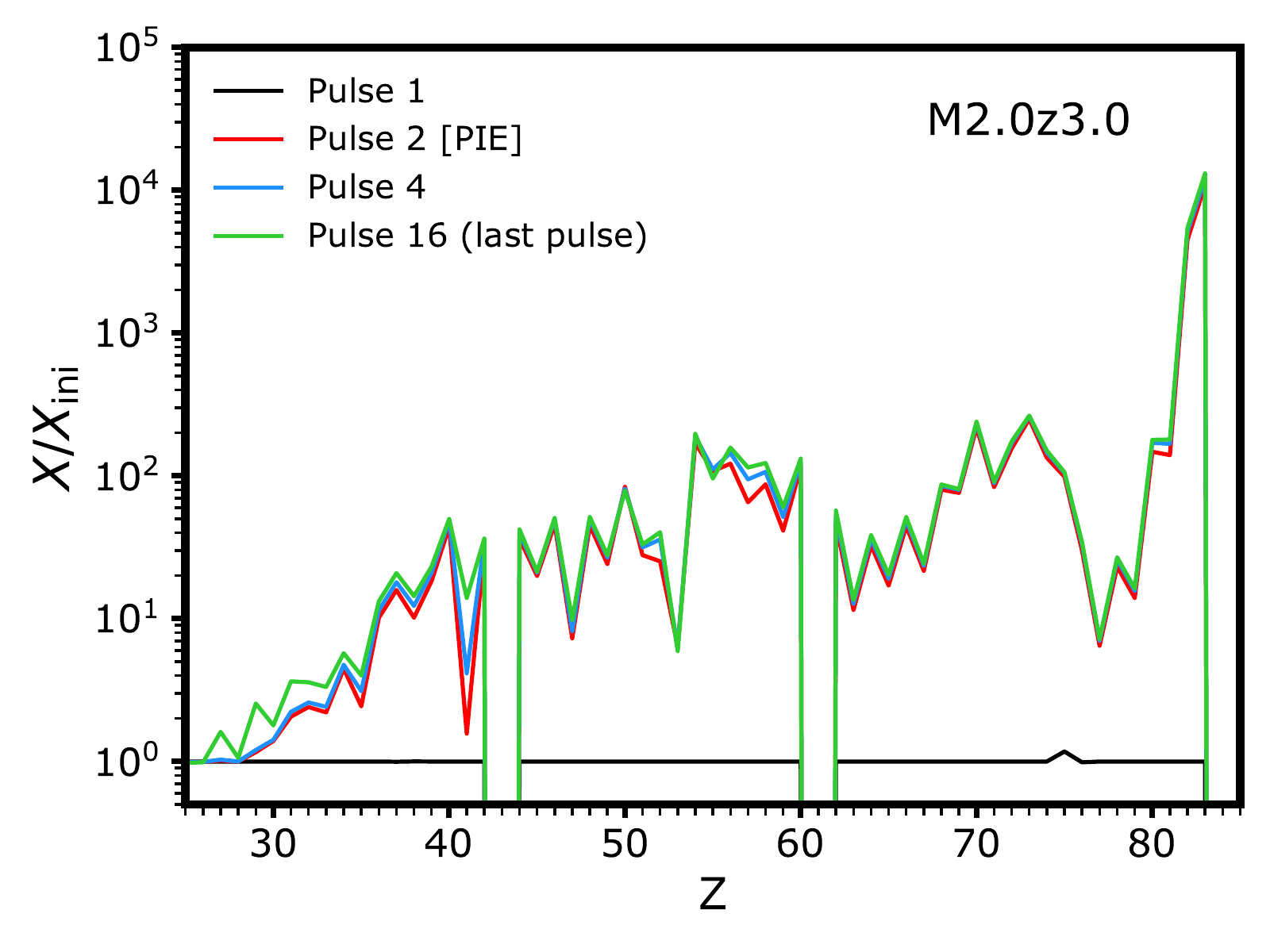}
\includegraphics[width=0.33\columnwidth]{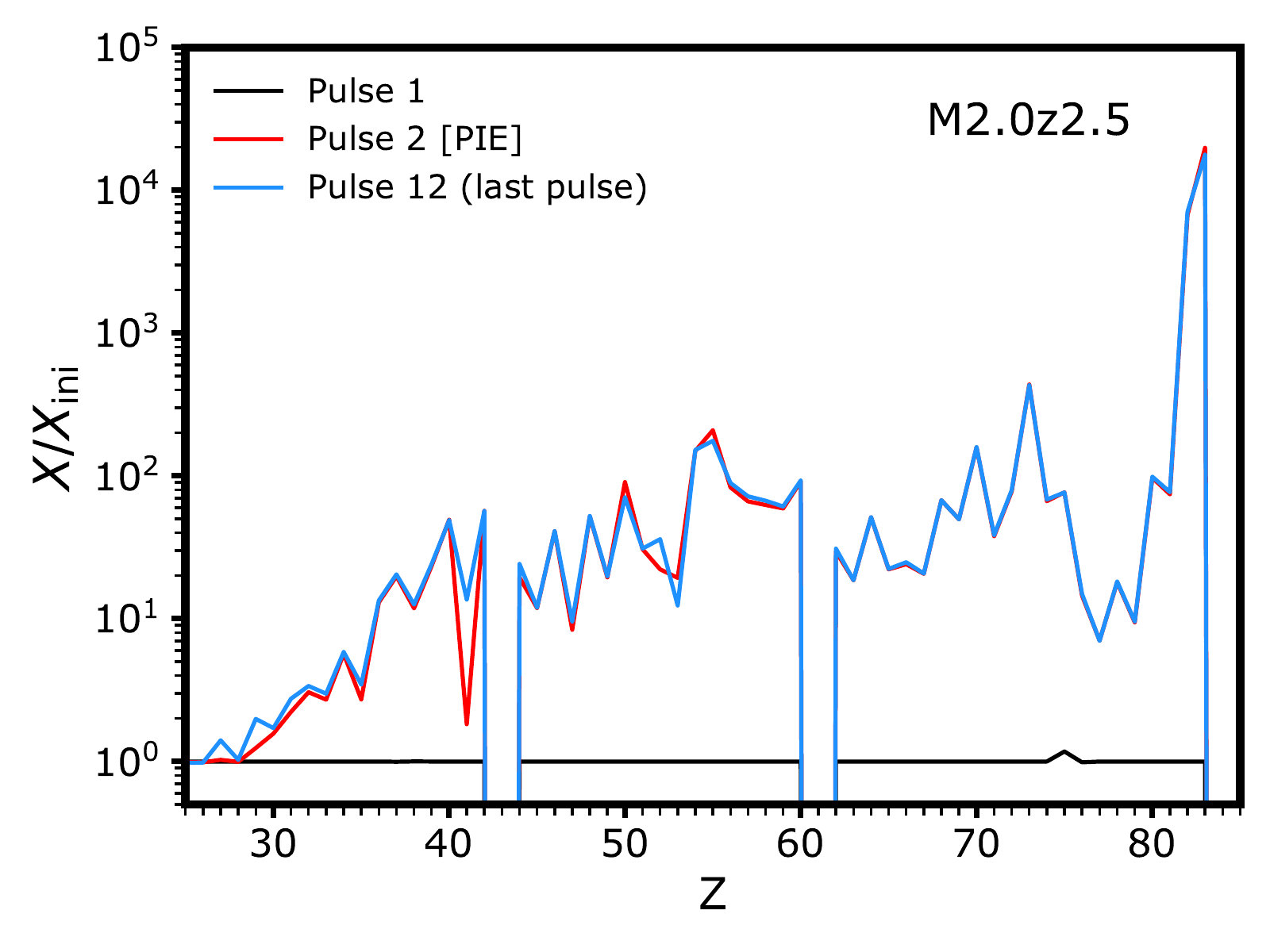}
\includegraphics[width=0.33\columnwidth]{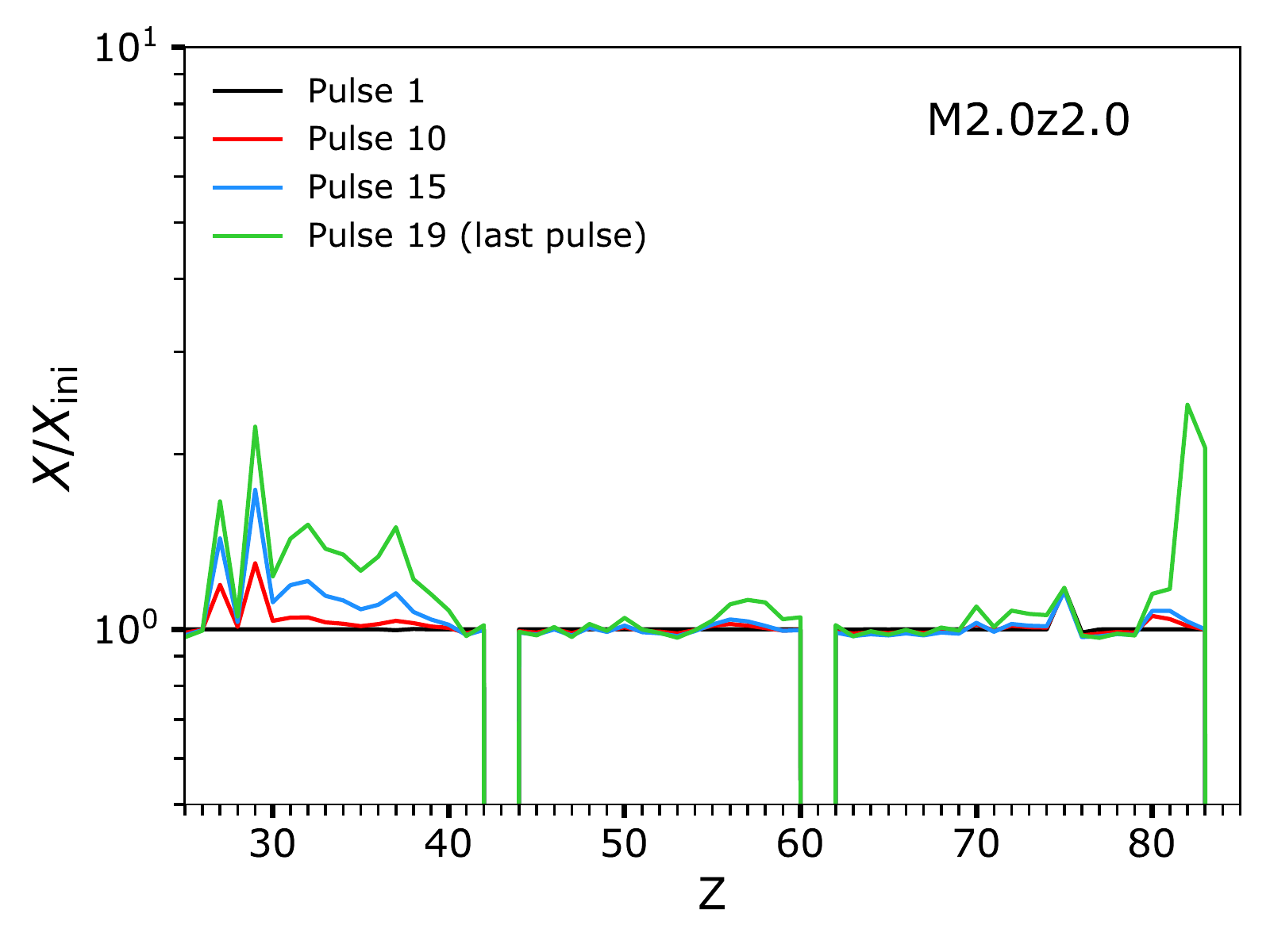}
  \end{minipage}
  \begin{minipage}[c]{2\columnwidth}
\includegraphics[width=0.33\columnwidth]{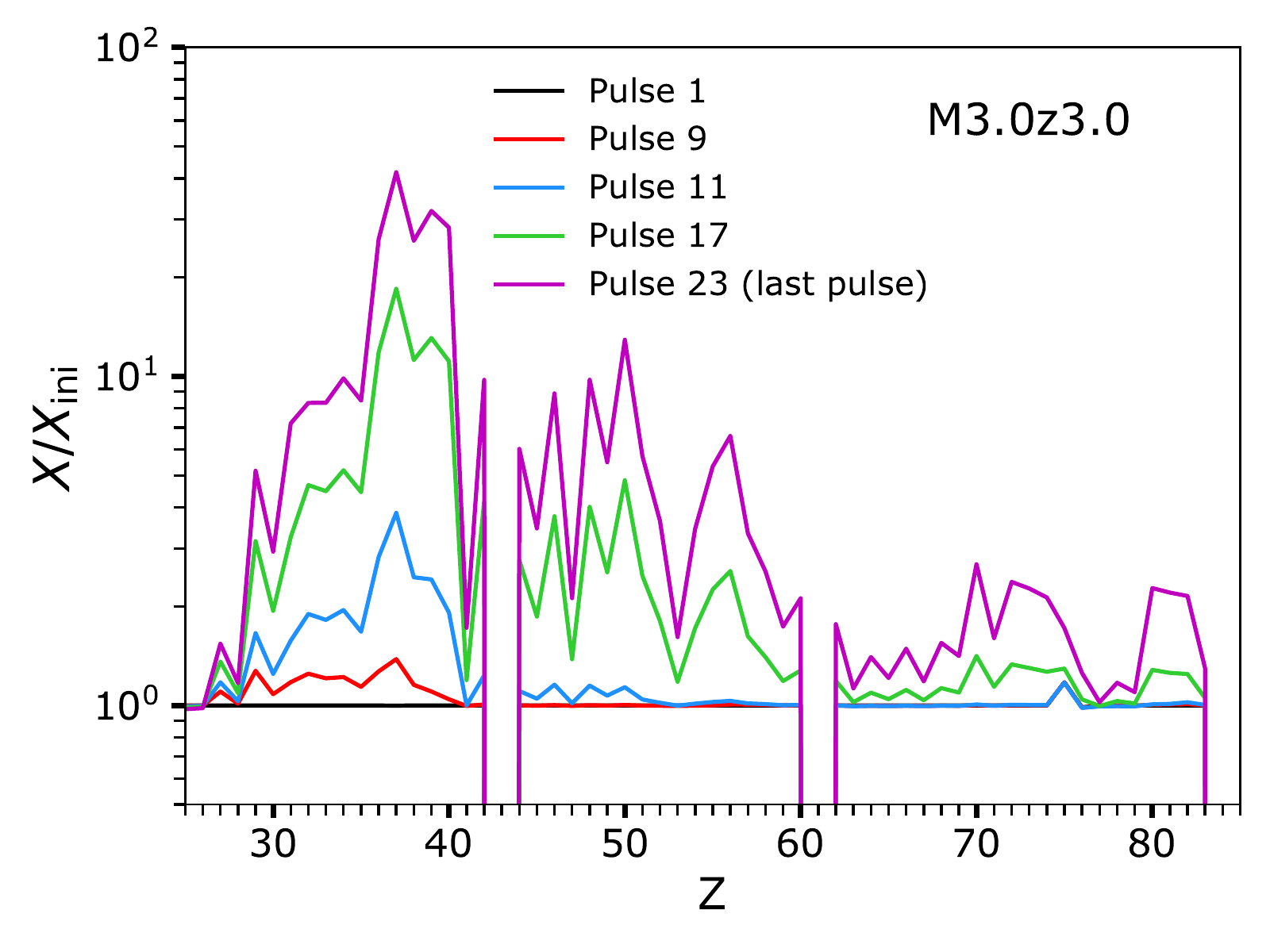}
\includegraphics[width=0.33\columnwidth]{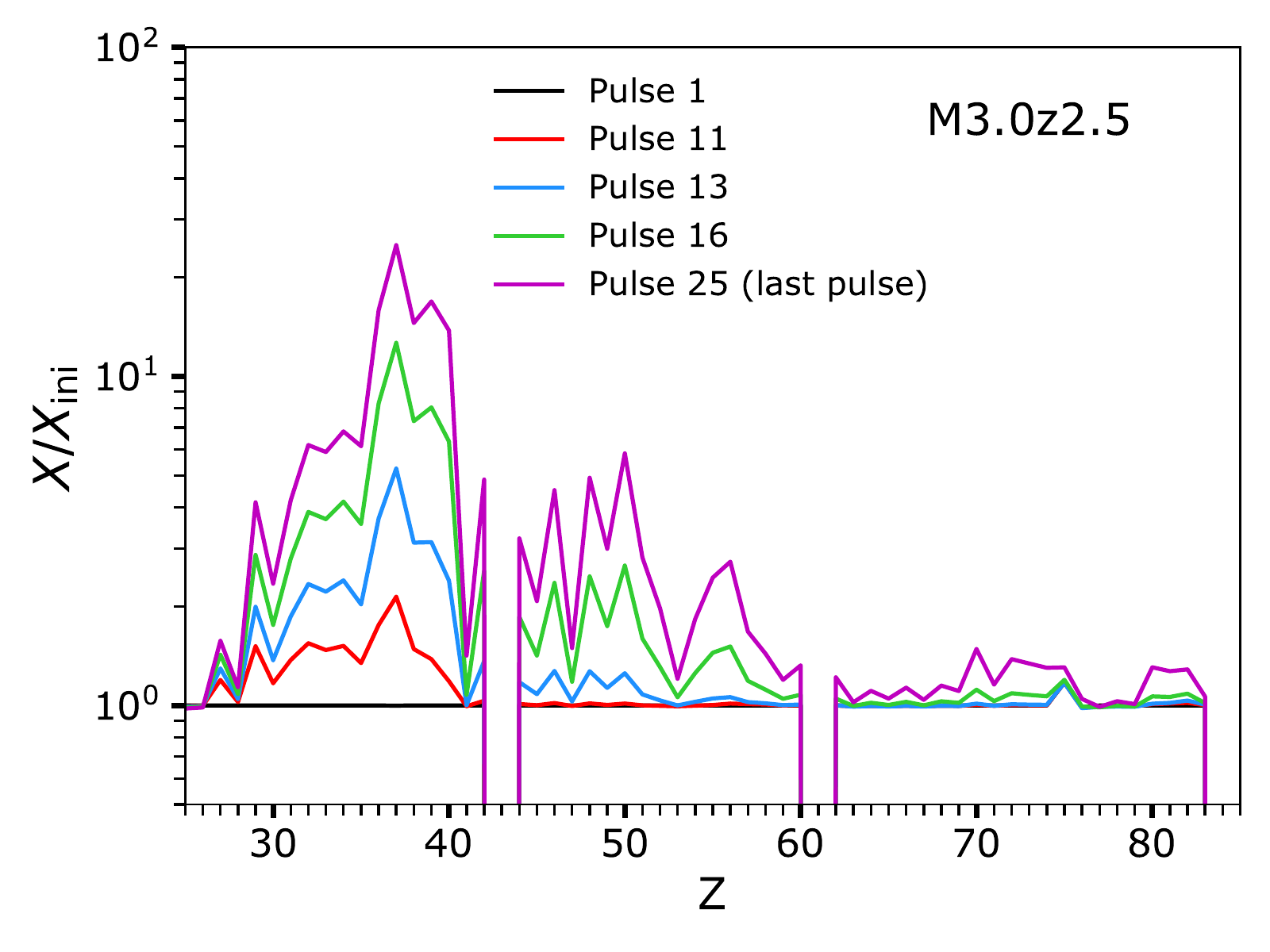}
\includegraphics[width=0.33\columnwidth]{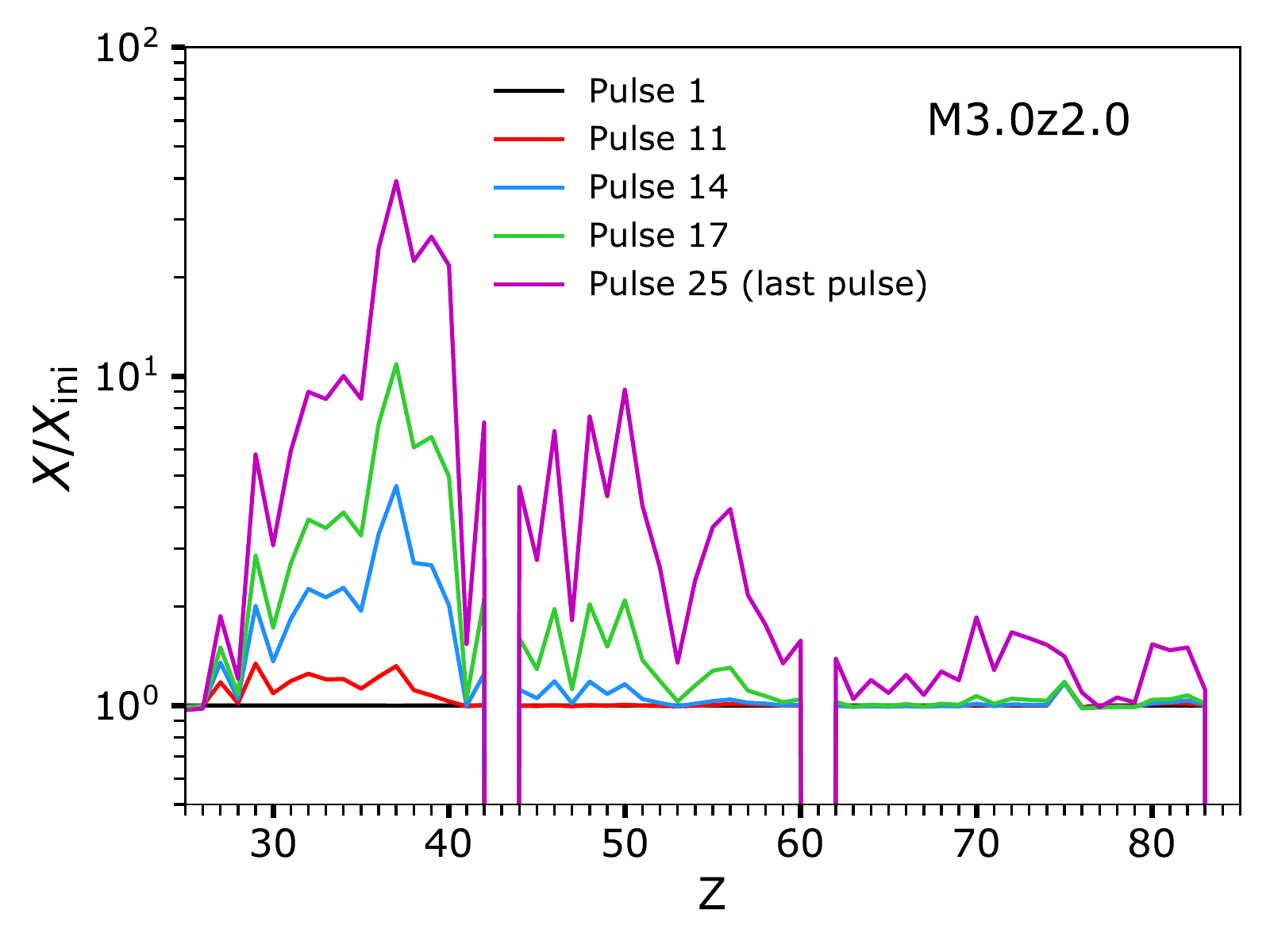}
  \end{minipage}
\caption{Surface elemental mass fractions normalized by their initial mass fractions for the nine models with $M_{\rm ini} = 1, 2$ and $3 M_{\odot}$, and [Fe/H] $=-2, -2.5$ and $-3$.
The initial mass is increasing from the top to the bottom panels and the initial metallicity is increasing from the left to the right panels. 
The patterns correspond to the surface abundances after the indicated pulse number.
The bold frames highlights models experiencing a PIE.  
Unstable isotopes are not beta-decayed and the unstable $Z=43$ and $Z=61$ elements are not considered since $X_{\rm ini} = 0$ for those elements.
}
\label{fig:enrpulse}
\end{figure*}

 \begin{figure}[h!]
\includegraphics[width=\columnwidth]{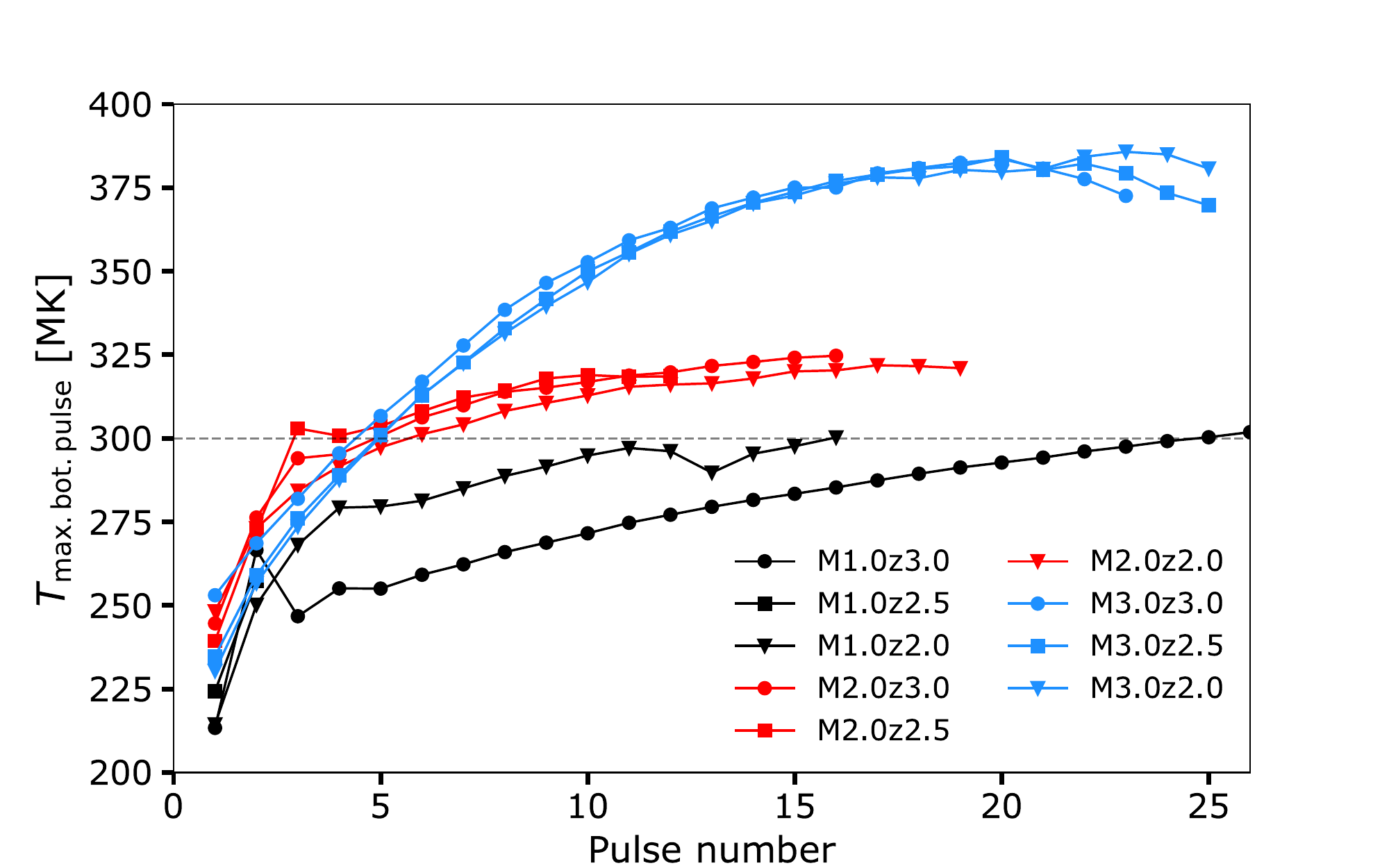}
\includegraphics[width=\columnwidth]{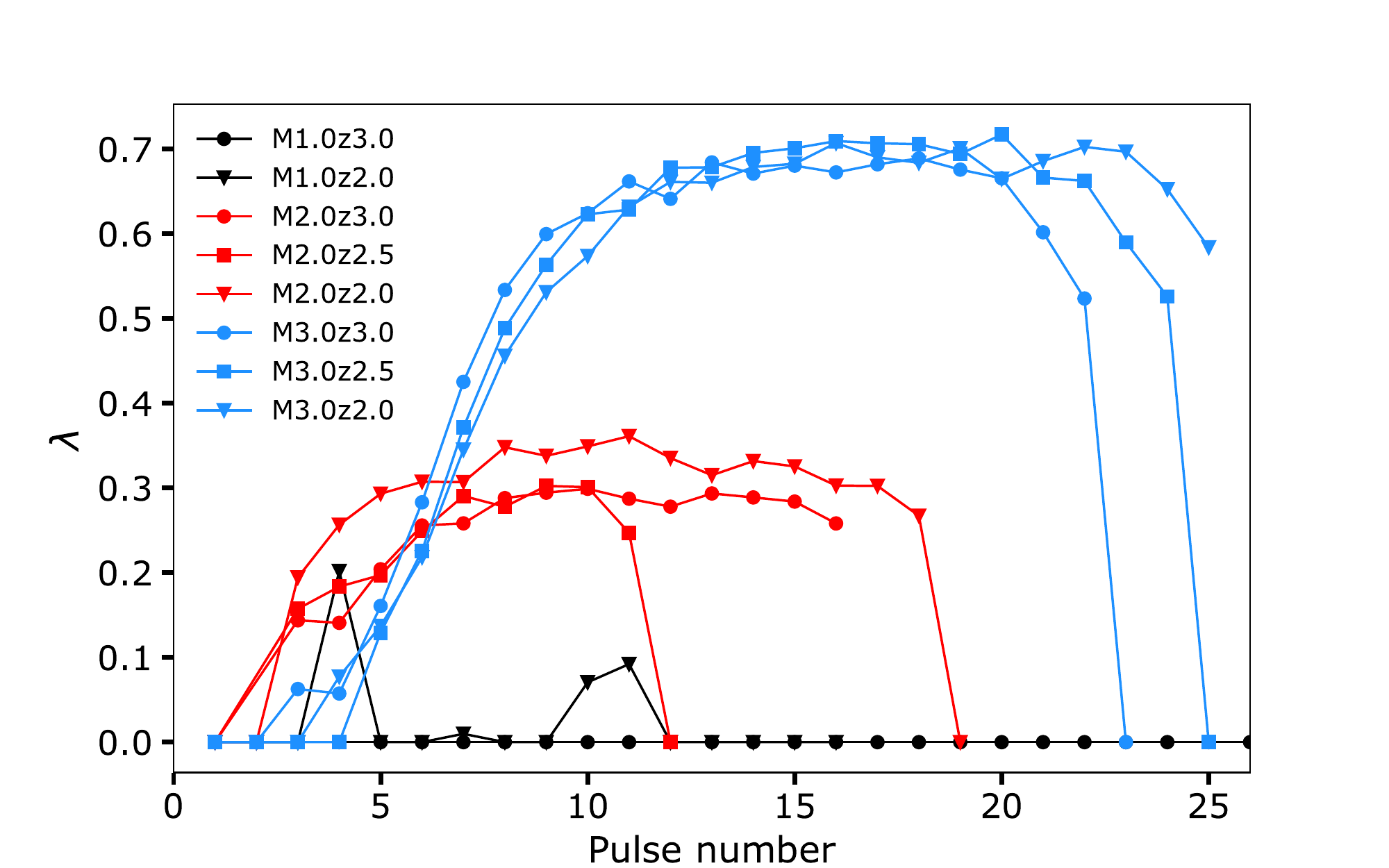}
\includegraphics[width=\columnwidth]{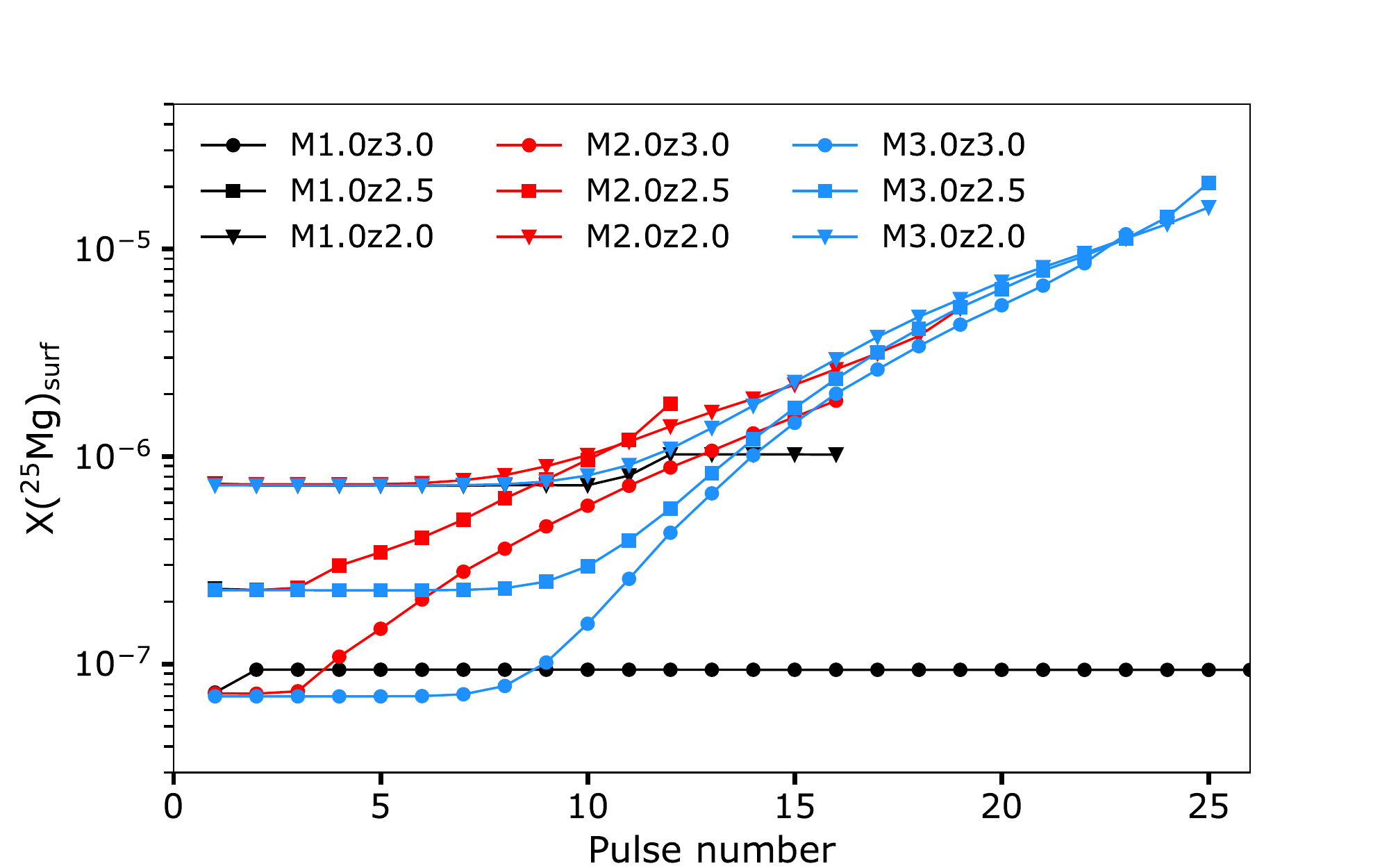}
\caption{Evolution of various quantities as a function of the pulse number. \textit{Top panel:} Maximum temperature reached at the bottom of the pulse. The horizontal dashed line shows the temperature above which the $^{22}$Ne($\alpha,n$)$^{25}$Mg reaction becomes efficient. \textit{Middle panel:} Third dredge up efficiency parameter $\lambda$ (see text for the definition). The $\lambda$ parameter for the pulses where a PIE occurs is not shown since it cannot be properly defined. \textit{Bottom panel:} Surface $^{25}$Mg mass fraction. 
}
\label{fig:ne22surf}
\end{figure}


\section{Nucleosynthesis, surface enrichment, and yields}
\label{sect:nucleo}

\subsection{Pulse splitting and the impact of the convective velocity}
\label{sect:conv}

As discussed in Sect.~\ref{sect:splitnosplit}, a split can take place at the location where $\tau_{\rm C}$ the timescale associated with the reaction \iso{12}C($p,\gamma$)\iso{13}N is similar to $\tau_{\rm conv}^{\rm loc}$ the local transport timescale of protons by convection. To better understand this aspect, we recomputed the M1.0z3.0 model starting just before the PIE while arbitrarily changing the convective velocity $v_{\rm conv}$.
This numerical experiment allows for the role of convective velocity on the splitting of the convective pulse to be understood.
We divided $v_{\rm conv}$ by a factor of 10, 100, and 1000.

The split occurs at $M_{\rm r}=0.516$~\Msun\ for the $v_{\rm c}/10$ model and at $M_{\rm r}=0.527$~\Msun\ for the $v_{\rm c}/100$ model (Fig.~\ref{fig:vc_kip}). 
The standard M1.0z3.0 model does not experience a proper split (as explained in Sect.~\ref{sect:splitnosplit}) and the $v_{\rm c}/1000$ model follows a standard evolution.
It is clear that the lower the convective velocity the higher in mass the split occurs.
This is because a lower $v_{\rm conv}$ leads to a higher $\tau_{\rm conv}$ and since the split occurs where $\tau_{\rm C} \sim \tau_{\rm conv}^{\rm loc}$, if $\tau_{\rm conv}$ is larger, this location corresponds to a higher mass coordinate, where the temperature is lower and thus the reaction rate of \iso{12}C($p,\gamma$) weaker (i.e.  $\tau_{\rm C}$ is smaller).

In the $v_{\rm conv}/1000$ model, protons do not go deep enough in the pulse. They do not release enough energy through $^{12}$C($p,\gamma$)$^{13}$N to make a clear second energy peak in the pulse (as in Fig.~\ref{fig:fluxes4} top panel).
This prevents the splitting of the pulse.
Also, the turnover timescale $\tau_{\rm conv}$ is about 1~yr, which becomes non-negligible compared to the pulse evolution timescale ($10-100$~yr). The pulse recedes in mass before significant protons could be engulfed and burnt.

The maximal neutron densities obtained in the four cases mentioned above  are shown in Fig.~\ref{fig:vc_Nn} and the resulting surface abundances after the PIE are shown in Fig.~\ref{fig:vc_ab}.
The lower the convective velocity, the lower the neutron density and the less dramatic the surface enrichment in heavy elements. If $v_{\rm conv}$ is too small, protons do not have time before the split occurs to reach the hot layers of the pulse where $^{13}$C($\alpha$,$n$)$^{16}$O is efficiently activated. 
Nevertheless, the convective velocity has to be decreased by a factor of more than 10 before seeing any effect on the nucleosynthesis. The $v_{\rm conv}/10$ model (red pattern in Fig.~\ref{fig:vc_ab}) leads to a similar surface enrichment compared to the standard model (black pattern). 

We also consider a case where $v_{\rm conv}$ multiplied by a factor of 10, according to the results of a 3D simulation by \cite{stancliffe11}. 
Increasing the convective velocity beyond the standard value has a small impact  (Figs.~\ref{fig:vc_Nn} and \ref{fig:vc_ab}). The maximal neutron density profile is similar to the standard case, although it reaches slightly higher values and the abundances are lower by a factor of typically $5-7$ for $60 \leq Z<82$ (except for Tantalum at $Z=73$, where the abundance is lower by a factor of 14). 

These results show that the i-process nucleosynthesis and surface enrichment following a PIE is a robust feature that does not depend strongly on the convective velocity. 
Nevertheless, we are aware that the physics of convection (hence, of PIE) cannot be fully captured by 1D models on the basis of the MLT formalism.

\subsection{Surface enrichment after a PIE}
\label{sect:surfenr}

Once a PIE has occurred, the important question of whether or not the nucleosynthesis products are brought up to the surface remains.
The merging of the convective pulse with the convective envelope will bring some pulse material up to the surface.
However, if the pulse splits, the lower part of the pulse after the split will remain locked deep into the star (this material can nevertheless reach the surface during subsequent thermal pulses, if any, followed by third dredge ups).
In this case, one major question is whether or not i-process nucleosynthesis happens before the split, both before and after the split, or only after the split. If it happens only before the split for instance, the surface will be fully enriched by i-process products.
To address this point, we define the following enrichment ratio, $e:$
\begin{equation}
e =  \frac{t_{\rm split} - t_{\rm Nnmax}}{\langle \tau_{\rm conv} \rangle}  
\label{eq:eratio}
,\end{equation}
where $t_{\rm split}$ is the time of the split, $t_{\rm Nnmax}$ the time corresponding to the maximum neutron density, and $\langle \tau_{\rm conv} \rangle$ is the time-averaged convective turnover timescale between $t_{\rm Nnmax}$ and $t_{\rm split}$, which can be written as:
\begin{equation}
\langle \tau_{\rm conv} \rangle = \frac{1}{t_{\rm split} - t_{\rm Nnmax}}  \int_{t_{\rm split}}^{t_{\rm Nnmax}} \tau_{\rm conv} \, (t) \, dt.
\end{equation}
We assume that most of the i-process nucleosynthesis takes place at $t = t_{\rm Nnmax}$. 
If $e \gg 1 $, the i-process material has enough time to homogenize in the thermal pulse before the split. In this case, the i-process products are fully transported to the surface. 
If $0 < e < 1 $, the split would occur before the pulse homogenization. 
A negative $e$ means that the neutron density peak takes place after the splitting. 
In this case, the i-process material is locked into the lower part of the pulse and will not enrich the surface.

In all our models experiencing a PIE and a split, $e \gg 1$. 
For the M2.0z3.0 model for instance, $t_{\rm split} - t_{\rm Nnmax} = 0.23$~yr $= 1985$~hr and the mean turnover timescale in the convective pulse during this interval is 3.95 hr. This leads to $e=503$.
For the M1.0z2.5 model,  $t_{\rm split} - t_{\rm Nnmax} = 0.014$~yr $= 125$~hr, $\tau_\mathrm{conv} \approx 1.70$~hr leading to\footnote{In Paper I, we reported a value of $t_{\rm split} - t_{\rm Nnmax} = 0.002$~yr for this model (the caption of Figure 2). This leads to $e \sim 15$. The difference is due to the fact that in the present paper, we recomputed this model with the latest version of {\sf STAREVOL} for a sake of homogeneity. Slight changes were noticed between the old and new M1.0z2.5 model. } $e=74$.
The enrichment factors $e$ for our models are reported in Table~\ref{table:2}.

\subsection{The $^{13}$N($n,p$)$^{13}$C reaction}
\label{sect:reac}

Because protons mix into an He-burning region, PIEs involve different reactions from both H-burning at very high temperatures and overall He-burning. 
As can be seen in the top panels of Figs.~\ref{fig:fluxes4} and \ref{fig:fluxes5}, the proton abundance decreases with decreasing $M_{\rm r}$ until a point where it increases again, close to the bottom of the pulse.
The protons here are not those that are ingested and transported downwards in the pulse.
They come from the $^{13}$N($n,p$)$^{13}$C reaction and, to a smaller extent, from $^{14}$N($n,p$)$^{14}$C. 
At $200-250$~MK, the rate of $^{13}$N($n,p$)$^{13}$C is about 500 times higher than the rate of $^{14}$N($n,p$)$^{14}$C.
The neutrons required for these reactions mainly come from the $^{13}$C($\alpha,n$)$^{16}$O reaction. 
The $^{13}$N isotope is synthesized upwards in the pulse by $^{12}$C($p,\gamma$)$^{13}$N and diffuses downwards until it reaches neutron-rich layers. Although $^{13}$N decays to $^{13}$C via $\beta^+$ in only 10 minutes, this timescale is comparable to the time it takes for $^{13}$N to be transported from the middle to the bottom of the pulse. 
This can be seen in Fig.~\ref{fig:fluxes5} (top and bottom panel) where it is clear that the $^{13}$N($\beta+$)$^{13}$C reaction is active, even at the bottom of the convective pulse.
At the end, the production of protons at the bottom of the pulse reactivates the $^{12}$C($p,\gamma$)$^{13}$N reaction.
Although the production of protons remains rather small, it is much stronger in the M1.0z3.0 model (Fig.~\ref{fig:fluxes5}, top panel) than in the M2.0z3.0 model (Fig.~\ref{fig:fluxes4}, top panel). 

The rate of the $^{13}$N($n,p$)$^{13}$C reaction is derived as the reverse rate of $^{13}$C($p,n$)$^{13}$N, available in both the compilations of \citet{caughlan88} and \citet[][NACRE]{angulo99}. 
In this work we use the rate from \cite{angulo99}.

As a numerical test, we re-computed a PIE model with the rate of $^{13}$N($n,p$)$^{13}$C divided or multiplied by a factor of 10. We found that the surface abundances of trans-iron elements after the PIE are impacted by a factor of about 3 at most.  
Therefore, although this reaction appears to be important for i-process nucleosynthesis, both as a poisoning reaction and as an energy source, its variation by a factor of 10 does not impact the outcome of our models significantly. 
A more detailed analysis of the existing uncertainties and their subsequent impact on nucleosynthesis is required to draw firmer conclusions.

 \begin{figure}[h!]
\includegraphics[scale=0.54, trim = 0cm 1.5cm 0cm 0cm]{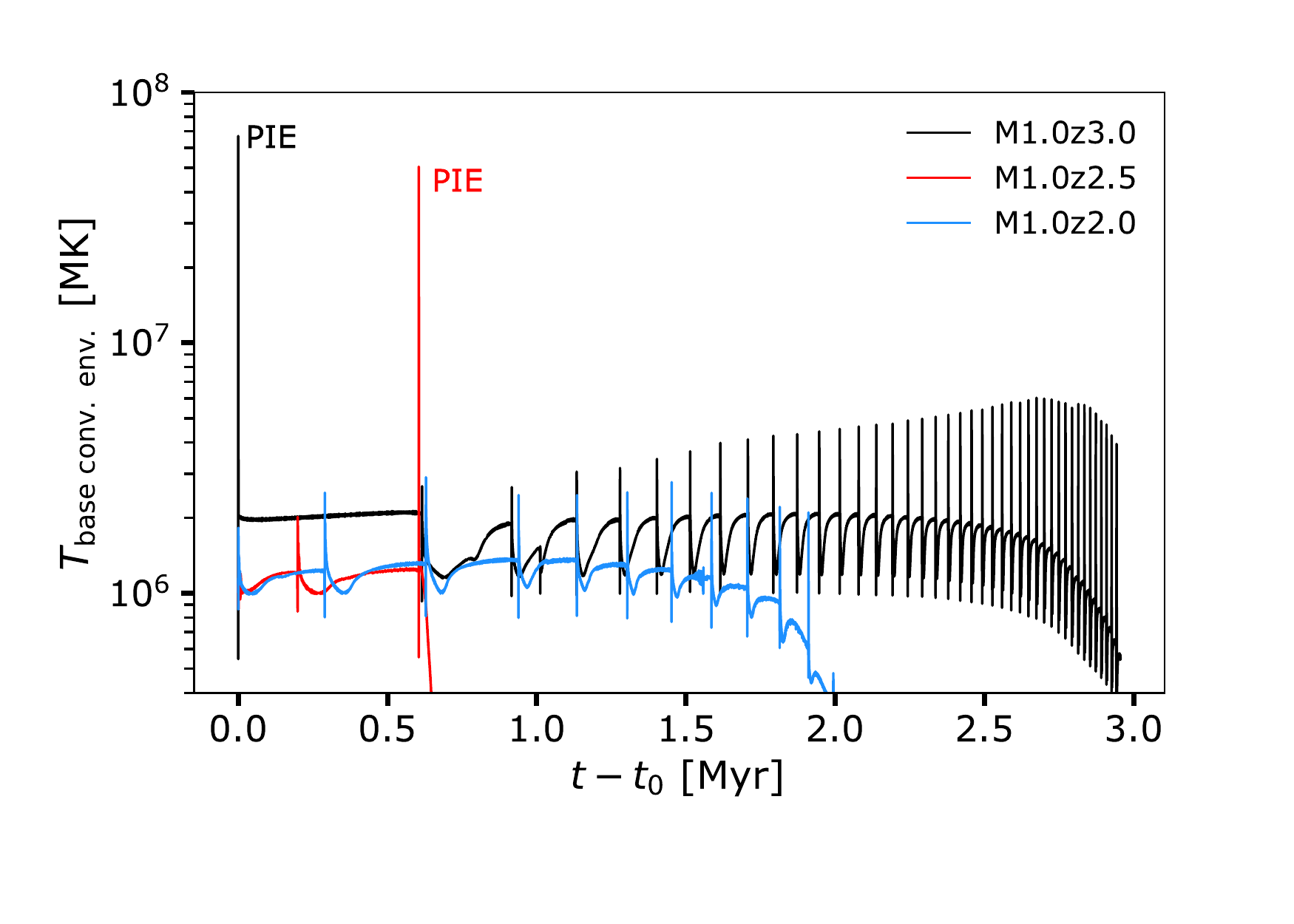}
\includegraphics[scale=0.54, trim = 0cm 1.5cm 0cm 0cm]{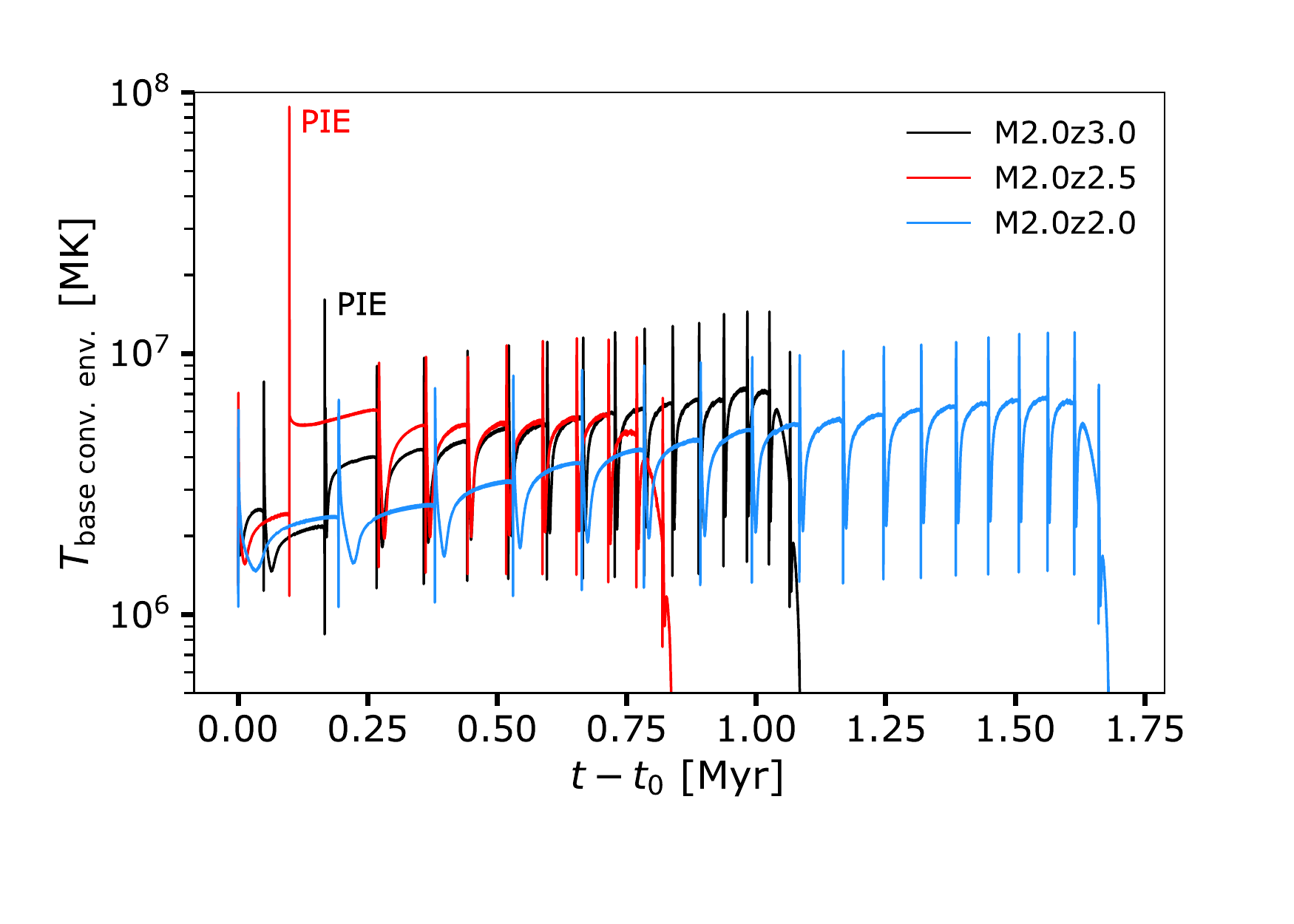}
\includegraphics[scale=0.54, trim = 0cm 1.5cm 0cm 0cm]{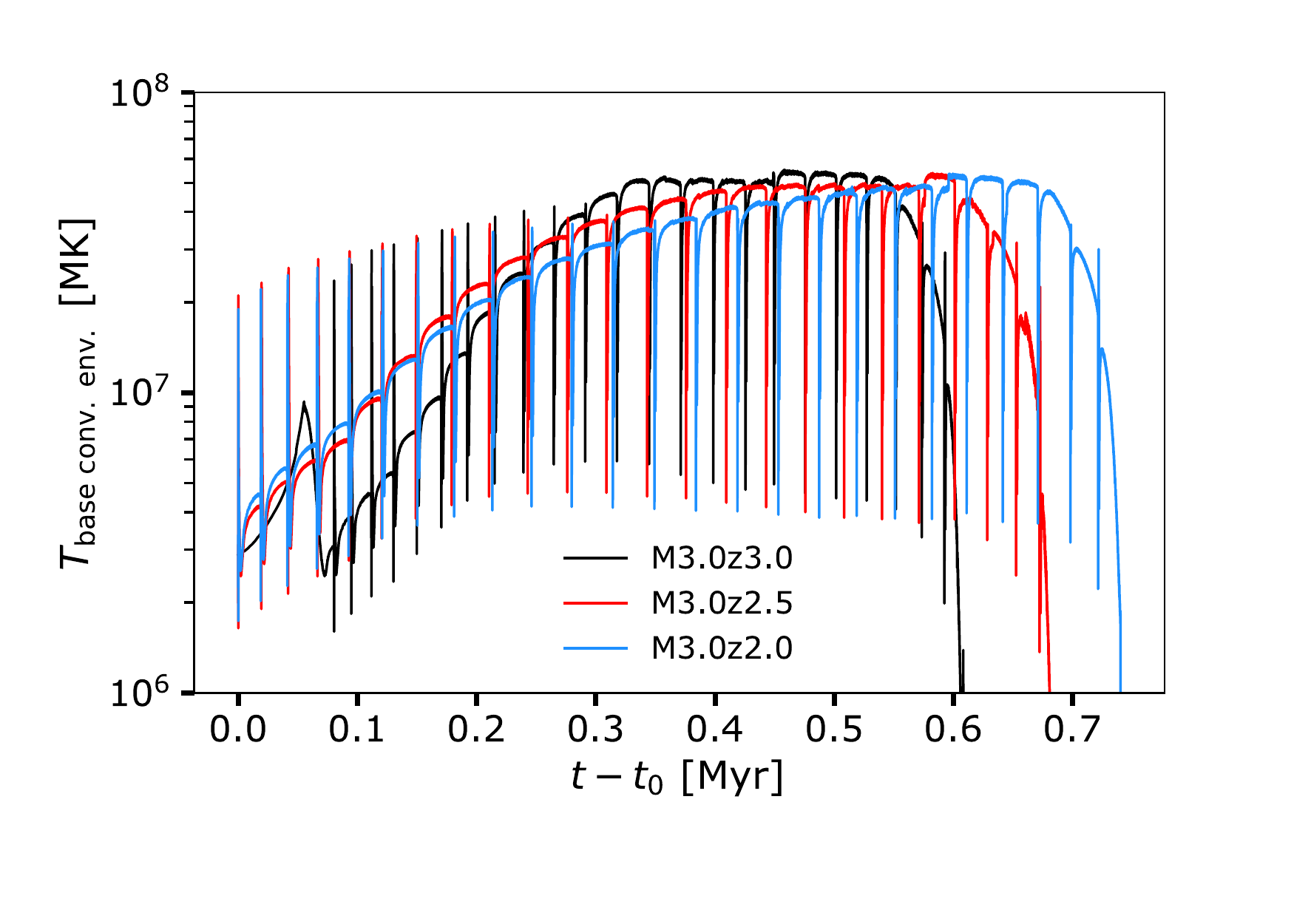}
\caption{Temperature at the bottom of the convective envelope during the AGB phase. Models of 1, 2, and 3 \Msun\, are shown in the top, middle, and bottom panels, respectively.
}
\label{fig:tbase}
\end{figure}

 \begin{figure}[h!]
\includegraphics[scale=0.46, trim = 0cm 0cm 0cm 0cm]{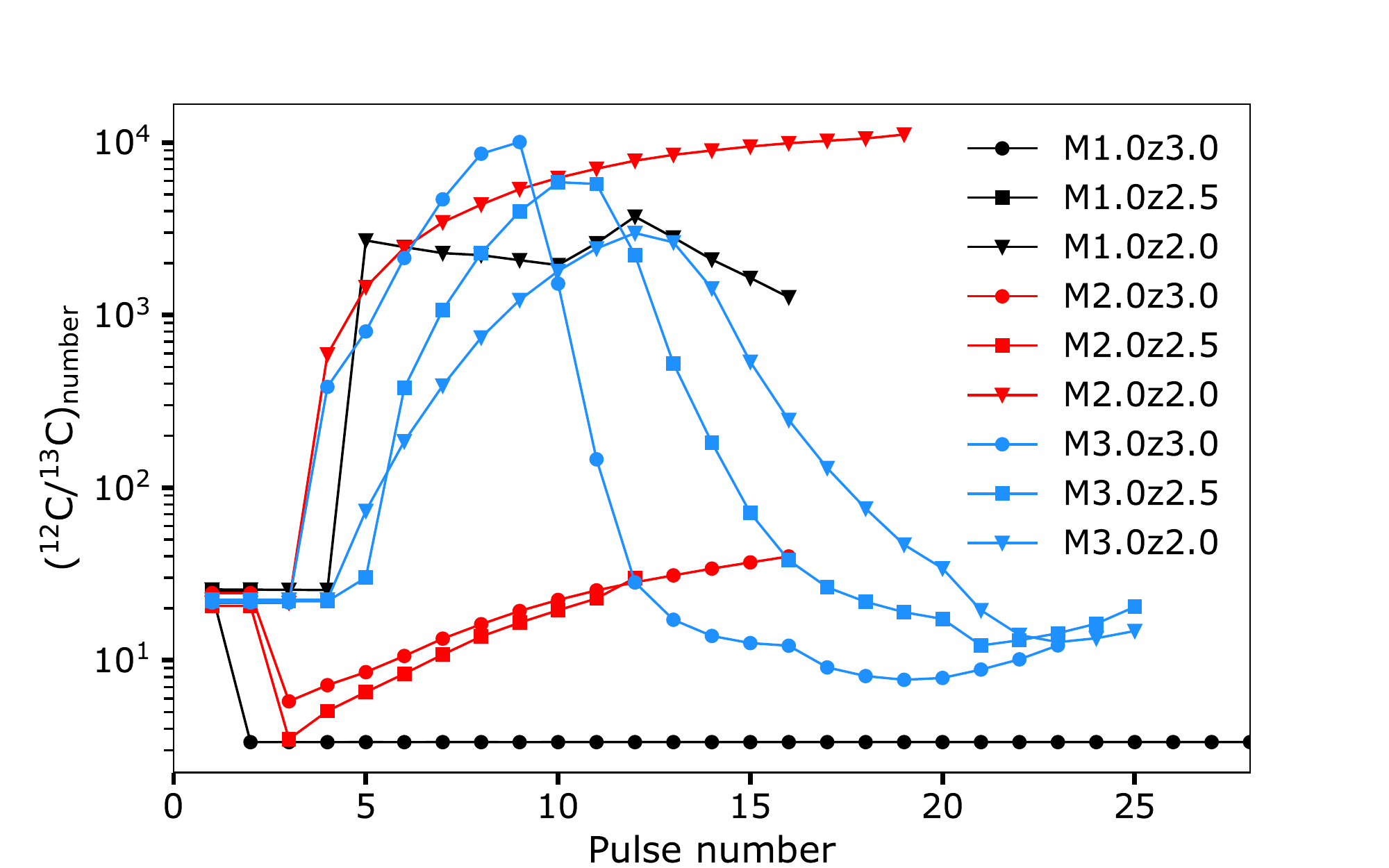}
\caption{
Surface $^{12}$C/$^{13}$C ratio in number as a function of the pulse number. 
The ratio is shown at the middle of each pulse (i.e. before the third dredge up, if any). 
}
\label{fig:CCall}
\end{figure}

\subsection{i-process nucleosynthesis}
\label{sect:ipronuc}

We highlight here the main features of the i-process nucleosynthesis in our models (see also  Sect.~\ref{sect:yields} for a discussion on the overall i-process yields).
For additional details, we refer to Paper I and II, where the i-process nucleosynthesis has already been extensively discussed. 

Our models with PIE experience a maximum neutron density ranging between $6.8 \times 10^{13}$ and $2.2 \times 10^{15}$~cm$^{-3}$ (Table~\ref{table:2}). 
The i-process paths for these two extreme models (M2.0z3.0 and M1.0z2.3) are shown in Fig.~\ref{fig:flow}. As expected, the M1.0z2.3 model with a higher neutron density follows a path which is farther away from the valley of $\beta$-stability. 
The i-process in our other models follows similar i-process path, in between the blue and green paths shown in Fig.~\ref{fig:flow}.

Figure~\ref{fig:enrpulse} shows that in the models experiencing a PIE (bold frames) the surface is enriched right after the PIE (when the pulse merges with the envelope) and is barely modified during the subsequent TP-AGB evolution. 
The rather weak convective s-process that may take place in the thermal pulses after the PIE (as discussed in Sect.~\ref{sec:convspro}) does not impact the surface abundances in these models.

Our i-process models show, on average, a rising distribution from Fe to Pb. 
Tantalum ($Z=73$) is largely overproduced except in the M2.0z3.0 model which experiences the weakest i-process (cf. Table~\ref{table:2}).
In Paper I, where only the M1.0z2.5 model was considered, we noticed that elements with $Z \lesssim 50$ were underproduced by $\sim 0-0.5$~dex compared to a standard s-process nucleosynthesis, while heavier elements with $Z \gtrsim 50$ were overproduced by $\sim 0-0.5$~dex. 
The important production of heavier elements is visible in the M1.0z3.0 and M1.0z2.5 models (Fig.~\ref{fig:enrpulse}), where we notice a jump between I ($Z=53$) and Xe ($Z=54$). 
In the 2 \Msun\ models, this feature is less pronounced. The difference between the elements with $Z \lesssim 50$ and $Z \gtrsim 50$ is also smaller. 

The progressive increase in Nb ($Z=41$) in the M2.0z3.0 model for instance is due to the decay of $^{93}$Zr via $^{93}$Zr($\beta^-$)$^{93}$Nb. The lifetime of $^{93}$Zr is 1.61~Myr which is comparable to the AGB lifetime (1.09 Myr for the M2.0z3.0 model, cf. Table~\ref{table:2}).

\subsection{The convective s-process}
\label{sec:convspro}

After about ten pulses, the temperature at the bottom of the pulse of our 3~\Msun\, models is high enough (at least 350~MK, Fig.~\ref{fig:ne22surf}, top panel) to efficiently activate the $^{22}$Ne($\alpha$,$n$)$^{25}$Mg reaction. 
This leads to the convective s-process \citep[e.g.][]{goriely05,karakas14}. Our 2~\Msun\, models barely reach the required temperature and therefore only experience a very weak convective s-process, as detailed below.

The middle panel of Fig.~\ref{fig:ne22surf} shows the $\lambda$ parameter which indicates the third dredge-up efficiency. It is defined as the ratio of the mass extent of the envelope in the pulse region to the mass increase in the H-free region during the interpulse \citep[e.g.][]{karakas02}. 
A high value corresponds to an efficient third dredge-up\footnote{We note that the $\lambda$ parameter is only defined for standard thermal pulses, not when a PIE occurs.}. 
The $\lambda$ parameter is greater than zero in our 2 and 3 \Msun\, models, which means that some of the pulse material reaches the stellar surface. 
In particular, this leads to an increase in $^{25}$Mg (Fig.~\ref{fig:ne22surf}, bottom panel) which is a signature of the operation of the $^{22}$Ne($\alpha$,$n$)$^{25}$Mg reaction in the convective pulse. 
The surface enrichment in $^{25}$Mg is smaller for the 2 \Msun\, models (red patterns) compared to the 3 \Msun\, models (blue patterns) due to the lower pulse temperatures of 325~MK and smaller dredge-up efficiency ($\lambda$) in the 2 \Msun\, models.

In the 3 \Msun\, model, the convective s-process leaves a clear chemical signature at the surface. After about ten pulses, the envelope already shows enrichment in light s-elements (Fig.~\ref{fig:enrpulse}, bottom panels). The dilution of the pulse material in the large convective envelope does not lead to overproduction factor greater than about 50.

The situation is different in the 2 \Msun\, models. 
The M2.0z2.0 model (which does not experience any PIE) shows a very weak signature of the convective s-process at its surface. The overproduction factors does not exceed a factor of $2-3$ (Fig.~\ref{fig:enrpulse}, middle right panel). 
The temperature of 325~MK at the bottom pulse in this model is too low to leave a noticeable s-process signature at the surface.  
The M2.0z3.0 and M2.0z2.5 models experience a PIE during the second pulse which strongly impacts the envelope composition. The subsequent dredge-up events have little effect on the surface abundances which remain almost unaffected (Fig.~\ref{fig:enrpulse}, middle panels).

\begin{table}
\scriptsize{
\caption{Surface $^{12}$C/$^{13}$C, C/N, O/N, and C/O ratios in number just after the PIE.
\label{table:3}
}
\begin{center}
\resizebox{7.0cm}{!} {
\begin{tabular}{lcccc} 
\hline
& $^{12}$C/$^{13}$C  & C/N & O/N & C/O        \\
\hline
M1.0z3.0               & 3.4  & 0.2 & 0.3 & 0.8        \\
M1.0z3.0$\alpha$ & 4.1  & 2.3 & 0.6 & 3.8        \\
M1.0z2.5               & 4.6  & 2.5 & 1.0 & 2.6        \\
M1.0z2.3               & 4.6  & 3.4 & 0.8 & 4.0        \\
\hline
M2.0z3.0               & 5.8  & 8.2 & 1.0 & 7.8        \\
M2.0z2.5               & 3.5  & 0.9 & 1.4 & 0.7        \\
\hline 
\end{tabular}
}
\end{center}
}
\end{table}

\subsection{Hot hydrogen-burning}
\label{sect:hotcno}

Hydrogen-burning through the CNO cycle operates efficiently during the PIE in the thermal pulse as well as at the bottom of the convective envelope for our 3 \Msun\, models (hot bottom-burning). We briefly review  its impact on the surface abundances below.

\subsubsection{In the pulse during a PIE }

The degree of CNO processing during a PIE varies from one model to another. 
In Table~\ref{table:3}, we report some surface abundance ratios after the PIE. 
This pollution results from the interplay between He-burning, H-burning at high temperatures, and dilution with the material in the stellar envelope.
All models but M2.0z3.0 show a $^{12}$C/$^{13}$C ratio very close to the CNO equilibrium value of $3-4$. 
The model that experiences the highest degree of CNO processing during the PIE is the M1.0z3.0 model (as seen from the low  $^{12}$C/$^{13}$C, C/N, and O/N ratios) because of the absence of a proper split in this model (cf. Sect.~\ref{sect:splitnosplit}). As a result, the temperature at the bottom of the pulse remains high for a longer period allowing for further CNO processing (hence, the strength of H-burning reactions). 
In the other models, when the split occurs, the temperature at the bottom of the upper part of the pulse strongly decreases, so that H-burning is slowed down. 
The drop in temperature depends on where the splits occurs. The pulse of the M2.0z2.5 model splits close to the bottom of the pulse. Consequently, the CNO cycle remains efficient in the upper part of the pulse of this model. This results in a stronger CNO cycle signature at the surface (Table~\ref{table:3}). 
In all the models (except M1.0z3.0 and M2.0z2.5), the C/N and O/N ratios are greater than 1, which means that the CNO cycle did not reach equilibrium otherwise $^{14}$N would have been the most abundant isotope and both C/N~$<1$ and O/N~$<1$.
In the end, we found that models with no or deep split are more prone to show a very high degree of CNO processing at their surface. 

After a PIE, however, the AGB phase may resume and the surface composition can still change. In the M2.0z3.0 and M2.0z2.5 models for instance, after the decrease in the $^{12}$C/$^{13}$C ratio caused by the PIE, the ratio progressively increases again as a result of the enrichment of the envelope with the pulse material (especially $^{12}$C), thanks to the operation of the third dredge up (Fig.~\ref{fig:CCall}).

\subsubsection{Hot bottom-burning}

The CNO cycle is also activated if the temperature at the base of the convective envelope rises above $\sim 50$~MK \citep[e.g.][]{karakas14}. 
In PIE models, right after the merging of the pulse with the envelope, the temperature at the bottom of the envelope can reach $\sim 90$~MK (Fig.~\ref{fig:tbase}), but this lasts for a short period of time (less than 1~yr in our M1.0z3.0 model, Fig.~\ref{fig:ratio_bot}), so that it barely affects the envelope composition. 

Our 3 \Msun\, models experience hot bottom-burning, although the temperature at the base of its envelope barely reaches 50~MK (Fig.~\ref{fig:tbase}, bottom panel). 
It nevertheless leaves a clear signature at the stellar surface: a progressive decrease in the $^{12}$C/$^{13}$C ratio (Fig.~\ref{fig:CCall} top panel) and to a lesser extent, of the C/N ratio.
The moderate envelope temperature prevents a full activation of the  Ne-Na and Mg-Al cycles.

\subsection{Effect of $\alpha-$enhancement}
\label{sect:alpha2}

\begin{figure}[t]
\includegraphics[width=\columnwidth]{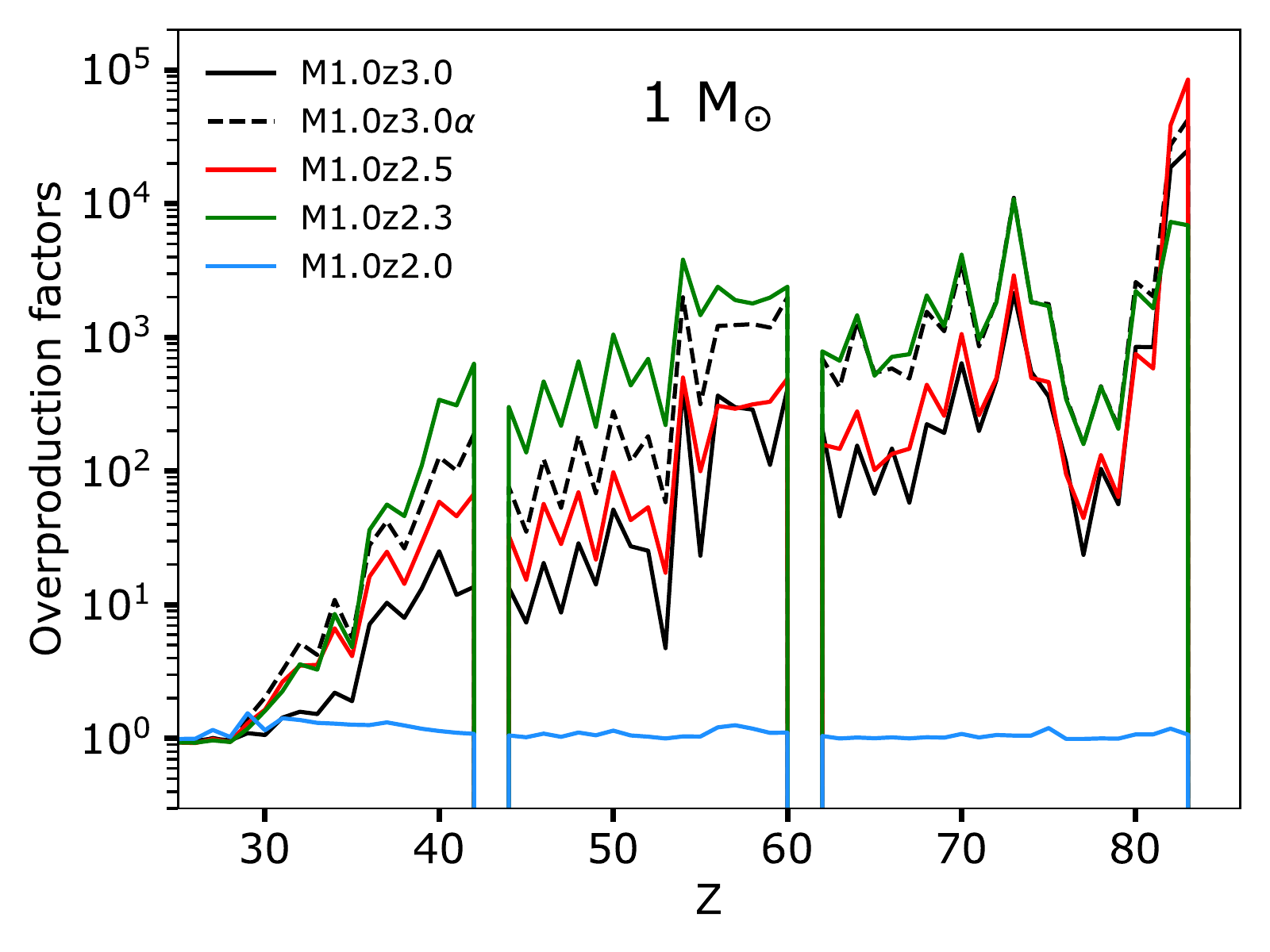}
\includegraphics[width=\columnwidth]{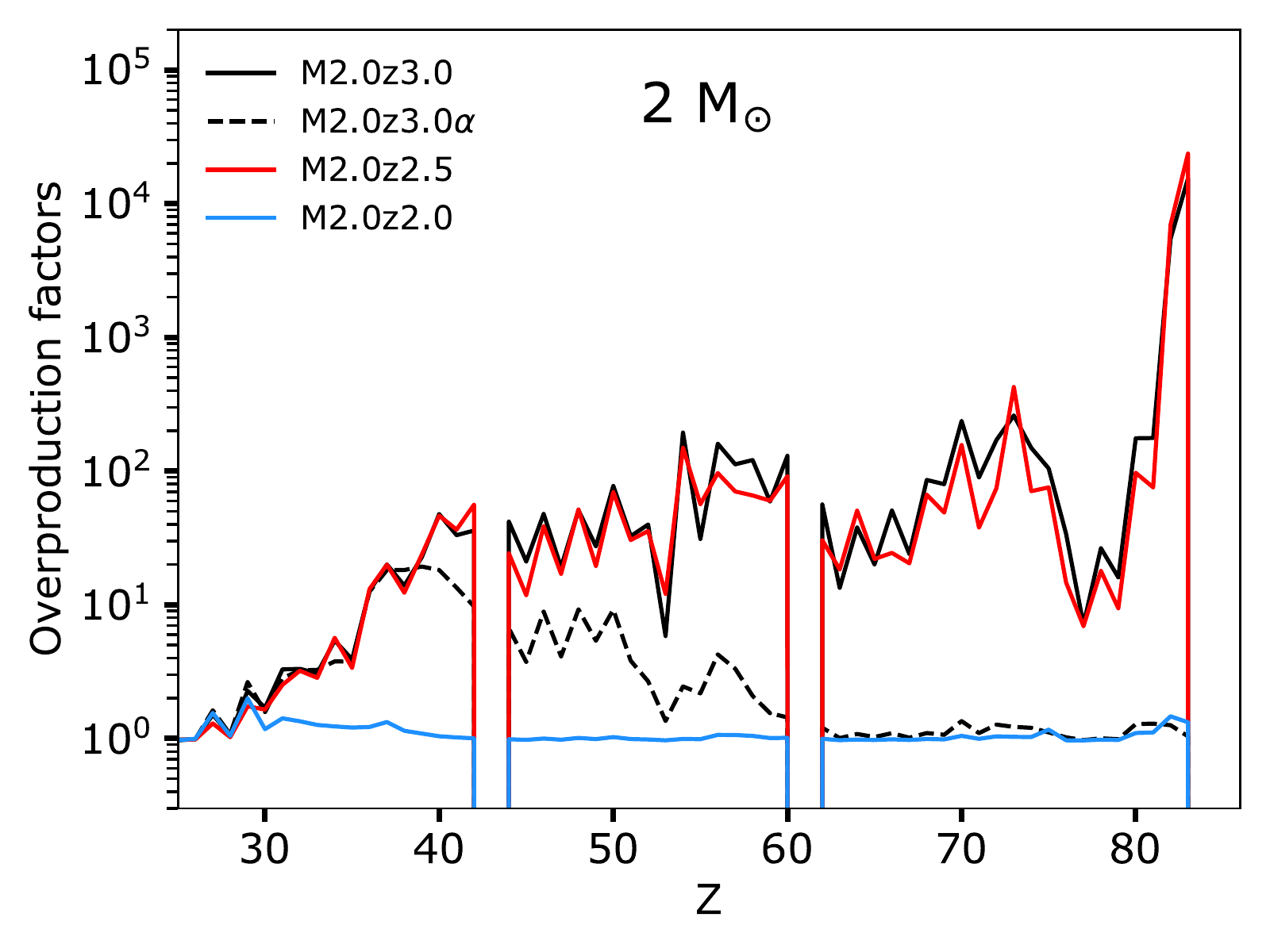}
\includegraphics[width=\columnwidth]{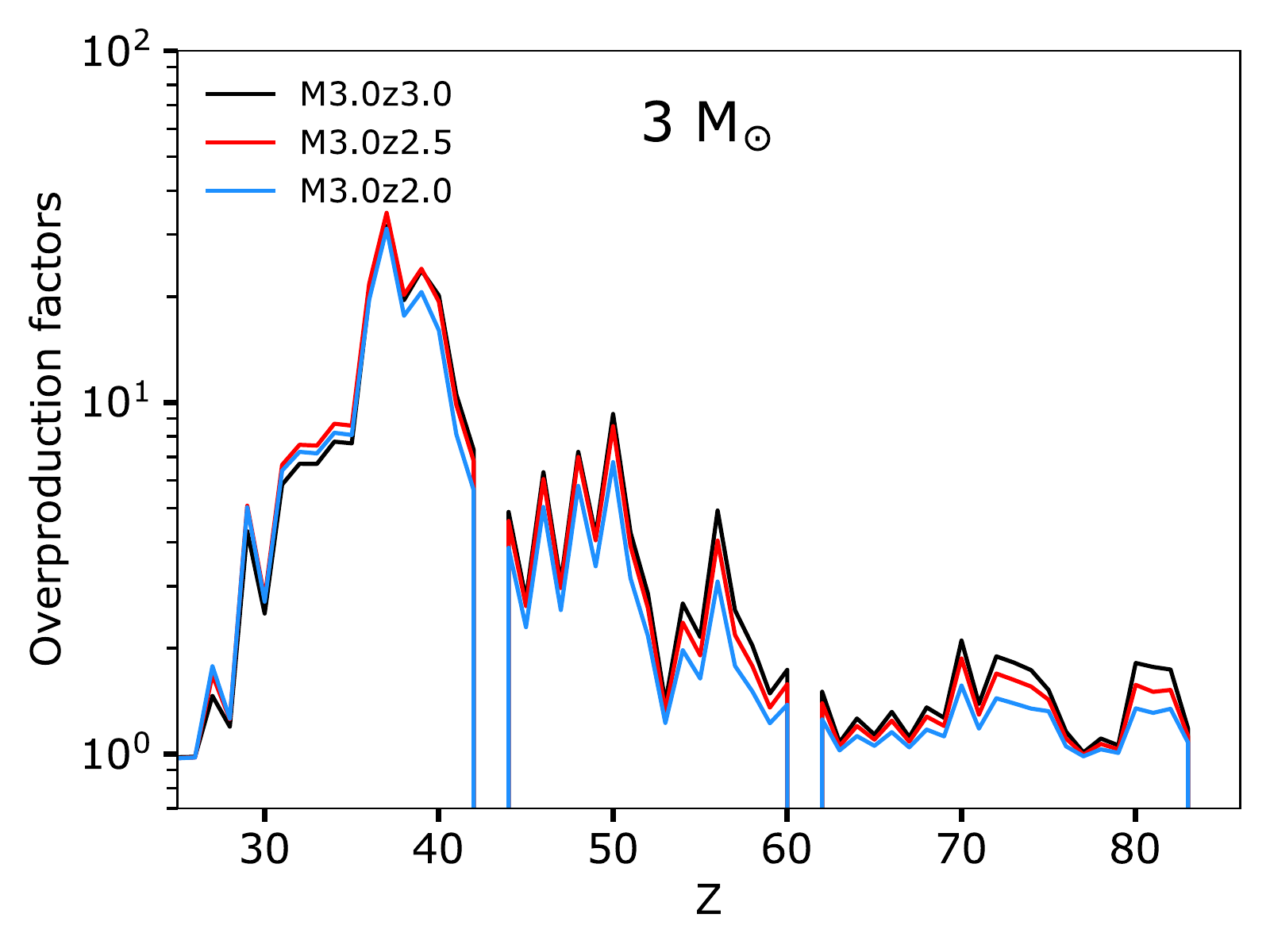}
\caption{Global overproduction factors (Eq.~\ref{eq:op}) for our 1 \Msun\ models (top panel), 2 \Msun\ models (middle panel), and 3 \Msun\ models (bottom panel). 
}
\label{fig:xfe}
\end{figure}

The M1.0z3.0$\alpha$ model was computed with an $\alpha$-enhanced mixture (cf. Sect.~\ref{sect:inputs}). 
This model does not behave similarly to the M1.0z3.0 model but, instead, it resembles the M1.0z2.5 and M1.0z2.3 models of higher [Fe/H]. It experiences a PIE during the second pulse and loses all its mass quickly, aborting a standard TP-AGB phase. 
This is because its metallicity in mass fraction is similar to those of the M1.0z2.5 and M1.0z2.3 models. 
The chemical yields of this model follow the same trend as the other 1 \Msun\, models experiencing a PIE (Fig.~\ref{fig:xfe}).

Unlike the M2.0z3.0 model, the M2.0z3.0$\alpha$ model does not experience a standard PIE (as the models in Table~\ref{table:2}), as the PIE is prevented due to its higher metallicity (Table~\ref{table:1}). 
This model lies just above the light grey zone in Fig.~\ref{fig:mzplot}. 
However, during the second and third thermal pulses, this model experiences weak proton ingestion events with maximal neutron densities of $10^{11} - 10^{12}$~cm$^{-3}$. Unlike a standard PIE, the AGB structure and evolution is not affected. 
During these weak PIEs, first-peak heavy elements are synthesized in the pulse and later brought up to the surface by the third dredge-up. The resulting yields are shown by the dashed line in the middle panel of Fig.~\ref{fig:xfe}. 
We note that, like the other 2~\Msun\, models, the M2.0z3.0$\alpha$ model does not experience an efficient convective s-process because the temperature at the bottom of the pulse is too low (about 325~MK at maximum; see also Sect.~\ref{sec:convspro}).

Our findings are consistent with those of \cite{cristallo16} who also concluded that the $\alpha$-enhancement tends to suppress the PIE. 
The higher C and O abundances in the H-shell lead to lower H-burning temperatures and, thus, to an increase in the entropy difference between the H- and He-burning shells. 
In these conditions, the PIE is hampered.

\subsection{Helium-rich ejecta}
\label{sect:he4}

\begin{figure}[t]
\includegraphics[width=\columnwidth]{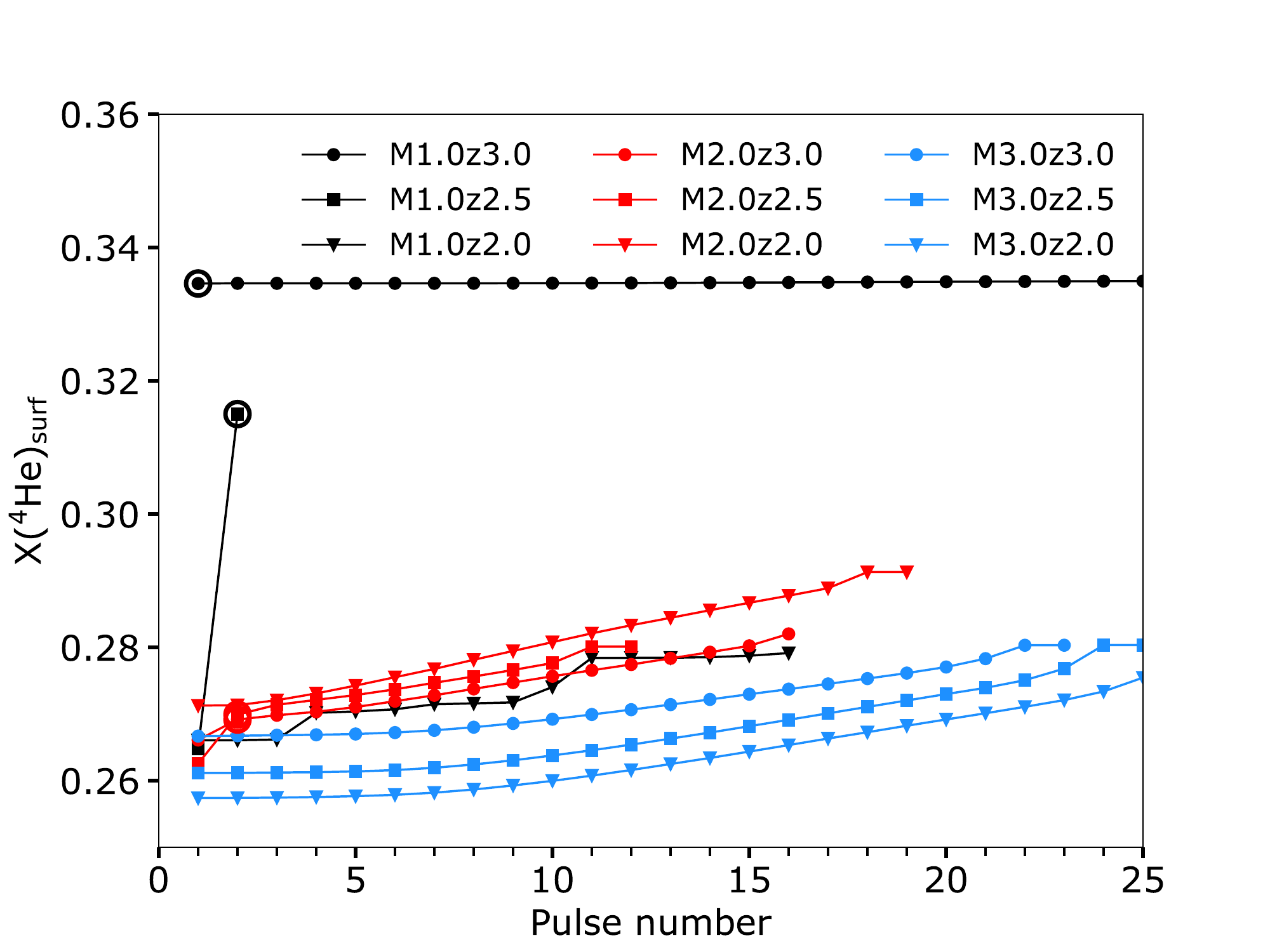}
\caption{
Surface $^{4}$He mass fraction as a function of the pulse number. 
The mass fraction is shown after each thermal pulse, just after the third dredge up, if any.
}
\label{fig:he4s}
\end{figure}

After a standard thermal pulse, some material of the pulse is mixed in the convective envelope provided the third dredge up occurs. 
In contrast, after a PIE, the whole helium-rich convective pulse merges with the hydrogen-rich convective envelope. 
The envelope He enrichment is clearly visible in some of our models (e.g. the M1.0z3.0 and M1.0z2.5 models, Fig.~\ref{fig:he4s}). Right after the PIE, the surface helium mass fraction rises up to about 0.34 (0.32) in the M1.0z3.0 and M1.0z2.5 models, respectively. 
Although present, this effect is much less pronounced in the M2.0z3.0 and M2.0z2.5 models because the helium contained in the pulse is diluted in a much bigger envelope and the pulse is less massive.

Although the detecton of helium is very challenging, there have been several recent direct detections of helium enrichment in stars. 
\cite{pasquini11} managed to estimate the abundance for two giants stars ([Fe/H] = $-1.22$ and $-1.08$) in the globular cluster NGC 2808. 
They found an enrichment of 0.17 in helium mass fraction, meaning that if one of the two star has a solar He abundance, the other would have  X($^{4}$He)~$\sim 0.42$.  
\cite{dupree13} reported helium abundances of X($^{4}$He) $< 0.22$ and $0.39 <$~X($^{4}$He)~$< 0.44$ in two giant stars ([Fe/H] = $-1.86$ and $-1.79$) belonging to the globular cluster $\omega$ Centauri. 
In NGC 2808, \cite{marino14} found X($^{4}$He)~$= 0.34$ in a sample of horizontal branch stars with metallicities $-1.5<$~[Fe/H]~$<-1$. 
The origin of this helium-rich population in Globular Clusters is debated. Possible polluters are intermediate-mass AGB stars \citep[between $\sim 3$ and $\sim 7$~\Msun][]{ventura09}, super-AGB stars \citep{Pumo2008,Siess2010} massive rotating stars \citep{decressin07}, or massive binary stars \citep{demink09}. 
The present work suggests that low-mass AGB stars experiencing a PIE may also be interesting candidates. 
Nevertheless, PIEs do not seem to occur in AGB stars of metallicity higher than [Fe/H]~$\simeq -2.3$ (Fig.~\ref{fig:mzplot}), while some helium-rich stars in globular clusters have [Fe/H]~$\simeq -1$ (cf. previous discussion). 
We recall, however, that Fig.~\ref{fig:mzplot} mostly considers models without extra mixing processes. It remains to be seen if AGB models including overshoot could experience PIEs at [Fe/H]~$\simeq -1$. 
This possibility will be explored in a future work.

 \begin{figure*}[t]
\includegraphics[scale=0.9, trim = 3.5cm 1cm 2cm 1cm]{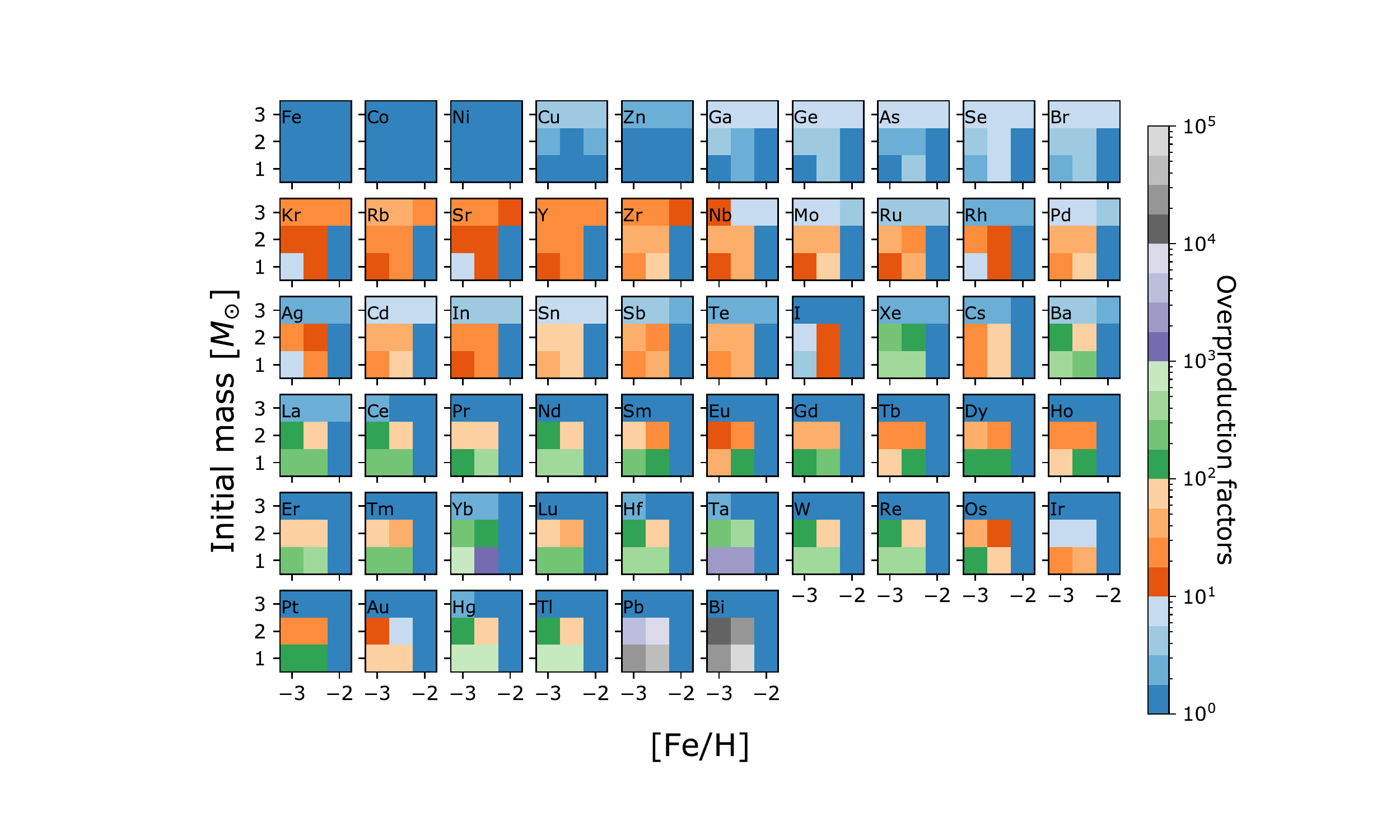}
\caption{ Global overproduction factors (Eq.~\ref{eq:op}) for the nine models with $M = 1$, 2 and 3 \Msun\, and [Fe/H]~$=-3$, $-2.5$ and $-2$. 
In each panel (corresponding to a given element), the x-axis (y-axis) shows the initial metallicity (initial mass) of the AGB model. 
The overproduction factors are color-coded according to the colorbar on the right.
}
\label{fig:yieldsMASSMET}
\end{figure*}

\subsection{Yields and i-process signatures}
\label{sect:yields}

We computed the yields and associated over-prodution factors of our models according to Eq.~\ref{eq:yie} and Eq.~\ref{eq:op}.
Results are displayed in Figs.~\ref{fig:xfe} and \ref{fig:yieldsMASSMET}. 

At [Fe/H]~$=-2$, neither the 1 \Msun\,, nor the 2 \Msun\, show any enhancement in heavy elements (blue patterns). Unlike all other 1 and 2 \Msun\, models, these two models do not experience any PIE and their pulse temperature is two low to experience significant convective s-process (cf. Sect.~\ref{sec:convspro}). We also recall that in the absence of additional mixing below the convective envelope, our models do not experience radiative s-process nucleosynthesis during the interpulse phase.

As discussed in Sect.~\ref{sec:convspro}, the 3 \Msun\, models experience an efficient convective s-process which leads to the production of the light s-process elements Sr, Y, and Zr with similar overproduction factors (Fig.~\ref{fig:xfe}, bottom panel). The other models have indeed experienced a PIE and the yields result from a complex interplay between the  characteristics of the PIE (mass of ingested protons, pulse temperature, maximum neutron density, split, etc.), the efficiency of the third dredge up (the $\lambda$ parameter), and the dilution of the pulse material into the envelope of variable size and composition.

Despite these differences, all these i-process models show a similar rising distribution from Fe to Bi (Fig.~\ref{fig:xfe}, top, and middle panels).
The overproduction factors of the 2~\Msun\ models are lower on average than those of the 1~\Msun\ models because of the larger dilution of the pulse material.
Models at metallicity [Fe/H]~$-3.0$ (solid black) and $-2.5$ (solid red) give similar overproduction factors but the yields in \Msun\ are $\sim 3$ times higher in the [Fe/H]~$-2.5$ models because of the difference in metallicity. 
The M1.0z2.3 model (green) shows higher overproduction factors except for Pb and Bi. The highest overproduction factor is found for Tantalum ($Z=73$) in this model.

Elements above Nb are only synthesized by low-mass  low-metallicity models (Fig.~\ref{fig:yieldsMASSMET}). In contrast, lighter elements are also synthesized by 3 \Msun\ models, thanks to the convective s-process (cf. Sect.~\ref{sec:convspro}). 
Nuclei between Ba and Bi are more produced in the 1 \Msun\ models than in 2 \Msun\ models.
Xenon (Xe), ytterbium (Yb), tantalum (Ta), lead (Pb), and bismuth (Bi) are among the most produced species. 
A key feature of these models is the high xenon over iodine ratio (also visible in Fig.~\ref{fig:xfe}, top and middle panels, at $Z=53-54$).
Below iodine ($Z=53$), the overproduction factors do not exceed $10^2$  (blue and orange squares). Above xenon ($Z=54$), the overproduction factors are often in the range of $10^2-10^3$ (green) and sometimes higher (purple and grey).
The high overproduction of elements with $Z>53$ compared to elements with $Z\leq 53$ is a significantly different feature that is observed between the i- and s-processes \citep[as shown in][their Figure 7]{choplin21}.



\section{Summary and conclusions}
\label{sect:concl}

In this third paper of the series, we study the i-process originating from a PIE in AGB stars of 1, 2, and 3~\Msun\, at metallicities of [Fe/H]~$=-3$, $-2.5$, $-2.3,$ and $-2$. 
We used the code {\sf STAREVOL} with a network of 1160 species and provide the first i-process yields\footnote{The yields are available online at \url{http://www.astro.ulb.ac.be/~siess/Site/StellarModels}} from a grid of AGB models experiencing PIEs. 
We note that extra mixing was not considered in these calculations.

We find that PIEs happen in six out of our 12 AGB stars models, preferentially at low-mass and low-metallicity. PIEs arise in 1 and 2~\Msun\, AGB models with [Fe/H]~$=-3$, $-2.5,$ and $-2.3$ during the first or second thermal pulse. 
In these models, $\simeq 10^{-6} - 10^{-4}$~\Msun\, of hydrogen is ingested in the convective pulse. 
This activates hydrogen-burning at very high temperature, triggers i-process nucleosynthesis characterized by neutron densities of  $\simeq 10^{14} -10^{15}$~cm$^{-3}$. This event strongly impacts the stellar structure and subsequent AGB evolution.

In our $1$~\Msun\, models with [Fe/H]~$=-2.5$ and $-2.3$, the AGB phase ends right after the PIE because of the strong effect of molecular CO opacities that leads to the quick loss of the whole convective envelope before any further thermal pulse can develop. 
Our $1$~\Msun\, model with [Fe/H]~$=-3$ is special in the sense that the PIE happens during the very first thermal pulse, at relatively low temperature. This leads to a different evolutionary pathway: the surface is less enriched, the molecular CO opacities smaller and hence the mass loss weaker. 
This results in the resumption of the AGB phase with $\sim 40$ further thermal pulses after the PIE. 
In our $2$~\Msun\, models, the PIE pulse material is more diluted in the more massive envelope, reducing the surface enrichment and allowing for the development of about ten subsequent thermal pulses. 

All our PIE models, except the $1$~\Msun\, model with [Fe/H]~$=-3,$ experience a proper split of the convective pulse. 
We highlighted the dependence of the pulse splitting on the convective velocity (the lower the convective velocity, the higher in mass the split occurs) and we show that (at least in our models), the resulting surface enrichment does not depend on the convective velocity unless it is unrealistically low. 
This result suggests that PIEs can happen similarly under various convective recipes or assumptions. 
In addition, we showed that in all our PIE models, there are more than 50 turnover timescales (the $e$ factor, cf. Eq.~\ref{eq:eratio} and Table~\ref{table:2}) between i-process nucleosynthesis and the split. This means that the surface of our AGB models is fully enriched in i-process products right after the PIE.

The yields of our PIE models show similar rising trends with overproduction factors of up to $10^5$ for Pb and Bi.
The elements Xe, Yb, Ta, Pb, and Bi show the highest overproduction factors. 
Also, overproduction factors of elements with $56<Z<83$ are higher in 1 \Msun\ than in 2 \Msun\, AGB models. Our $1$~\Msun\, PIE models are also strong producers of helium. 
Finally, we found that our $1$~\Msun\, model with [Fe/H]~$=-3$ and $\alpha$-enhanced mixture behaves like our $1$~\Msun\, model with [Fe/H]~$=-2.5$ and $=-2.3$ of similar metallicity  (although a lower [Fe/H]). 

Our 3~\Msun\, models do not experience i-process but convective s-process and efficient third dredge up events. Their yields consequently show overproduction from the first neutron peak with factors of up to about 40.

This study has shown that AGB stars of low mass and low metallicity can undergo PIE and eject heavy elements made from i-process nucleosynthesis. The evolutionary pathway after a PIE depends on the details of the PIE itself (e.g. pulse where it occurs, mass of protons ingested).
It is interesting to note that our AGB models make heavy elements without any parametrised extra mixing process such as overshoot and without the artificial inclusion of a $^{13}$C-pocket. 
Nevertheless, similar models with extra mixing deserve to be computed and analyzed. 
Overshooting could take place at the bottom of the envelope and at the bottom and top of the convective pulse. As it enhances   mixing, it may facilitate ingestion of protons into the convective pulse. But the modelling of this process also include badly constrained parameters; the impact of overshoot on proton ingestion events is therefore complex and deserves a detailed study. This will be explored in a future work.


\section*{Acknowledgments}
This work was supported by the Fonds de la Recherche Scientifique-FNRS under Grant No IISN 4.4502.19. 
L.S. and S.G. are senior FRS-F.N.R.S. research associates.



\bibliographystyle{aa}
\bibliography{astro.bib}


\end{document}